\documentclass[11pt]{article}
\usepackage{graphicx}

\usepackage{natbib}
\usepackage{amsfonts}
\usepackage{amsmath}
\usepackage{amssymb}
\usepackage{url}
\usepackage{fancyhdr}
\usepackage{indentfirst}
\usepackage{enumerate}
\usepackage{titlesec}
\usepackage{amsthm}
\usepackage{dsfont}
\usepackage{multicol}
\usepackage{blindtext}
\usepackage{geometry}
\usepackage{appendix}
\usepackage{algorithm}
\usepackage{algpseudocode}
\usepackage{cases}
\usepackage{bm}
\usepackage{tikz-qtree}
\usepackage{tikz}
\usepackage[onehalfspacing]{setspace}

\usepackage[colorlinks=true,citecolor=blue]{hyperref}

\theoremstyle{definition}

\newtheorem{example}{Example}

\newtheorem{definition}{Definition}

\newtheorem{verification}{Verification}

\newtheoremstyle{propertystyle}
  {\topsep}      
  {\topsep}      
  {\normalfont}  
  {}             
  {\bfseries}    
  {:}            
  {0.5em}        
  {}             

\theoremstyle{propertystyle}
\newtheorem{property}{Property}

\theoremstyle{plain}
\newtheorem{theorem}{Theorem}
\newtheorem{lemma}{Lemma}
\newtheorem{proposition}{Proposition}
\newtheorem{corollary}{Corollary}
\theoremstyle{remark}
\newtheorem{remark}{Remark}

\theoremstyle{definition}
\def\id{\mathds{1}}

\def\N{\mathbb{N}}

\def\R{\mathbb{R}}

\def\d{\mathrm{d}}

\title{A unified decision model for Bayesian ambiguity and non-affine statistics}

\author{Zachary Van Oosten\thanks{Department of Decision Sciences, HEC Montréal, Canada. {\scriptsize\url{zachary.van-oosten@hec.ca}}} \and Ruodu Wang\thanks{Department of Statistics and Actuarial Science, University of Waterloo,  Canada. {\scriptsize\url{wang@uwaterloo.ca}}}}
\date{\today}

\begin{document}
	\maketitle
	\begin{abstract}
This paper studies decision functionals under ambiguity when pure-risk evaluation is determined by a general, possibly non-affine statistic. We represent ambiguity perception by a Bayesian second-order model over first-order models and encode ambiguity attitude with a separate parameter. We formulate four consistency and monotonicity properties for decision functionals combining these objects. We show that second-order averaging does not generally preserve these properties for non-affine statistics, including expected shortfall, the industry-standard risk measure. To address this limitation, we introduce an ambiguity iron, a mapping from a second-order model and an ambiguity attitude to an ambiguity set of first-order models. Worst-case evaluation over this set preserves the chosen statistic as the pure-risk criterion. We construct an ambiguity iron by distorting the second-order model and show that it produces convex and compact sets of predictive models. We derive formulas for several classes of statistics and, in parametric settings, a sample-based approximation for the distorted expectation. Portfolio examples illustrate how ambiguity perception and attitude affect evaluations and allocations. Under posterior concentration and continuity of the parametrization, the ambiguity sets contract, and the associated decision functionals converge to their pure-risk values. We also describe a KL-based robustification for possible model misspecification.

	\end{abstract}

\noindent\rule{\textwidth}{0.5pt}

\section{Introduction}

Decision functionals assign numerical values to choice objects under uncertainty. 
In this paper, choice objects are real-valued acts. 
We begin with a simple construction of decision models, where the decision maker's preferences depend on two ingredients: a statistic that represents how the DM evaluates distributions and a first-order model that specifies the probability description of future states. An act is evaluated by applying the statistic to its distribution under the first-order model. This construction separates an act's probabilistic description from its evaluation and appears, for example, in portfolio selection \citep{M52}, statistical decision theory \citep{S54}, and financial risk management \citep{ADEH99}.

This construction presumes that the decision maker (DM) can identify a unique first-order model. Following \cite{K21}, we distinguish between \emph{pure risk}, where this model is known, and \emph{ambiguity}, where it cannot be uniquely identified. The distinction is reflected in the behavior described by \cite{E61}. Under ambiguity, a decision functional should incorporate both the perception of ambiguity and the DM's reaction to it. These components need not coincide: the same perception of ambiguity may combine with different ambiguity attitudes, while the same attitude may apply to different perceptions.

Two common approaches emphasize different parts of this problem. Distributionally robust optimization (DRO) represents ambiguity by a set of first-order models, called an ambiguity set, and evaluates an act by its worst-case value over that set. Rooted in the maxmin expected utility model of \cite{GS89}, this approach is used in portfolio selection \citep{GOO03,NPS08,ZF09}, risk measurement \citep{DKW19,FLW24, MWW26}, robust optimization \citep{DY10,BTDDMR13}, and actuarial science \citep{BLTY19,BP19}.
It allows the same ambiguity set to be combined with different statistics. The set must nevertheless be specified, and the construction alone does not determine how it changes with new information. A Bayesian approach instead represents uncertainty about the first-order model with a second-order model: a probability measure over first-order models. Bayesian updating then provides a learning mechanism. An ambiguity-neutral benchmark averages the second-order model to obtain the predictive first-order model and evaluates acts under that model. This benchmark is used in financial risk management \citep{A08,AKT17}, actuarial science \citep{K92,C00}, and portfolio allocation \citep{KS96,FH22}. However, it does not encode attitudes toward ambiguity.

The smooth ambiguity model of \cite{KMM05} combines second-order beliefs with ambiguity-sensitive evaluation. In its expected-loss formulation, the inner loss function represents pure-risk preferences, the second-order model represents ambiguity perception, and an outer loss function represents ambiguity attitude. This separation has proved useful in economic and financial applications \citep{G11,JM12,MMR13}. However, its benchmark properties rely on using the expected-loss statistic (which is expected utility when modeling wealth). Direct substitution of different statistics therefore need not preserve the desirable properties, as we show in this paper.

To the best of our knowledge, no existing model can simultaneously accommodate a general pure-risk statistic, represent ambiguity perception, capture ambiguity attitudes separately, and support Bayesian updating. Our goal is to construct ambiguity-sensitive decision functionals that retain these properties while allowing for general statistics.

We fix a statistic representing pure-risk preferences and formulate four properties for decision functionals indexed by a second-order model and an ambiguity attitude. Pure-risk consistency requires that the functional reduces to the statistic when the second-order model is degenerate. Predictive ambiguity neutrality requires evaluation under the predictive model at the ambiguity-neutral benchmark. Attitude monotonicity orders functionals as the DM becomes more ambiguity averse. Strong ambiguity aversion requires monotonicity when second-order beliefs become more dispersed in convex order. Together, these properties separate pure-risk preferences, ambiguity perception, and ambiguity attitude.

Our first set of results clarifies the obstruction created by non-affine statistics. Within the reduction rules considered, second-order averaging satisfies ambiguity neutrality only when the pure-risk statistic is expected loss, and the reduction is the predictive model. Moreover, averaging model-specific expected shortfalls moves in the direction opposite to strong ambiguity aversion as second-order beliefs become more dispersed. We therefore introduce an \emph{ambiguity iron} that maps a second-order model and an ambiguity attitude to an ambiguity set. Applying the original statistic to every model in this set and taking the worst-case value yields a DRO decision functional. The defining properties of an ambiguity iron ensure the four properties above.

We develop a specific ambiguity iron by distorting second-order models. A concave distortion determines a collection of second-order models dominated by the corresponding distorted probability, and the ambiguity set consists of their predictive first-order models. These distortions represent ambiguity attitudes. We show that this construction defines an ambiguity iron and produces convex, compact ambiguity sets. In a binary-state example, the ambiguity set is an interval of predictive probabilities whose width increases with ambiguity aversion. The example also distinguishes this interval, constructed for decision-making, from a Bayesian credible interval, constructed for posterior coverage.

We next study computation, learning, and model misspecification. We derive alternative formulas for the resulting worst-case functionals and, under a parametric representation, an approximation based on order statistics drawn from the second-order model. We specialize these formulas to optimized certainty equivalents, expected loss, and value-at-risk, and illustrate them in a portfolio problem using four statistics. We then show that posterior concentration, together with continuity of the parametrization in total variation, makes the induced ambiguity sets contract and the decision functionals converge to their pure-risk values. A separate KL-based robustification addresses model misspecification, thereby distinguishing misspecification from ambiguity perception and attitude.

The remainder of the paper is organized as follows. Section \ref{sec:2} introduces the main objects and the predictive model. Section \ref{sec:3} formulates the four properties and examines the limitations of second-order integral aggregation. Section \ref{sec:4} introduces ambiguity irons, develops the distortion construction, and derives alternative formulas for several statistics. Section \ref{sec:7} studies unambiguous events and acts, including epistemic distributional invariance. Section \ref{sec:5} presents the parametric representation, sampling-based approximation, and numerical illustrations. Section \ref{sec:6} studies Bayesian consistency and model misspecification. The appendices contain the proofs, additional results, and supplementary figures.

\section{Preliminaries}

\label{sec:2}

Under pure risk, evaluating an act requires both a statistic and a first-order model: the model determines the act's distribution, while the statistic specifies how to evaluate that distribution. Under ambiguity, a second-order model represents uncertainty about the first-order model. This section introduces these objects and distinguishes their roles in the decision problem. We also define the predictive model obtained by averaging first-order models under the second-order model. This reduction will provide the benchmark for ambiguity-neutral evaluation.

\subsection{General notation}

Given measurable spaces $(\Omega,\mathcal{F})$ and $(\Psi,\mathcal{G})$, a measurable mapping $X:\Omega\to \Psi$, and a probability measure $\mu$ on $(\Omega,\mathcal{F})$, the \emph{pushforward probability measure} $X_{\#}\mu$ on $(\Psi,\mathcal{G})$ is defined by
$$X_{\#}\mu(A)=\mu(X^{-1}(A)),~~A\in \mathcal{G}.$$

A \emph{Polish measurable space} is a measurable space $(\Psi,\mathcal{G})$ where $\Psi$ is a Polish topological space and $\mathcal{G}$ is its Borel $\sigma$-algebra. We will sometimes denote the Borel $\sigma$-algebra by $\mathcal{B}(\Psi)$.  Given a Polish measurable space $(\Psi,\mathcal{G})$ and a probability measure $\mu$ on $(\Psi,\mathcal{G})$, the \emph{support} of $\mu$ is the smallest closed set $C\subseteq \Psi$ such that $\mu(C)=1$. Given a Polish measurable space $(\Psi,\mathcal{G})$ and a sequence of probability measures $(\mu_n)_{n\in \N}$ on $(\Psi,\mathcal{G})$, we say that $(\mu_n)_{n\in \N}$ \emph{converges in distribution} to some probability measure $\mu$ on $(\Psi,\mathcal{G})$, denoted by $\mu_n\Rightarrow \mu$, if for all continuous and bounded $f:\Psi\to\mathbb{R}$ we have
$\lim_{n\to\infty}\int_\Psi f\d\mu_n=\int_\Psi f \d\mu.$

\subsection{Acts, decision functionals and first-order models}

 Throughout the paper, we fix a Polish measurable space $(\Omega,\mathcal{F})$, which is interpreted as the set of future states of the world. An \emph{act} is a real-valued measurable function on $(\Omega,\mathcal{F})$. In this paper, an act is to be viewed in the loss-profit sense (positive values correspond to losses or bad outcomes). Acts are the main objects of choice studied in this paper and, depending on the context, may represent financial losses, insurable risks, or unrealized endowments. We denote by $\mathcal{X}$ the set of all acts, by $\mathcal{X}_b$ the set of all bounded acts, by $\mathcal{C}$ the set of all continuous acts, and by $\mathcal{C}_b$ the set of all continuous bounded acts. Given $X,Y\in \mathcal{X}$, we write $X\geq Y$ if $X(\omega)\geq Y(\omega)$ for all $\omega\in \Omega$.
    \begin{definition}
Let $\mathcal{Y}\subseteq\mathcal{X}$ contain all constant acts. A \emph{decision functional} is a map
$\rho:\mathcal{Y}\to\mathbb{R}\cup\{\infty\}$ for which $\rho(X)\geq\rho(Y)$ whenever $X,Y\in\mathcal{Y}$ satisfy $X\geq Y$, and $\rho(c)=c$ for every $c\in\mathbb{R}$, where $c$ is identified with the corresponding constant act.
\end{definition}

Decision functionals are operational tools for evaluating acts and making decisions. Their numerical output can serve as a stand-alone assessment or as part of an optimization problem. For example, the value of a decision functional may represent an economically meaningful quantity, such as a financial institution's capital requirement or the premium charged to a policyholder for insuring an uncertain loss. Alternatively, it may rank feasible choices, as in portfolio selection, where one chooses a portfolio by optimizing the value of the decision functional. In economics, decision functionals are used to represent preference relations and, more specifically, represent preferences by a functional that maps acts to their certainty equivalents.

Let $\Delta$ denote the set of probability measures on $(\Omega,\mathcal{F})$; we call its elements \emph{first-order models}. First-order models represent aleatoric uncertainty or \emph{pure-risk perception}, namely the inherent randomness in which state of the world will be realized. Given $X\in \mathcal{X}$ and $\mu\in \Delta$, we use $\mathbb{E}^{\mu}[X]$ to denote the Lebesgue integral of $X$ under $\mu$. Given $X\in \mathcal{X}$, define the function $\psi_X:\Delta\to\mathbb{R}\cup\{-\infty,\infty\}$ by $\psi_X(\mu)=\mathbb{E}^{\mu}[X]$.\footnote{When $\psi_X(\mu)$ is undefined, set it equal to zero.} We equip $\Delta$ with the coarsest topology $\tau$ where the maps $\{\psi_X\}_{X\in \mathcal{C}_b}$ are continuous, i.e., the weak topology. Given $\mu,\nu\in \Delta$ and $\lambda\in [0,1]$, their $\lambda$-mixture, denoted by $\lambda\mu+(1-\lambda)\nu$, is the first-order model defined by
$$\left[\lambda\mu+(1-\lambda)\nu\right](A)=\lambda\mu(A)+(1-\lambda)\nu(A),~~A\in \mathcal{F}.$$
Equipped with this operation, $\Delta$ is a mixture space. Throughout, all references to convex subsets of $\Delta$ and convex, concave, or affine functions defined on $\Delta$ are understood with respect to this mixture structure.

A DM is said to be in the \emph{pure-risk setting} when they can, either through personal beliefs or statistical inference, identify a unique first-order model $\mu\in\Delta$ that they regard as the probability description governing the realization of future outcomes. Alternatively, we can say that the DM \emph{perceives no ambiguity}. In this setting, decisions are evaluated using functionals that depend on an act only through its distribution under $\mu$. We now introduce the notation needed to formalize this approach.

Denote by $\mathcal{D}$ the set of all Borel probability measures on $\mathbb{R}$; we will refer to $\mathcal{D}$ as the set of \emph{distributions}. Given $Q\in \mathcal{D}$, we define the \emph{survival function of $Q$} by $S_Q(x)=Q((x,\infty))$ for all $x\in \mathbb{R}$. For $Q,P\in \mathcal{D}$, we say that $Q$ is larger than $P$ in \emph{stochastic order}, denoted by  $Q\geq_{\mathrm{st}}P$, if $S_Q(x)\geq S_P(x)$ for all $x\in \mathbb{R}$. We have that $\geq_{\mathrm{st}}$ is a partial order on $\mathcal{D}$. Given $Q,P\in \mathcal{D}$ and $\lambda\in [0,1]$, their $\lambda$-mixture, denoted by $\lambda Q+(1-\lambda)P$, is the distribution defined by
$$\left[\lambda Q+(1-\lambda)P\right](B)=\lambda Q(B)+(1-\lambda) P(B),~~B\in \mathcal{B}(\mathbb{R}).$$
Equipped with this operation, $\mathcal{D}$ is a mixture space. Throughout, all references to convex, concave, or affine functions defined on $\mathcal{D}$ are understood with respect to this mixture structure.

\begin{definition}
    Let $\mathcal{E}\subseteq \mathcal{D}$ contain all Dirac distributions. A \emph{statistic} is a map $\gamma:\mathcal{E}\to\mathbb{R}\cup\{\infty\}$ for which $\gamma(Q)\geq\gamma(P)$ whenever $Q,P\in \mathcal{E}$ satisfy $Q\geq_{\mathrm{st}} P$, and $\gamma(\delta_c)=c$ for every $c\in \mathbb{R}$, where $\delta_c$ denotes the Dirac measure at $c$.
\end{definition}

We use the term \emph{statistic} for a functional that summarizes a distribution by a single value.\footnote{This terminology is used, for example, by \cite{MPST24}. Other common terms for such mappings include ``statistical functionals'' and ``law-invariant risk measures''.} The following are several important statistics from the literature. For each statistic, let $\mathcal{E}\subseteq\mathcal{D}$ denote a domain on which it is well defined; we leave the precise choice of $\mathcal{E}$ implicit.
    \begin{enumerate}[(i)]
        \item the expected value statistic
    $$\mathbb{E}(Q)=\int_{\mathbb{R}}xQ(\d x),~~Q\in \mathcal{E};$$
    \item the expected-loss statistic of \cite{vNM47}
    $$\mathbb{E}_{\ell}(Q)=\ell^{-1}\left(\int_{\mathbb{R}}\ell\d Q\right)=\ell^{-1}\left(\mathbb{E}(\ell_{\#}Q)\right),~~Q\in \mathcal{E},$$
    where $\ell:\mathbb{R}\to\mathbb{R}$ is strictly increasing and continuous, called a \emph{loss function};
    \item the dual statistic of \cite{Y87}
    $$\mathbb{E}^g(Q)=\int_0^\infty g(S_Q(x)) \d x+\int_{-\infty}^0 (g(S_Q(x))-1)\d x,~~Q\in \mathcal{E},$$
    where $g:[0,1]\to[0,1]$ is increasing, $g(0)=0$, and $g(1)=1$, called a \emph{distortion function}. 
    \end{enumerate}

    \begin{remark}
        We use the expected loss convention rather than the usual expected utility because we are working with loss-profit acts. We can switch conventions by taking $u(x)=-\ell(-x).$
    \end{remark}
    
    Under the distortion $g_\alpha(x)=\frac{x}{1-\alpha}\wedge 1$, where $\alpha\in [0,1)$, the dual statistic $\mathbb{E}^{g_\alpha}$ coincides with \emph{expected shortfall} of \cite{AT02}, which we denote by $\mathrm{ES}_{\alpha}$. Expected shortfall is widely used in financial practice and serves as the standard risk measure for market risk within the banking regulatory framework of \cite{BASEL19}. Under the distortion $h_{\alpha}=\id_{(1-\alpha,1]}$, where $\alpha \in (0,1)$, the dual statistic $\mathbb{E}^{h_\alpha}$ coincides with the \emph{value-at-risk}, which we denote by $\mathrm{VaR}_\alpha$. 

We can now formalize decision-making under pure risk. Recall that, in this setting, the DM identifies a unique first-order model $\mu\in\Delta$. Given a statistic $\gamma:\mathcal{E}\to\mathbb{R}\cup\{\infty\}$ also specified by the DM, acts are evaluated using the decision functional $$\rho(X)=\gamma(X|\mu):=\gamma(X_{\#}\mu),~~X\in \mathcal{Y},$$
where the relevant domains $\mathcal{Y}$ and $\mathcal{E}$ are chosen so that the decision functional is well defined.
Thus, each act is evaluated solely through the distribution it induces under the DM's uniquely identified first-order model. Decision functionals produced in this way are said to be \emph{law invariant}; see \cite{MFE15}.

\subsection{Second-order models and predictive reduction}

A natural way to move beyond the pure-risk setting is to replace the uniquely identified first-order model with a collection of models that the DM regards as plausible. The following example illustrates this approach.

\begin{example}\label{ex:DRO}
In many applications, the DM specifies a collection of plausible first-order models $\mathcal{P}\subseteq\Delta$, called an \emph{ambiguity set}. Distributionally robust optimization (DRO) then evaluates an act according to the \emph{worst-case decision functional}
$$\rho(X)=\sup_{\nu\in\mathcal{P}}
\gamma(X|\nu),~~
X\in\mathcal{Y},
$$
where $\gamma$ is the statistic governing the DM's evaluation in the pure-risk setting. Functionals of this form stem from the maxmin expected utility model of \cite{GS89}.
\end{example}

 In this paper, we take a different approach. Rather than requiring the DM to specify an ambiguity set directly, we represent uncertainty about the appropriate first-order model probabilistically. We equip $\Delta$ with the $\sigma$-algebra generated by $\tau$, denoted by $\Sigma$. Since $(\Omega,\mathcal{F})$ is a Polish measurable space, we know that $(\Delta,\Sigma)$ is a Polish measurable space; see \citet[Theorem 6.8]{B13}. We will use $\Pi$ to denote the set of probability measures on $(\Delta,\Sigma)$, which we call the \emph{second-order models}. Second-order models represent epistemic uncertainty or \emph{ambiguity perception}, namely uncertainty regarding which first-order model provides the appropriate probabilistic description of the world. In a Bayesian statistical context, both priors and posteriors are second-order models. This is why we use the term ``second-order model'' rather than ``Bayesian prior'': the latter might suggest that our framework applies only before data collection, whereas it also applies after Bayesian updating and moving to the posterior.

For $\pi\in\Pi$, let $L^1(\pi)$ denote the set of acts $X\in \mathcal{X}$ satisfying $\int_{\Delta}\mathbb{E}^{\nu}[|X|]\pi(\d\nu)<\infty$. We verify that $\psi_X$ is $\Sigma$-measurable for all $X\in \mathcal{X}$ in Appendix \ref{app:proof}. For $\pi\in \Pi$, we define the \emph{predictive model} $\mu_{\pi}\in \Delta$ by
\begin{equation*}
\mu_{\pi}(A) = \int_{\Delta} \nu(A)\pi(\d\nu),~~A\in\mathcal{F}.
\end{equation*}
An application of the dominated-convergence theorem will verify that $\mu_{\pi}\in \Delta$. We verify in Appendix \ref{app:proof} that,  for $\pi\in\Pi$ and $X\in L^1(\pi)$, we have
    \begin{equation}\label{eq:pred}
        \int_{\Delta}\mathbb{E}(X|\nu)\pi(\d\nu)=\int_{\Delta}\mathbb{E}^{\nu}[X]\pi(\d\nu)=\mathbb{E}^{\mu_{\pi}}[X]=\mathbb{E}(X|\mu_\pi).
    \end{equation}
\begin{example}\label{ex:affine}
    Consider a DM who would evaluate acts using the expected value statistic $\mathbb{E}$ in the pure-risk setting, but who perceives ambiguity and represents it through a nondegenerate second-order model $\pi$. The DM may aggregate their model-specific evaluations by integrating them with respect to the second-order model $\pi$. In the present case, this yields the decision functional
    $$\rho(X)=\int_{\Delta}\mathbb{E}(X|\nu)\pi(\d\nu)\overset{\eqref{eq:pred}}{=}\mathbb{E}(X|\mu_\pi),~~X\in \mathcal{Y}.$$
    The resulting decisions therefore coincide with those of a DM in the pure-risk setting who evaluates acts using the same statistic (the expectation statistic) and takes the predictive model $\mu_\pi$ as their unique first-order model.
 We call this behavior \emph{ambiguity neutrality}: although the DM perceives ambiguity, their decisions can be represented as if they did not and instead relied on a single first-order model. This reduction relies on the expectation statistic being affine. For a statistic that is not affine, the integral of model-specific evaluations generally cannot be represented by applying the same statistic to the distribution of acts under the predictive model, or, more generally, to any single first-order model.
\end{example}

\section{Decision functionals that incorporate second-order models}

\label{sec:3}

We now ask how a decision functional should incorporate second-order beliefs when the statistic governing pure-risk preferences is fixed. We first introduce the convex order to compare the dispersion of second-order models. To provide a general basis for assessing different decision functionals, we then formulate four natural properties for such decision functionals. In particular, we identify restrictions on expected-loss aggregation and show why averaging model-specific expected shortfalls can produce a response opposite to ambiguity aversion.

\subsection{The convex order on second-order models}

We begin by introducing the \emph{convex order} on second-order models that formalizes the idea that one model represents a greater degree of perceived ambiguity than another. To this end, define the binary relation $\geq_{\mathrm{cx}}$ on $\Pi$ by
$$\tilde{\pi}\geq_{\mathrm{cx}}\pi\iff \int_{\Delta}f\d\tilde{\pi}\geq \int_{\Delta}f\d\pi,~~\tilde{\pi},\pi\in \Pi$$
for every convex $\tau$-continuous function $f:\Delta\to \mathbb{R}$ for which both integrals are well defined.

\begin{theorem}\label{theo:big}
    The binary relation $\geq_{\mathrm{cx}}$ is a partial order.
\end{theorem}

This construction extends the familiar convex order on distributions in the pure-risk setting; see \cite{RS70}. Whereas the classical convex order compares dispersion over outcomes, this convex order compares the dispersion of second-order models over first-order models. When $\Delta$ is compact, i.e., when $\Omega$ is compact, this relation coincides with the Choquet order discussed in \citet[Chapter 15]{P02}. It is also closely connected to Blackwell's comparison of experiments for a finite family of statistical models. \cite{B53} shows that one experiment is more informative than another if and only if the posterior beliefs over these models induced by the former dominate those induced by the latter in convex, or equivalently Choquet, order. Here, we use the same geometric order for a different purpose: a more dispersed second-order model is interpreted as perceiving greater ambiguity. 

\begin{proposition}\label{prop:check}
    For all $\pi\in \Pi$, $\pi\geq_{\mathrm{cx}}\delta_{\mu_\pi}$.
\end{proposition}

Thus, among all second-order beliefs with predictive model $\mu\in \Delta$, $\delta_{\mu}$ is the least element under the convex order. Given the second-order model $\pi$, although $\pi$ and $\delta_{\mu_\pi}$ assign the same average probability to every event, the former may distribute probability across several first-order models, whereas the latter places all mass on the single first-order model $\mu_\pi$. The order therefore interprets dispersion around a fixed predictive model as perceived ambiguity.

\begin{example}\label{ex:ber}
Consider the binary-state case $\Omega=\{\omega_1,\omega_2\}$. In this setting, each first-order model $\mu\in\Delta$ can be uniquely identified by $p=\mu(\{\omega_1\})$. Consequently, the set of second-order models can be identified with distributions supported on $[0,1]$. Under this identification, the convex order satisfies
$$
\tilde{\pi}\geq_{\mathrm{cx}}\pi
~~\iff~~
\int_{[0,1]}h(p)\tilde\pi(\d p)\geq \int_{[0,1]}h(p)\pi(\d p)
$$
for every continuous convex function $h:[0,1]\to\mathbb{R}$. For example, suppose the initial second-order model, or prior, is the standard uniform distribution $\pi_0=\operatorname{U}(0,1)=\operatorname{Beta}(1,1)$. Under independent and identically distributed sampling, observing $s$ realizations of $\omega_1$ and $f$ realizations of $\omega_2$ yields, through Bayesian updating, the posterior second-order model
$\operatorname{Beta}(1+s,1+f).$
Along the balanced sample path $$(\omega_1,\omega_1,\omega_2,\omega_2,\omega_1,\omega_2,\omega_2,\omega_1,\omega_1,\omega_2),$$
the posterior distributions after four and ten observations are, respectively,
$
\pi_4=\operatorname{Beta}(3,3)$ and
$\pi_{10}=\operatorname{Beta}(6,6).
$
Since $\pi_0$, $\pi_4$, and $\pi_{10}$ all have mean $1/2$, they induce the same predictive model $(1/2,1/2)\in\Delta$. At the same time, the posterior becomes progressively less dispersed as we collect additional balanced observations; see Figure \ref{fig:beta}. In particular,
$$
\pi_0\geq_{\mathrm{cx}}\pi_4
\geq_{\mathrm{cx}}\pi_{10}
\geq_{\mathrm{cx}}\delta_{1/2}.
$$
Thus, along a balanced part of the sample path, Bayesian updating leaves the predictive model unchanged while progressively reducing the second-order model's dispersion. The common-mean condition is essential for this comparison. In the unbalanced part of the sample path, the posterior mean differs from 1/2, and the resulting posterior need not be comparable under the convex order.

\begin{figure}[ht]
    \centering
    \includegraphics[width=0.8\linewidth]{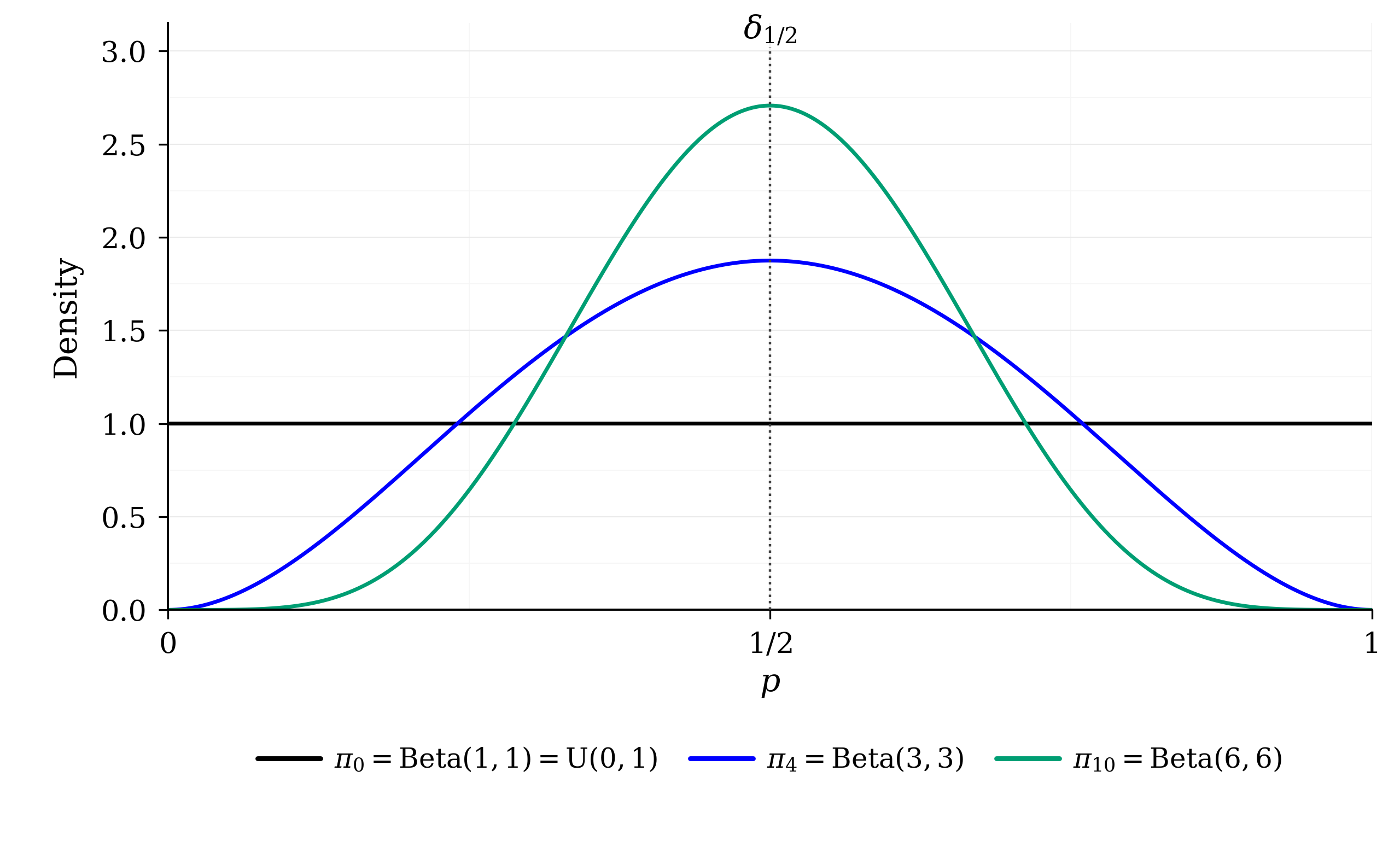}
    \caption{Balanced observations concentrate the posterior around
$p=1/2$.}
    \label{fig:beta}
\end{figure}
\end{example}

\subsection{Four properties}

Let $\mathcal{D}_c$ denote the set of all Borel probability measures on $\mathbb{R}$ with compact support. Throughout this section, following the convention in the risk-measure literature, statistics will have domain $\mathcal{E}=\mathcal{D}_c$ and decision functionals will have domain $\mathcal{Y}=\mathcal{X}_b$; see \cite{FS16}.

We begin by identifying general properties that a decision functional incorporating a second-order model should satisfy. Fix a statistic $\gamma$ that determines the DM's decision functional in the absence of ambiguity; we say that $\gamma$ encodes the DM's \emph{pure-risk preferences}. To introduce ambiguity, we distinguish between the DM's perception of ambiguity and the DM's attitude toward it. We represent the former with a second-order model $\pi\in\Pi$, while the latter is represented by an element of a partially ordered set $(\Lambda,\geq)$ with least element $\lambda_0$. In this section, we keep the partially ordered set $(\Lambda,\geq)$ abstract to keep the discussion general. We recall and introduce specific choices for this structure later.

The order on $\Lambda$ has a comparative interpretation: for $\lambda,\tilde{\lambda}\in\Lambda$, $\lambda\geq\tilde{\lambda}$ means that the attitude indexed by $\lambda$ is at least as ambiguity averse as the attitude indexed by $\tilde{\lambda}$. The least element $\lambda_0$ serves as the ambiguity-neutral benchmark. Thus, the partial order provides a comparative notion of ambiguity aversion, while $\lambda_0$ anchors that comparison and thereby yields an absolute classification of ambiguity attitudes. This construction follows the benchmark method underlying the comparative foundations of absolute ambiguity aversion in \cite{GM02} and \cite{E04}: ambiguity attitudes are first ordered comparatively, and an attitude is then classified in absolute terms by its position relative to ambiguity neutrality. Because $\lambda_0$ is the least element of $\Lambda$, the present formulation restricts attention to ambiguity-neutral and ambiguity-averse attitudes. The framework can be extended straightforwardly to accommodate ambiguity-seeking attitudes. 

We now formulate several desirable properties for decision functionals that combine these three primitives. We therefore consider a family
$$
\rho(\,\cdot\,;\pi,\lambda):\mathcal{X}_b\to\mathbb{R},
\qquad
(\pi,\lambda)\in\Pi\times\Lambda.
$$
This definition of the decision functional separates the DM's pure-risk preferences, encoded by $\gamma$, from their ambiguity perception $\pi$ and ambiguity attitude $\lambda$. The first two properties ensure consistency with the pure-risk and ambiguity-neutral benchmarks, while the final two require the functional to respect the comparative orders on ambiguity attitudes and ambiguity perceptions.

Consider the case when the DM's second-order model is given by $\delta_\mu$ for some $\mu\in \Delta$. Conceptually, this means the DM has no epistemic uncertainty and believes that the first-order model $\mu$ correctly describes the aleatoric uncertainty; that is, the DM perceives no ambiguity. In this case, independently of ambiguity attitudes, the DM should make decisions using their pure-risk preferences. This leads us to our first property.
\begin{property}[Pure-risk consistency]
    \label{property1} For all $\mu\in \Delta$, $\lambda\in \Lambda$, and $X\in \mathcal{X}_b$, $$\rho(X;\delta_\mu,\lambda)=\gamma(X|\mu).$$
\end{property}

Because the least element $\lambda_0$ serves as the benchmark for ambiguity neutrality, we need a canonical procedure to reduce a second-order model to a single first-order model. Motivated by Example \ref{ex:affine}, we take the corresponding predictive model to provide this canonical reduction. This reduction is common in the literature and has been used in financial risk management \citep{A08, AKT17}, actuarial science \citep{K92, C00}, and portfolio allocation problems \citep{KS96, FH22}. Under ambiguity neutrality, the resulting evaluation should be given by pure-risk preferences after reduction. The following property formalizes these considerations.
\begin{property}[Predictive ambiguity neutrality]\label{property2}
 For all $\pi\in \Pi$ and $X\in \mathcal{X}_b$,
 $$\rho(X;\pi,\lambda_0)=\gamma(X|\mu_{\pi}).$$
\end{property}

In the decision-theory literature, one preference relation is more ambiguity averse than another if, for every act (in the profit-loss sense), the certainty equivalent under the former is at most as large as that under the latter. Generalizing this idea to arbitrary decision functionals, while accounting for loss-profit acts, yields the following property.

\begin{property}[Attitude monotonicity]\label{property3}
     For all $\pi\in \Pi$, $X\in \mathcal{X}_b$, and $\lambda,\tilde\lambda\in \Lambda$ satisfying $\lambda\geq \tilde\lambda$,$$\rho(X;\pi,\lambda)\geq \rho(X;\pi,\tilde\lambda).$$
\end{property}

An immediate consequence of Properties \ref{property2} and \ref{property3} is that, for every $\pi\in\Pi$, $\lambda\in\Lambda$, and $X\in\mathcal{X}_b$,
\begin{equation}\label{eq:waa}
    \rho(X;\pi,\lambda)
\geq
\rho(X;\pi,\lambda_0)=\gamma(X|\mu_\pi).
\end{equation}
Thus, holding the ambiguity perception $\pi$ fixed, every attitude $\lambda\in \Lambda$ yields an evaluation at least as conservative as the ambiguity-neutral benchmark. Thus, every attitude $\lambda\in\Lambda$ is ambiguity averse relative to the predictive benchmark. This follows the benchmark-comparison approach of 
\cite{GM02}, while allowing the benchmark to retain a general pure-risk statistic. Properties \ref{property2} and \ref{property3} thereby give the partially ordered set $(\Lambda,\geq)$ its intended interpretation.

The inequality in \eqref{eq:waa} is a benchmark comparison: under Property \ref{property1}, it compares the evaluation under the ambiguity perception $\pi$ to the pure-risk evaluation under the predictive model. In this respect, it parallels the classical notion of weak risk aversion from \cite{P64} in the pure-risk setting, which compares a distribution of outcomes with the certain outcome equal to its expectation. The passage from weak to strong risk aversion in \cite{RS70} replaces this single comparison with monotonicity under the traditional convex order. Since the convex order on $\Pi$ provides the corresponding notion of increased dispersion among second-order models, the following property imposes an analogous monotonicity requirement on ambiguity perceptions.

\begin{property}[Strong ambiguity aversion]\label{property4}
     For all $\lambda\in \Lambda$, $X\in \mathcal{X}_b$, and $\pi,\tilde\pi\in \Pi$ satisfying $\pi\geq_{\mathrm{cx}}\tilde\pi$,
     $$\rho(X;\pi,\lambda)\geq \rho(X;\tilde\pi,\lambda).$$
\end{property}

A consequence of Proposition \ref{prop:check} is that Properties \ref{property1} and \ref{property4} imply the ambiguity-aversion inequality in \eqref{eq:waa}. Therefore, Property \ref{property4}, in combination with Property \ref{property1}, strengthens the benchmark notion of ambiguity aversion in the same way that strong risk aversion strengthens weak risk aversion, which motivates the terminology.

In the next subsection, we turn to the decision functionals associated with the smooth model from \cite{KMM05} and show how the model provides a concrete specification of the partially ordered set $(\Lambda,\geq)$ and a family of decision functionals satisfying the above properties. However, we also discuss this model's inability to represent general pure-risk preferences.

\subsection{Limits of second-order integral aggregation}

We now provide a first concrete specification of the framework developed in the preceding subsection. In particular, we take the statistic representing the DM's pure-risk preferences to be an expected-loss statistic. That is, for some loss function $ h$, we have 
$\gamma=\mathbb{E}_{ h}$. Throughout this subsection, we normalize all loss functions $\ell$ so that $\ell(0)=0$ and $\ell(1)=1$. This normalization entails no loss of generality, since loss functions that differ by a positive affine transformation generate the same expected-loss statistic.

We model ambiguity attitudes by the set
$$
\Lambda_{ h}=\left\{
\phi\circ h:
\phi \text{ is a convex loss function}
\right\}.
$$
We equip $\Lambda_{ h}$ with the partial order
$$
\ell\geq\tilde{\ell}
\iff
\ell=\phi\circ\tilde{\ell}~~
\text{for some convex loss function }\phi.
$$
It is clear that $ h$ is the least element of $\Lambda_{ h}$.

For each $\pi\in\Pi$ and $\ell\in\Lambda_{ h}$, consider the loss-based formulation of the smooth decision functional of \cite{KMM05},
\begin{equation}
\label{eq:smooth}
\rho(X;\pi,\ell)=\ell^{-1}\left(\int_{\Delta}\ell\left(
\mathbb{E}_{ h}(X|\nu)\right)\pi(\d\nu)\right),~~
X\in\mathcal{X}_b.
\end{equation}
These decision functionals have been used extensively in the literature (\citealp{G11,JM12,MMR13}) because they clearly separate pure risk, ambiguity perception, and ambiguity attitude. The next result shows that this family of decision functionals realizes all the properties introduced in the preceding subsection.

\begin{proposition}
\label{prop:smooth-properties}
The family of decision functionals defined by \eqref{eq:smooth} satisfies
Properties \ref{property1}--\ref{property4}.
\end{proposition}

Although the smooth model provides a concrete family satisfying Properties \ref{property1}--\ref{property4}, this conclusion depends on the assumption that the pure-risk preferences are given by an expected-loss statistic. To see this, observe that
$$\int_{\Delta} h\left(
\mathbb{E}_{ h}(X|\nu)\right)\pi(\d\nu)=\int_{\Delta}\mathbb{E}\left( h(X)|\nu\right)\pi(\d\nu)\overset{\eqref{eq:pred}}{=}\mathbb{E}( h(X)|\mu_\pi),~~X\in \mathcal{X}_b.$$
Therefore, Property \ref{property2} relies on the fact that the statistic is expected loss. Second, composing the affine map $\mu\mapsto \mathbb{E}( h(X)|\mu)$, where $X\in \mathcal{X}_b$, with a convex loss function produces a convex function on first-order models, which allows the smooth functional to respect the convex order on second-order models and thereby satisfy Property~\ref{property4}. A general statistic need not have either property, as we show below.

This restriction is important in applications because the statistic governing pure-risk preferences should be decoupled from ambiguity perception and attitude. For example, under the perception of no ambiguity, a bank may be required to calculate regulatory capital requirements using expected shortfall. Therefore, a Bayesian extension of expected shortfall should incorporate a second-order model while still reducing to expected shortfall both when the DM perceives no ambiguity (Property \ref{property1}) and when the DM is ambiguity neutral (Property \ref{property2}). It should also satisfy Properties \ref{property3} and \ref{property4} if we want to consider ambiguity-averse behavior. 

To illustrate the above points more concretely, if one wanted to create a Bayesian extension of the expected shortfall, a natural first attempt would be to replace the $\mathbb{E}_{ h}$ in \eqref{eq:smooth} with $\mathrm{ES}_{\alpha}$. This substitution is appealing because decision functionals originating in decision theory are often adapted to other areas by replacing the expected-loss statistic with the relevant domain-specific statistic; the derivation of the DRO paradigm from the maxmin expected utility model of \cite{GS89} is a prominent example. However, such a direct substitution is incompatible with ambiguity neutrality, as the following theorem shows.

One might wonder whether this incompatibility depends on reducing each second-order model to its predictive first-order model. To show that the result is independent of this particular choice of reduction, we consider a broader class of mappings from second-order models to first-order models.

\begin{definition}\label{def:reduc}
    A mapping $\mathfrak{R}:\Pi\to\Delta$ is called a \emph{reduction rule} if, for every $\pi\in\Pi$, $A\in\mathcal F$, and $a\in[0,1]$,
    if $\pi(\nu\in \Delta:\nu(A)=a)=1$ then $\left[\mathfrak{R}(\pi)\right](A)=a.$
\end{definition}

The predictive reduction is a particular reduction rule, but
Definition \ref{def:reduc} permits more general reductions of a
second-order model to a first-order model. The following theorem shows that
this additional freedom does not rescue the second-order aggregation procedure found in the smooth model.
Indeed, among all statistics and all reduction rules, the only case in which
loss-based second-order aggregation can always be represented as evaluation
of the same statistic under a reduced first-order model is the expected-loss
statistic together with the predictive reduction.

\begin{theorem}\label{theo:mainmot}
Assume that $\Omega$ is uncountable. Let $\ell$ be a loss function,
$\gamma$ be a statistic, and $\mathfrak{R}:\Pi\to\Delta$ be a reduction rule.
Then the following statements are equivalent:
\begin{enumerate}[(i)]
    \item For every $\pi\in\Pi$ and $X\in\mathcal{X}_b$,
    \begin{equation}\label{eq:mainmot}
        \ell^{-1}\left(
        \int_{\Delta}
        \ell\left(\gamma(X|\nu)\right)\,
        \pi(\d\nu)
        \right)
        =
        \gamma\left(X|\mathfrak{R}(\pi)\right).
    \end{equation}

    \item The statistic $\gamma$ is the expected-loss statistic $\mathbb{E}_\ell$ and $\mathfrak{R}$ is the predictive reduction.
\end{enumerate}
\end{theorem}

Theorem \ref{theo:mainmot} therefore shows that the expected-loss
statistic is not incidental to the smooth model: it is the unique
pure-risk statistic compatible with this form of second-order aggregation
and first-order reduction. Consequently, accommodating general statistics
while preserving ambiguity neutrality requires a different
procedure.

Finally, expected shortfall has another distinctive feature under second-order averaging. For $X\in \mathcal{X}$, define the mapping $\iota_X:\Delta\to \mathcal{D}$ by $\iota_X(\mu)=X_{\#}\mu$. We verify in Appendix \ref{app:3} that, for every $X\in \mathcal{X}_b$, the mapping $\mathrm{ES}_{\alpha}\circ\iota_X$ is $\Sigma$-measurable.

\begin{proposition}\label{prop:need3}
    If $\tilde\pi,\pi\in \Pi$ satisfy $\tilde\pi\geq_{\mathrm{cx}}\pi$, then
    $$\int_{\Delta}\mathrm{ES}_{\alpha}(X|\nu)\tilde{\pi}(\d\nu)\leq \int_{\Delta}\mathrm{ES}_{\alpha}(X|\nu)\pi(\d\nu),~~X\in \mathcal{X}_b.$$
\end{proposition}

 This inequality goes in the opposite direction of Property \ref{property4}. In particular, by Proposition \ref{prop:check},
$$\int_\Delta \mathrm{ES}_{\alpha}(X|\nu)\pi(\d\nu)\leq \mathrm{ES}_{\alpha}(X|\mu_{\pi}),~~X\in \mathcal{X}_b.$$
Consequently, if banks were permitted to incorporate a second-order model in calculating regulatory capital requirements, they would have an incentive to adopt a highly dispersed second-order model and compute the average expected shortfall under that model, rather than determine regulatory capital requirements based on the predictive model. Therefore, the resulting decision functional will incentivize ambiguity-seeking behavior. From a regulatory perspective, this is highly problematic, as banks should exhibit ambiguity aversion to strengthen the financial system's resilience during periods of heightened economic ambiguity. This observation is not unique to expected shortfall, as any concave statistic (see \cite{AS13,DKW19,WWW20}) would have the same interpretation.
 
\section{Ambiguity irons}

The limitations of second-order integral aggregation motivate a construction in which beliefs and ambiguity attitudes determine an ambiguity set. We call this transformation an \emph{ambiguity iron}. The resulting decision functional is a worst-case decision functional (see Example \ref{ex:DRO}) under this ambiguity set. Conditions on the transformation ensure that this decision functional satisfies the four properties introduced above. We then construct an ambiguity iron using distortions of second-order models, study the resulting ambiguity sets, and derive alternative formulas for the statistic when it is given by an optimized certainty equivalent, expected loss, or value-at-risk.

\label{sec:4}

\subsection{General definition}

In this section, we define ambiguity irons. Ambiguity irons transform second-order models into an ambiguity set. As we discuss later, this transformation accounts for ambiguity attitudes. We will use ambiguity irons to construct decision functionals within the DRO paradigm. We will see that this idea naturally leads to decision functionals that satisfy Properties \ref{property1}--\ref{property4}. 

Let $(\Lambda,\geq)$ be a partially ordered set with least element $\lambda_0$. As in the preceding sections, $\Lambda$ indexes ambiguity attitudes, the partial order $\geq$ represents comparative ambiguity aversion, and $\lambda_0$ serves as the ambiguity-neutral benchmark. The defining properties of an ambiguity iron stated below parallel Properties~\ref{property1}--\ref{property4} and admit analogous interpretations.

\begin{definition}\label{def:bus-map}
A map $\mathcal{U}:\Pi\times \Lambda \to 2^{\Delta}$
is called an \emph{ambiguity iron} if it satisfies:
\begin{enumerate}[(1)]
\item[--] Degeneracy consistency: for all $\mu\in\Delta$ and $\lambda\in\Lambda$,
$\mathcal{U}(\delta_\mu,\lambda)=\{\mu\}.$
\item[--] Predictive ambiguity neutrality: for all $\pi\in\Pi$,
$\mathcal{U}(\pi,\lambda_0)=\{\mu_{\pi}\}.$
\item[--] Attitude monotonicity: for all $\pi\in \Pi$ and $\lambda,\tilde\lambda\in \Lambda$ satisfying $\lambda\geq \tilde\lambda$,
$\mathcal{U}(\pi,\lambda)\supseteq \mathcal{U}(\pi,\tilde\lambda).$
\item[--] Strong ambiguity aversion: for all $\lambda\in \Lambda$ and $\pi,\tilde\pi\in \Pi$ satisfying $\pi\geq_{\mathrm{cx}}\tilde\pi$, 
$\mathcal{U}(\pi,\lambda)\supseteq \mathcal{U}(\tilde{\pi},\lambda).$
\end{enumerate}
\end{definition}

 A concept similar to ambiguity irons appears in \cite{GHTV08}, where the authors introduce \emph{selected probability-possibility sets}. Like our ambiguity irons, their selected probability-possibility sets take an object representing ambiguity perception as an input and output an ambiguity set. The key distinction lies in how they model ambiguity perception: we use a second-order model as input, whereas they use the initial structured ambiguity set, i.e., the ambiguity set consistent with the constraints of the decision problem. In that work, they likewise interpret this reduction as encoding attitudes toward ambiguity.

Ambiguity sets constructed from second-order models have recently been discussed in \cite{DOD24,DOD25}. Taking $\Lambda=[0,\infty)$, these papers introduce the mappings $\mathcal{U}_{\mathrm{PP}}$ and $\mathcal{U}_{\mathrm{PE}}$, both with domain $\Pi\times [0,\infty)$, defined by
\begin{align*}
    \mathcal{U}_{\mathrm{PP}}(\pi,\epsilon)&=\{\mu\in \Delta~|~\mathrm{KL}(\mu|\mu_{\pi})\leq\epsilon\}\\\mathcal{U}_{\mathrm{PE}}(\pi,\epsilon)&=\left\{\mu\in \Delta~\bigg|~\int_{\Delta}\mathrm{KL}(\mu|\nu)\pi(\d\nu)\leq\epsilon\right\},
\end{align*}
where $\mathrm{KL}$ denotes the Kullback--Leibler divergence. We verify in Appendix \ref{app:distortion} that the mapping $\nu\mapsto \mathrm{KL}(\mu|\nu)$ is $\Sigma$-measurable for all $\mu\in \Delta$. The mappings $\mathcal{U}_{\mathrm{PP}}$ and $\mathcal{U}_{\mathrm{PE}}$ have a similar structure to our ambiguity irons, but neither satisfies all the properties of an ambiguity iron. For example, neither satisfies degeneracy consistency. This failure follows from the fact that $\mathcal{U}_{\mathrm{PP}}$ and $\mathcal{U}_{\mathrm{PE}}$ were introduced to address problems related to model misspecification; see \cite{SZL23} and \cite{CHMM25}. We will discuss model misspecification later in Section \ref{sec:miss}.

Given a statistic $\gamma$, an ambiguity iron $\mathcal{U}$, $\pi\in \Pi$, and $\lambda\in \Lambda$, define the worst-case decision functional $\rho(\,\cdot\,;\pi,\lambda):\mathcal{X}_b\to \mathbb{R}$ by
\begin{equation}\label{eq:OurModel}
    \rho(X;\pi,\lambda)=\sup_{\nu\in \mathcal{U}(\pi,\lambda)}\gamma(X|\nu),~~X\in \mathcal{X}_b.
\end{equation}
The family of decision functionals defined in \eqref{eq:OurModel} satisfies Properties \ref{property1}--\ref{property4}. In the next subsection, we provide a tractable example of an ambiguity iron.

\subsection{The distortion ambiguity iron}

Let $\Lambda_{\mathrm{D}}$ denote the set of all continuous concave distortion functions. Define the partial order $\geq$ on $\Lambda_{\mathrm{D}}$ by pointwise dominance: $g\geq \tilde g$ if and only if $g(x)\ge \tilde g(x) \text{ for all }x\in[0,1].$ The partially ordered set $(\Lambda_{\mathrm{D}},\geq)$ has a least element $g_0$, which is given by $g_0(x)=x$ for all $x\in[0,1]$. Given $\pi\in \Pi$ and $g\in \Lambda_{\mathrm{D}}$, define the collection of second-order models
$$\mathcal{R}(\pi,g) = \left\{ \xi\in\Pi~|~ \xi(B)\leq g(\pi(B)) \ \text{for all }B\in\Sigma \right\}.$$
Since $g$ is concave and continuous, the capacity $g\circ\pi$ is submodular and continuous; see \cite{MM04}. Furthermore, this implies $\mathcal{R}(\pi,g)\neq \varnothing$ and \begin{equation}\label{eq:Choquet}
    \max_{\xi\in \mathcal{R}(\pi,g)}\int_{\Delta}f(\nu)\xi(\d\nu)=\int_0^\infty g(\pi(f>x))\d x+\int_{-\infty}^0(g(\pi(f>x))-1)\d x=:\int_{\Delta} f(\nu) g\circ\pi(\d\nu),
\end{equation} where $f:\Delta\to\mathbb{R}$ is $\Sigma$-measurable, whenever the right-hand side is well-defined (i.e., at least one of the two integrals is finite). The right-hand side of \eqref{eq:Choquet} is the Choquet integral; see \cite{D94}.

Define the mapping $\mathcal{S}:\Pi\times \Lambda_{\mathrm{D}}\to 2^{\Delta}$ by
\begin{equation}\label{eq:S-pi-g}
\mathcal{S}(\pi,g)=\left\{\mu_{\xi} ~|~ \xi\in\mathcal{R}(\pi,g)\right\}.
\end{equation}

\begin{theorem}\label{theo:Main}
The mapping $\mathcal{S}$ from \eqref{eq:S-pi-g} is an ambiguity iron.
\end{theorem}

Before turning to applications of the distortion ambiguity iron, we record two useful structural properties of the induced ambiguity sets.

\begin{proposition}\label{prop:compact}
    Given $\pi\in \Pi$ and $g\in \Lambda_{\mathrm{D}}$, we have that $\mathcal{S}(\pi,g)$ is convex and $\tau$-compact.
\end{proposition}

\begin{example}\label{example:beta}
    Assume the same set-up as Example \ref{ex:ber}. For $\beta\in [0,1)$, consider the distortion function $g_\beta(x)=\frac{x}{1-\beta}\wedge 1$. Since $g_{\beta}\geq g_{\beta'}$ if and only if $\beta\geq \beta'$, the parameter $\beta$ corresponds to ambiguity-aversion attitudes, with larger values of $\beta$ corresponding to greater ambiguity aversion. For $a,b>0$, denote by $\pi_{a,b}$ the second-order model given by $\operatorname{Beta}(a,b)$. Since $\mathcal{S}(\pi_{a,b},g_\beta)$ is compact and convex by Proposition \ref{prop:compact}, it follows that $$\mathcal{S}(a,b,\beta):=\mathcal{S}(\pi_{a,b},g_\beta)=[p_{l}(a,b,\beta),p_{u}(a,b,\beta)],~~\text{where}~~0\leq p_{l}(a,b,\beta)\leq p_{u}(a,b,\beta)\leq 1.$$ We verify in the Appendix \ref{app:distortion} that
    $$p_{l}(a,b,\beta)=\frac{1}{1-\beta}\int_0^{1-\beta}\mathrm{VaR}_\alpha(\pi_{a,b})\d\alpha~~~~\text{and}~~~~p_{u}(a,b,\beta)=\frac{1}{1-\beta}\int_\beta^{1}\mathrm{VaR}_\alpha(\pi_{a,b})\d\alpha.$$
    In the simple case when $a=b=1$, we have that $\pi_{1,1}=\operatorname{U}(0,1)$ and a simple calculation shows that
    $$\mathcal{S}(1,1,\beta)=\left[\frac{1-\beta}{2},\frac{1+\beta}{2}\right].$$
    An important interval to compare with the ambiguity-iron interval is the standard Bayesian credible interval for beta distributions. This interval is given by
    $$\mathcal{Q}(a,b,\beta)=\left[\mathrm{VaR}_{(1-\beta)/2}(\pi_{a,b}),\mathrm{VaR}_{(1+\beta)/2}(\pi_{a,b})\right].$$

    \begin{figure}[ht]
    \centering
    \includegraphics[width=1\linewidth]{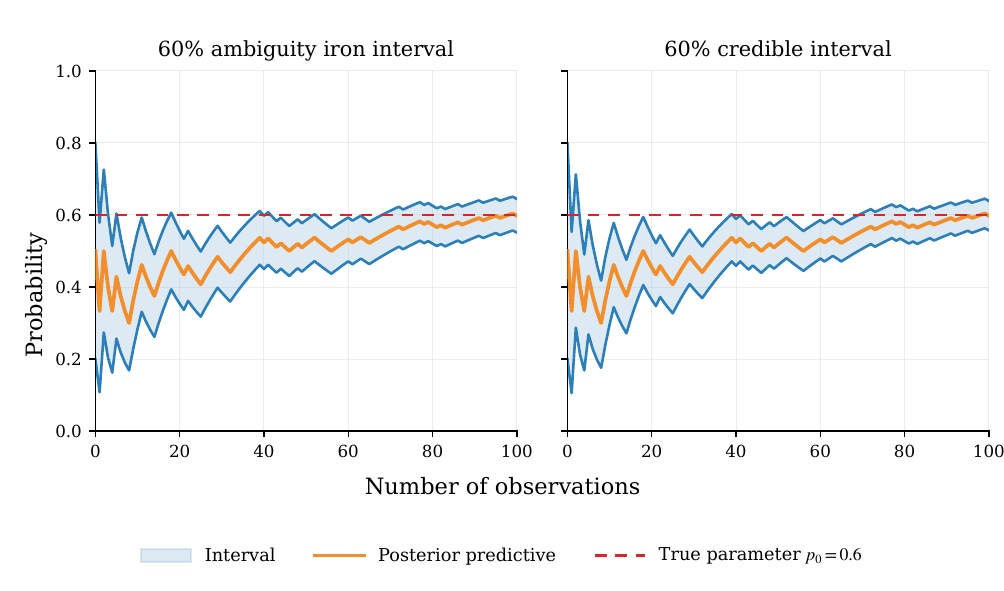}
    \caption{Sequential method comparison}
    \label{fig:ber}
\end{figure}
    
    \noindent The Bayesian credible interval has the benefit that $\pi_{a,b}(p\in \mathcal{Q}(a,b,\beta))=\beta$. In the case where $a=b=1$, we have that $\mathcal{S}(1,1,\beta)=\mathcal{Q}(1,1,\beta)$. Figure \ref{fig:ber} reports the results of a numerical experiment in which we begin with the uniform prior $\operatorname{Beta}(1,1)$, observe a sequence of independent realizations generated by the fixed first-order model $\mu=(0.6,0.4)$, and update the posterior after each observation. At each stage, we plot the corresponding predictive model, the interval generated by the ambiguity iron, and the associated Bayesian credible interval, all constructed using $\beta=0.6$. The two constructions produce similar intervals in this simple binary setting, with only minor visible differences. Rather than diminishing the value of the ambiguity-iron approach, this similarity shows that it recovers behavior consistent with familiar Bayesian methods in a setting where those methods are easy to implement. In higher-dimensional problems, optimization over Bayesian credible regions can become computationally difficult. The representations developed in Section \ref{sec:5} provide sampling-based methods for evaluating the decision functionals induced by the distortion ambiguity iron.
\end{example}

\subsection{Alternative formulas}

In this section, we present alternative representations of \eqref{eq:OurModel} for several families of statistics. We also allow the decision functionals to be defined on domains larger than $\mathcal{X}_b$, which is more suitable for the numerical analysis in Section \ref{sec:5}. To accommodate this larger domain, let $\mathcal{D}_1$ denote the subset of $\mathcal{D}$ consisting of distributions with finite first moment, which will serve as the domain of the statistics considered in this section.

A key consideration in the results that follow is ensuring the required integrability conditions hold when changing first-order models. Accordingly, we denote by $\Lambda_{\mathrm{D}}^*$ the subset of $\Lambda_{\mathrm{D}}$ consisting of distortion functions $g$ for which there exists $a\in\mathbb{N}$ such that
$$
g(x)\leq ax,\qquad x\in[0,1].
$$

\paragraph*{Optimized certainty equivalents.}
We first derive an alternative formula when the statistic is an optimized certainty equivalent; see \cite{BTT07}.
Let $h:\R\to\R$ be convex, continuous, and increasing satisfying $h(0)=0$ and $h(x)\geq x$ for all $x\in\mathbb{R}$. Define the statistic
$$\mathrm{OCE}_h(Q)
=
\inf_{m\in\R}\left\{ m +\int_{\mathbb{R}}h(x-m)Q(\d x)\right\},~~Q\in \mathcal{D}_1.$$

The expected shortfall statistic $\mathrm{ES}_\alpha$ provides a classic example of an optimized certainty equivalent. It arises when $h(x)=x_+/(1-\alpha)$ for some $\alpha\in [0,1)$. In this case, the optimized certainty equivalent representation coincides with the well-known formulation of expected shortfall in \cite{RU02}.

We have the following alternative formula for \eqref{eq:OurModel} under the distortion ambiguity iron when the statistic is given by an optimized certainty equivalent.

\begin{theorem}
\label{theo:minmax}
Let $\pi\in\Pi$, $g\in\Lambda_{\mathrm{D}}^*$,
and $X\in L^1(\pi)$. If
$
h(X-m)\in L^1(\pi)$ for every $m\in\mathbb{R}$, then
$$
\sup_{\nu\in\mathcal{S}(\pi,g)}
\mathrm{OCE}_h(X|\nu)
=
\inf_{m\in\mathbb{R}}
\left\{
m+\int_\Delta
\mathbb{E}^{\nu}[h(X-m)]\,g\circ\pi(\d\nu)
\right\}.
$$
\end{theorem}

Theorem \ref{theo:minmax} is a Bayesian counterpart to the robust-optimized certainty-equivalent constructions studied in \cite{BDT20} and \cite{WMH24}. \cite{BDT20} replace the reference expectation in the classical OCE with a nonlinear expectation and derive tractable representations when model uncertainty is described, in particular, through Wasserstein penalties around a fixed baseline first-order model. \cite{WMH24} consider the more general functional
$$\inf_{m\in\mathbb R}\{m+\rho(h(X-m))\},$$ where $\rho$ is a convex risk measure. Theorem \ref{theo:minmax} shows that, under the distortion ambiguity iron, the worst-case OCE admits a representation with a similar form to those derived in these works. The main distinction is that these papers impose the ambiguity specifications directly through ambiguity sets or penalty functions; they do not generate them from a second-order model over first-order models. 

\paragraph{Expected loss.}
The expected-loss case admits a particularly simple representation, which can be approximated using Theorem \ref{theo:num}. In contrast to optimized certainty equivalents, it does not require an additional optimization: after evaluating the transformed loss under each first-order
model, the resulting values can be aggregated directly through the Choquet integral. Let
$\ell:\mathbb{R}\to\mathbb{R}$ be a convex loss function, $\pi\in\Pi$,
$g\in\Lambda_{\mathrm{D}}^*$, and $X\in L^1(\pi)$ such that $\ell(X)\in L^1(\pi)$. We have
$$\sup_{\nu\in \mathcal{S}(\pi,g)}\mathbb{E}_{\ell}(X|\nu)=\ell^{-1}\left(\int_{\Delta}\mathbb{E}^{\nu}[\ell(X)]g\circ\pi(\d\nu)\right).$$
This decision functional is the loss-based certainty-equivalent version of the second-order dual
expected utility model of \cite{G14}, in which model-specific expected losses are aggregated using
the dual model of \cite{Y87} rather than through a nonlinear second-order utility.

This representation also connects directly to the Bayesian risk optimization framework of \cite{WZZ18}. When the second-order model $\pi$ is taken to be a posterior distribution, the quantity inside the inverse of the loss function is a functional applied to the posterior distribution of model-specific expected losses, of the type studied in that work.

When the loss function is given by $\ell_\eta(x)=e^{\eta x}$ for all $x\in \mathbb{R}$, where $\eta>0$, the statistic $\mathbb{E}_{\ell_\eta}$ coincides with the entropic risk measure $\mathrm{ER}_{\eta}$; see \cite{FS02}.

\paragraph{Value-at-risk.} The final statistic for which we derive an alternative representation is value-at-risk. Unlike the preceding statistics, we cannot obtain its worst-case evaluation by calculating a worst-case expected loss. Theorem \ref{prop:var} shows that worst-case value-at-risk is computed by first aggregating the model-specific survival functions and then inverting the resulting aggregate, similarly to \cite{LMW24}. 

\begin{theorem}
\label{prop:var}
Let $\pi\in\Pi$, $g\in\Lambda_{\mathrm{D}}$, and
$\alpha\in(0,1)$. Then 
$$
\sup_{\nu\in\mathcal{S}(\pi,g)}
\mathrm{VaR}_{\alpha}(X|\nu)
=
\inf\left\{
x\in\mathbb{R}:
\int_\Delta
\nu(X>x)\,g\circ\pi(\d\nu)
\leq1-\alpha
\right\},~~X\in \mathcal{X}.
$$
\end{theorem}

Notably, unlike the preceding expected-loss-based formulas, Theorem \ref{prop:var} does not require a moment condition on acts. The worst-case value-at-risk remains finite because the distorted survival function converges to zero as the threshold increases. 

\section{Unambiguous events and unambiguous acts}

A nondegenerate second-order model need not make every act ambiguous. Different first-order models may disagree about some events while inducing the same distribution for a particular act. This section identifies acts and events that are unambiguous in this sense. For the distortion ambiguity iron, we show that evaluating such an act reduces to applying the statistic to its common distribution, independently of ambiguity attitude. We also establish an invariance result for acts inducing the same epistemic distributional beliefs. These results clarify which uncertainty the decision functional responds to and explain the treatment of unambiguous positions in the numerical examples that follow.

\label{sec:7}

Let $\tau_\mathcal{D}$ denote the topology on $\mathcal{D}$ associated with convergence in distribution and let $\Sigma_{\mathcal{D}}$ be the $\sigma$-algebra generated by $\tau_{\mathcal{D}}$.
It follows that $(\mathcal{D},\Sigma_{\mathcal{D}})$ is a Polish measurable space.

\begin{proposition}\label{prop:meas}
   Given $X\in \mathcal{X}$, the mapping $\iota_X:\Delta\to \mathcal{D}$ is $\Sigma/\Sigma_{\mathcal{D}}$-measurable.  
\end{proposition}

Given $\pi\in \Pi$ and $X\in \mathcal{X}$, the probability measure $(\iota_X)_{\#}\pi$ on $(\mathcal{D},\Sigma_{\mathcal{D}})$ represents the epistemic belief regarding the distribution of $X$. Given $\pi\in \Pi$, define $$\mathfrak{C}(\pi)=\{X\in \mathcal{X}:(\iota_X)_{\#}\pi=\delta_Q\text{ for some }Q\in \mathcal{D}\}.$$
Intuitively, $\mathfrak{C}(\pi)$ represents the set of acts with an unambiguous distribution under the second-order model $\pi\in\Pi$. Additionally, for $\pi\in \Pi$, define the set 
$$\mathfrak{U}(\pi)=\{A\in \mathcal{F}:\id_A\in \mathfrak{C}(\pi)\}.$$
We can view $\mathfrak{U}(\pi)$ as the set of unambiguous events. 

\begin{proposition}\label{prop:lambda}
    Given $\pi\in \Pi$, the set $\mathfrak{U}(\pi)$ is a $\lambda$-system, that is,
\begin{enumerate}[(i)]
    \item $\Omega\in \mathfrak{U}(\pi)$;
    \item if $A\in \mathfrak{U}(\pi)$ then $A^c\in \mathfrak{U}(\pi)$;
    \item if the sequence $(A_n)_{n\in \N}\subseteq \mathfrak{U}(\pi)$ is pairwise disjoint then $\bigcup_{n=1}^{\infty}A_n\in \mathfrak{U}(\pi)$.
\end{enumerate}
Moreover, $\mathfrak{C}(\pi)=\{X\in \mathcal{X}~|~X^{-1}(B)\in \mathfrak{U}(\pi)\text{ for all }B\in \mathcal{B}(\mathbb{R})\}$.
\end{proposition}

The use of a $\lambda$-system for unambiguous events is common in the literature; see 
\cite{Z02}, \cite{M02}, \cite{E04}, and \cite{VW25}. Given $\pi\in \Pi$ and $X\in \mathfrak{C}(\pi)$, let $Q_X^{\pi}\in \mathcal{D}$ denote the distribution satisfying $(\iota_X)_{\#}\pi=\delta_{Q_X^{\pi}}$. The following proposition confirms that, when an act has an unambiguous distribution, the decision functional given in \eqref{eq:OurModel} under the distortion ambiguity iron reduces to calculating the statistic on this unambiguous distribution.

\begin{proposition}\label{prop:reduc}
    Let $\gamma$ be a statistic, $\pi\in \Pi$, and $g\in \Lambda_{\mathrm{D}}$. We have that
    $$\sup_{\nu\in \mathcal{S}(\pi,g)}\gamma(X|\nu)=\gamma(Q_X^{\pi}),~~X\in \mathfrak{C}(\pi).$$
\end{proposition}

The final result states that the decision functionals in \eqref{eq:OurModel} under the distortion ambiguity iron are invariant to epistemic distributional beliefs. 

\begin{theorem}\label{prop:LI}
Let $\pi,\xi\in \Pi$, $g\in \Lambda_{\mathrm{D}}$, and $\gamma$ be a statistic. We have
$$(\iota_X)_{\#}\pi=(\iota_Y)_{\#}\xi\implies \sup_{\mu\in \mathcal{S}(\pi,g)}\gamma(X|\mu)=\sup_{\mu\in \mathcal{S}(\xi,g)}\gamma(Y|\mu),~~X,Y\in \mathcal{X}.$$
\end{theorem}

It seems natural that any Bayesian decision model should be invariant to epistemic distributional beliefs. This requirement is analogous to law invariance in the pure-risk setting: evaluation should depend only on distributional beliefs.

\section{Parametrization and computation}

\label{sec:5}

We now turn to calculating the proposed decision functionals from samples of the second-order model. We first introduce a parametrization that represents first-order models by parameters without changing the ambiguity-set construction. We then show how to approximate the Choquet integrals appearing in the earlier representations using weighted order statistics. The convergence result applies to sampling schemes satisfying the stated averaging condition, including direct sampling and suitable MCMC schemes. Finally, portfolio examples illustrate how to implement these formulas and how ambiguity perception and ambiguity attitude affect the evaluation of financial positions.

\subsection{Parametrization}

To make these ideas more tractable, we work with the Polish measurable space $(\Theta,\mathcal{A})$, interpreted as the \emph{parameter space}. We also have a mapping $\mathfrak{P}:\Theta\to\Delta$, called a \emph{parametrization}. We will assume that
\begin{enumerate}[(1)]
    \item $\mathfrak{P}$ is $\mathcal{A}/\Sigma$-measurable;
    \item $\theta=\tilde{\theta}\iff \mathfrak{P}(\theta)=\mathfrak{P}(\tilde\theta)$;
    \item $\mathcal{A}=\mathfrak{P}^{-1}(\Sigma).$
\end{enumerate}
It is well-known (see \citet[Proposition 7.25]{BS96}) that Condition (1) is equivalent to requiring that, for all $A\in \mathcal{F}$, the mapping
$\theta\mapsto [\mathfrak{P}(\theta)](A)$ is $\mathcal{A}$-measurable. In other words, $\mathfrak{P}$ is a probability kernel. Condition (3) may be hard to check in practice; however, by the Lusin-Souslin theorem (\citealt[Theorem 15.1]{K12}), if Condition (2) holds and $\mathfrak{P}$ is continuous, Condition (3) holds.

We denote by $\Delta(\Theta)$ the set of probability measures on $(\Theta,\mathcal{A})$. Through the parametrization, an element $\pi\in \Delta(\Theta)$ induces a second-order model $\mathfrak{P}_{\#}\pi$. Therefore, we may view $\Delta(\Theta)$ as a subset of second-order models. Conceptually, there are two different ways to define ambiguity sets given $\pi\in \Delta(\Theta)$ and $g\in \Lambda_{\mathrm{D}}$. The first is to take $\mathcal{S}(\mathfrak{P}_{\#}\pi,g)$. The second begins by defining
$$\mathcal{R}_0(\pi,g)=\left\{\xi\in \Delta(\Theta)~|~\xi(A)\leq g(\pi(A))~\text{for all}~A\in \mathcal{A}\right\},$$
and then defining the ambiguity set $\mathcal{S}_0(\pi,g)$ as the set of all first-order models given by
$$\mu(A)=\int_{\Theta}\mu_{\theta}(A)\xi(\d\theta),~~A\in \mathcal{F},$$
for some $\xi\in \mathcal{R}_0(\pi,g)$,
where $\mu_{\theta}=\mathfrak{P}(\theta)$ for all $\theta\in\Theta$. The following proposition confirms that these methods produce the same ambiguity set.

\begin{proposition}\label{prop:para}
    Let $\pi\in \Delta(\Theta)$ and $g\in \Lambda_{\mathrm{D}}$. We have that
    $$\mathcal{R}(\mathfrak{P}_{\#}\pi,g)=\mathfrak{P}_{\#}(\mathcal{R}_0(\pi,g)):=\left\{\mathfrak{P}_{\#}\xi~\big|~\xi\in\mathcal{R}_0(\pi,g)\right\}~~\text{and}~~\mathcal{S}(\mathfrak{P}_{\#}\pi,g)=\mathcal{S}_0(\pi,g).$$
\end{proposition}
Therefore, with a slight abuse of notation, given $\pi\in \Delta(\Theta)$ and $g\in \Lambda_{\mathrm{D}}$, we denote the set $\mathcal{R}_0(\pi,g)$ by $\mathcal{R}(\pi,g)$ and the set $\mathcal{S}_0(\pi,g)$ by $\mathcal{S}(\pi,g)$, in order to maintain consistency with the notation used thus far. For $\pi\in\Delta(\Theta)$, we likewise write $L^1(\pi):=L^1(\mathfrak{P}_{\#}\pi)$.

\subsection{Convergence result}

Given $g\in \Lambda_{\mathrm{D}}^*$ and $\pi\in \Delta(\Theta)$, in the following algorithms, we will need to numerically approximate the decision functional 
$$\int_{\Theta}\mathbb{E}^{\mu_{\theta}}[X]g\circ\pi(\d\theta)=\sup_{\mu\in \mathcal{S}(\pi,g)}\mathbb{E}^{\mu}[X],~~X\in L^1(\pi).$$
Many of the numerical schemes found in the literature can produce a stochastic process $(\theta_n)_{n\in \N}$ with values in the parameter space $\Theta$ such that
\begin{equation}\label{eq:MCMC}
    \lim_{n\to\infty}\frac{1}{n}\sum_{k=1}^n f(\theta_k)=\int_{\Theta}f\d\pi~~\kappa\text{-a.s.},
\end{equation}
where $f:\Theta\to\mathbb{R}$ is measurable and satisfies $\int_{\Theta}|f|\d\pi<\infty$, and $\kappa\in \Delta(\Theta^{\N})$ is the probability measure induced by the stochastic process on path space. For example, many of the schemes used to sample from the posterior probability measure in Bayesian statistics satisfy this property; see \cite{MT12}. Additionally, if we are able to sample $\pi$ directly, we may take $\kappa=\pi^{\N}$, in which case, \eqref{eq:MCMC} follows directly from the law of large numbers. For the numerical examples in the following subsection, we assume that $\pi$ can be sampled directly. Nevertheless, we formulate the next theorem under the more general condition \eqref{eq:MCMC} in order to motivate the use of MCMC-type algorithms in settings where direct sampling from 
$\pi$ is not available.

\begin{theorem}\label{theo:num}

    Given a stochastic process $(\theta_n)_{n\in \N}$ satisfying \eqref{eq:MCMC}, $\pi\in \Delta(\Theta)$, $X\in L^1(\pi)$, and $g\in \Lambda_{\mathrm D}^*$, define $Y_n=\mathbb{E}^{\mu_{\theta_n}}[X]$. We have that
    $$\lim_{n\to\infty}\sum_{k=1}^n \left(g\left(\frac{n-k+1}{n}\right)-g\left(\frac{n-k}{n}\right)\right)Y_{k:n}=\int_{\Theta}\mathbb{E}^{\mu_{\theta}}[X]g\circ\pi(\d\theta),~~\kappa\text{-a.s.,}$$
    where $Y_{k:n}$ denotes the $k$th order statistic of the random vector $(Y_1,\dots,Y_n)$.
\end{theorem}

\subsection{Numerical illustrations}

In this section, we numerically implement the alternative formulas for \eqref{eq:OurModel} derived in Section \ref{sec:4}, using Theorem \ref{theo:num} to approximate the associated Choquet integrals. A portfolio optimization example illustrates how ambiguity perception and ambiguity attitude affect the resulting evaluations.

In all numerical implementations, we assume that $(\Omega,\mathcal{F})=(\mathbb{R}^2,\mathcal{B}(\mathbb{R}^2))$. The parameter space is taken to be
$\Theta=\mathbb{R}^2\times (0,\infty)^2\times (-1,1),$ with the parametrization
$$\Theta\ni(m_1,m_2,\sigma_1,\sigma_2,c)\mapsto \mathrm{MVN}_2\left(\begin{pmatrix}
    m_1\\m_2
\end{pmatrix},\begin{pmatrix}
    \sigma_1^2&c\sigma_1\sigma_2\\c\sigma_1\sigma_2&\sigma_2^2
\end{pmatrix}\right).$$
 Let $X_1$ and $X_2$ denote the acts given by the marginal projections, representing the single-period percentage losses of two financial positions. Our objective is to evaluate 
 \begin{equation}\label{eq:calc}
     \sup_{\nu \in \mathcal{S}(\pi,g_{\beta})} \gamma(wX_1+(1-w)X_2|\nu),
 \end{equation}
where $\gamma:\mathcal{D}_1\to\mathbb{R}\cup\{\infty\}$ is a statistic, $w\in [0,1]$, $\beta\in [0,1)$, and $g_\beta(x)=\frac{x}{1-\beta}\wedge 1$ for all $x\in [0,1]$. Since $g_{\beta}\geq g_{\beta'}\iff \beta\geq \beta'$, the parameter $\beta$ corresponds to ambiguity aversion attitudes, with larger values of $\beta$ corresponding to greater ambiguity aversion. Finally, in each numerical implementation, we consider the three second-order models listed in Table \ref{tab:1}, specifying the marginal distributions, as we assume the components are independent.
 
 Note that the distributions of the acts $X_1$ and $X_2$ under the predictive model are the same for both $\pi_1$ and $\pi_2$ in Table \ref{tab:1}. Additionally, under $\pi_1$, we perceive no ambiguity for the act $X_1$, as defined in Section \ref{sec:7}. Therefore, in the numerical implementations later, the decision functional in \eqref{eq:calc} should assign the same value to $X_1$ ($w=1$) and $X_2$ ($w=0$) under ambiguity neutrality ($\beta=0$). However, as we become more ambiguity averse ($\beta$ increases), the decision functional should assign a higher value to $X_2$ ($w=0$) than to $X_1$ ($w=1$).

\begin{table}[ht]
\centering
\caption{Second-order models used in the numerical implementations}\label{tab:1}
\vspace{0.5em}
\renewcommand{\arraystretch}{1.6}
\begin{tabular}{l|lll}
           & $\pi_1$                  & $\pi_2$                  & $\pi_3$                  \\ \hline
$m_1$      & $\delta_{-0.05}$  & $\mathrm{N}(-0.05,0.01)$ & $\mathrm{N}(-0.05,0.01)$        \\
$m_2$      & $\mathrm{N}(-0.05,0.02)$ & $\mathrm{N}(-0.05,0.02)$ &  $\mathrm{N}(-0.1,0.03)$\\
$\sigma_1$ &   $\delta_{0.01\sqrt{5}}$    & $\delta_{0.02}$     & $\delta_{0.02}$        \\
$\sigma_2$ &    $\delta_{0.01}$  & $\delta_{0.01}$     &    $\mathrm{HN}\left(0.02\sqrt{\frac{2}{\pi}}\right)$  \\
$c$        & $\mathrm{U}(-0.5,0.5)$    & $\mathrm{U}(-0.5,0.5)$  & $\mathrm{U}(-0.5,0.5)$
\end{tabular}
\end{table}

\paragraph*{Expected shortfall.}

\begin{figure}[ht]
    \centering
    \includegraphics[width=1\linewidth]{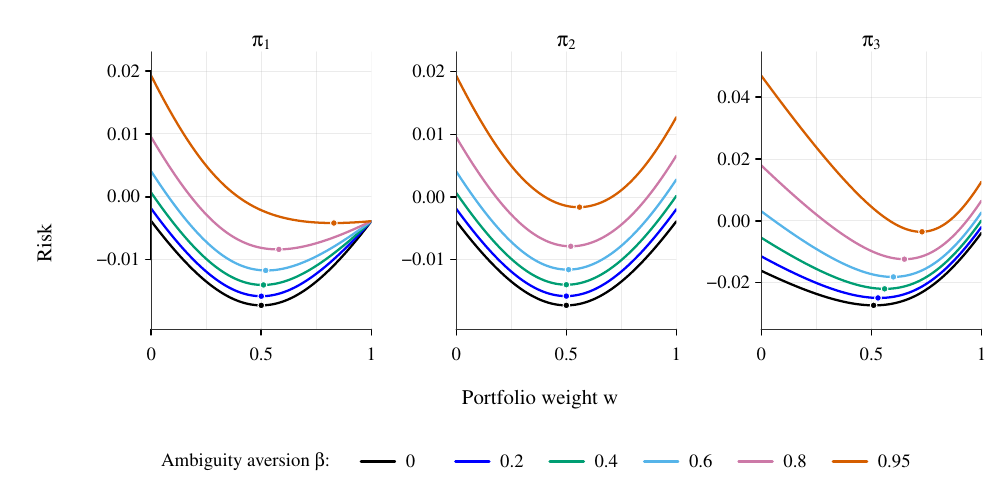}
    \caption{Worst-case expected shortfall}
    \label{fig:ES}
\end{figure}

Figure \ref{fig:ES} presents the values of the decision functional \eqref{eq:calc} when $\gamma=\mathrm{ES}_{0.95}$. Under the second-order model $\pi_1$ (left panel), the ambiguity-neutral case ($\beta = 0$) is minimized by an equal allocation between $X_1$ and $X_2$. This is consistent with the preference for equal diversification, since $X_1$ and $X_2$ have the same distribution under the predictive model. As $\beta$ increases, however, the minimizing portfolio shifts monotonically toward a full allocation in $X_1$. This is consistent with interpreting $\beta$ as a parameter of ambiguity aversion: under $\pi_1$, ambiguity is perceived in the distribution of $X_2$, but ambiguity is not perceived in the distribution of $X_1$.

This story changes in the center and right panels, corresponding to the second-order models $\pi_2$ and $\pi_3$, respectively. In both cases, ambiguity is perceived in the distributions of $X_1$ and $X_2$. For $\pi_2$, increasing $\beta$ produces little deviation from the fact that an equal allocation between $X_1$ and $X_2$ minimizes the decision functional. This is because the ambiguity associated with the two assets is roughly comparable. The case of $\pi_3$ differs markedly. There, the predictive distribution of $X_2$ is roughly as favorable as that of $X_1$, but $\pi_3$ perceives greater ambiguity in $X_2$ than in $X_1$ (e.g., the standard deviation of $X_2$ is ambiguous). As a result, when ambiguity aversion increases, the minimizer shifts toward a larger allocation in $X_1$.

\paragraph{Monotone mean--variance.} To further illustrate the scope of Theorem \ref{theo:minmax}, we consider the monotone mean--variance statistic from \cite{MMRT09}. This statistic is the optimized certainty equivalent generated by
$$
h_\zeta(x)=\zeta\left(x+\frac{1}{2\zeta}\right)_+^2-\frac{1}{4\zeta},~~x\in \mathbb{R},
$$
where $\zeta\in (0,\infty)$ is a risk aversion parameter. It provides a monotone counterpart to the classical mean--variance criterion from \cite{M52} while retaining its quadratic sensitivity to losses.
\begin{figure}[ht]
    \centering
    \includegraphics[width=1\linewidth]{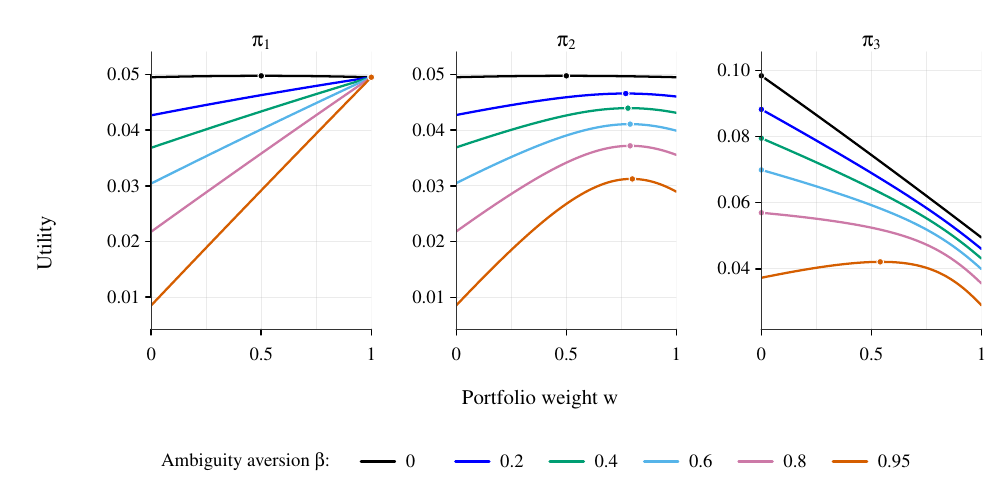}
    \caption{Worst-case utility under monotone mean--variance}
    \label{fig:mmv}
\end{figure}

Figure \ref{fig:mmv} plots the negative worst-case functional for $\zeta=2$. We plot the negative of the worst-case monotone mean--variance functional so that it is expressed in utility terms: maximizing this quantity is equivalent to minimizing worst-case risk and mimics the familiar maxmin utility formulation. As ambiguity aversion increases, the utility curves shift downward. The accompanying movement of the maximizing portfolio toward $X_1$, particularly under $\pi_1$ and $\pi_3$, shows that ambiguity aversion can materially alter the optimal allocation.

\paragraph{Entropic risk measure.}
 Figure \ref{fig:ER} in Appendix \ref{app:fig} presents the values of the decision functional \eqref{eq:calc} when $\gamma=\mathrm{ER}_{10}$. An interesting contrast between the numerical results in Figure \ref{fig:ES} and those in Figure \ref{fig:ER} arises under the second-order model $\pi_3$ (right panel). For low values of $\beta$, the minimizing portfolio fully allocates to $X_2$. As $\beta$ increases, however, the minimizing portfolio shifts toward $X_1$. Intuitively, this happens since the standard deviation of $X_2$ is ambiguous. This differs from the previous results because $\mathrm{ER}_{10}$ evaluated on the predictive distribution of $X_2$ is much smaller than when evaluated on the predictive distribution of $X_1$.

\paragraph{Value-at-risk.} 
Figure \ref{fig:VAR} in Appendix \ref{app:fig} reports the worst-case value-at-risk at level $\alpha=0.95$. At $\beta=0$, the criterion reduces to value-at-risk under the predictive model. Under $\pi_1$, increasing $\beta$ shifts the minimizing portfolio toward $X_1$, and all curves coincide at $w=1$ because the distribution of $X_1$ is unambiguous. The movement is more modest under $\pi_2$, where the ambiguity associated with the two assets is more comparable, while under $\pi_3$ the ambiguity in the mean and standard deviation of $X_2$ again shifts the minimizing allocation toward $X_1$. 

\section{Consistency and model misspecification}

\label{sec:6}

The preceding section studies whether posterior concentration leads the decision functional to recover the chosen statistic under the data-generating model. Under the stated continuity conditions, this holds when the posterior concentrates on the correct parameter. Under misspecification, however, posterior concentration can leave persistent model error. We therefore study a robustification of the statistic that accounts for misspecification separately from ambiguity perception and ambiguity attitude.

We will assume that the parametrization $\mathfrak{P}$ is continuous with respect to the total variation metric on $\Delta$, where
$$d_{\mathrm{TV}}(\mu,\tilde\mu)=\sup_{A\in \mathcal{F}}|\mu(A)-\tilde\mu(A)|,~~(\mu,\tilde\mu)\in\Delta^2.$$
This implies that the parametrization is continuous with respect to the weak topology $\tau$. This assumption is reasonable, as many regular parametric models satisfy it.

 In this section, statistics will have domain $\mathcal{E}=\mathcal{D}_c$ and decision functionals will have domain $\mathcal{Y}=\mathcal{X}_b$. A collection $\mathcal{Q}\subseteq \mathcal{D}_c$ is said to have \emph{uniformly bounded support} if there exists a compact set that contains the support of each $Q\in \mathcal{Q}$. We say that a statistic $\gamma$ is \emph{Lebesgue continuous} if for all sequences $(Q_n)_{n\in \N}\subseteq\mathcal{D}_c$ with uniformly bounded support satisfying $Q_n\Rightarrow Q$ for some $Q\in \mathcal{D}_c$, we have that $\lim_{n\to\infty}\gamma(Q_n)=\gamma(Q)$.

\subsection{Bayesian consistency}

In Bayesian statistics, consistency concerns the asymptotic behavior of the posterior distribution (second-order model). For example, suppose we observe an IID sequence of random variables $(\zeta_k)_{k\in \N}$ taking values in $\Omega$. Assume that $\zeta_k\sim \mu_{\theta_0}$ for some $\theta_0\in \Theta$. Under certain regularity conditions, see \cite{GVV17}, we have that the sequence of second-order models $(\pi_n)_{n\in \N}$ given by the posterior distribution ($\pi_n$ is obtained after realizing $(\zeta_k)_{k=1}^n$) satisfies $$\pi_n\Rightarrow \delta_{\theta_0}~~\mu_{\theta_0}^\infty\text{-a.s.}$$
The following theorem verifies that the decision functional \eqref{eq:OurModel}, under the distortion ambiguity iron, asymptotically agrees with the correct value.

\begin{theorem}\label{prop:consis}
    Let $(\pi_n)_{n\in \N}\subseteq \Delta(\Theta)$ be a sequence such that $\pi_n\Rightarrow\delta_{\theta_0}$ for some $\theta_0\in \Theta$. Let $g\in \Lambda_{\mathrm{D}}$ and $\gamma$ be a Lebesgue continuous statistic. Then, we have that
    $$\lim_{n\to\infty}\max_{\nu\in \mathcal{S}(\pi_n,g)}\gamma(X|\nu)=\gamma(X|\mu_{\theta_0}),~~X\in \mathcal{X}_b.$$
\end{theorem}

Finally, we have a result regarding the actual ambiguity sets. This result shows that, asymptotically, the first-order models in the ambiguity set are arbitrarily close to the true model. 

\begin{proposition}\label{prop:total}
    Let $(\pi_n)_{n\in \N}\subseteq \Delta(\Theta)$ be a sequence such that $\pi_n\Rightarrow\delta_{\theta_0}$ for some $\theta_0\in \Theta$ and $g\in \Lambda_{\mathrm{D}}$. Then, we have that
    $$\lim_{n\to\infty}\sup_{\nu\in \mathcal{S}(\pi_n,g)}d_{\mathrm{TV}}(\nu,\mu_{\theta_0})=0.$$
\end{proposition}

The next subsection discusses how concerns regarding model misspecification can be incorporated into these decision functionals.

\subsection{Model misspecification}
\label{sec:miss}

Model misspecification refers to the case in which the parametric model $(\Theta,\mathfrak{P})$ fails to contain the true first-order model that generates the random phenomenon.  In the Bayesian literature, there are results on the asymptotic behavior of the posterior distribution (second-order model). For example, as in the preceding subsection, suppose we observe an IID sequence of random variables $(\zeta_k)_{k\in \N}$ taking values in $\Omega$. We now consider the misspecified case, meaning that $\zeta_k\sim \mu_0$, but, unlike the preceding subsection, no parameter $\theta_0\in\Theta$ satisfies $\mu_0=\mu_{\theta_0}$. However, assume that the mapping
$$\Theta\to \mathbb{R}\cup\{\infty\}:\theta\mapsto \mathrm{KL}(\mu_0|\mu_\theta),$$
admits a unique minimizer. We denote this minimizer by $\theta_0$. Under certain regularity conditions, see \cite{B66}, we have that the sequence of second-order models $(\pi_n)_{n\in \N}$ given by the posterior distribution ($\pi_n$ is obtained after realizing $(\zeta_k)_{k=1}^n$) satisfies $$\pi_n\Rightarrow \delta_{\theta_0}~~\mu_0^\infty\text{-a.s.}$$
Therefore, a natural way for a DM to address model misspecification is to first robustify their statistic. The following proposition discusses how this can be done.

\begin{proposition}\label{prop:KL}
    Given a statistic $\gamma$, atomless $\mu\in \Delta$, and $\epsilon>0$, there exists a statistic $\gamma_{\epsilon}^{\mathrm{KL}}$ satisfying 
    $$\gamma_{\epsilon}^{\mathrm{KL}}(X|\mu)=\sup_{\nu\in \mathrm{KL}(\mu;\epsilon)}\gamma(X|\nu),~~X\in \mathcal{X}_b,$$
    where $\mathrm{KL}(\mu;\epsilon)=\{\nu\in \Delta~|~\mathrm{KL}(\nu|\mu)\leq \epsilon\}$. Moreover, if $\gamma$ is Lebesgue continuous, then $\gamma_{\epsilon}^{\mathrm{KL}}$ is Lebesgue continuous.
\end{proposition}

\begin{example}
    Given $\epsilon>0$, the statistic $\mathbb{E}^{\mathrm{KL}}_\epsilon$ is known as entropic value-at-risk; see \cite{FK11} and \cite{AJ12}.
\end{example}

Assuming the DM uses $\gamma$ to represent their pure-risk preferences, if they were worried about model misspecification, given $\pi\in \Delta(\Theta)$, $g\in \Lambda_{\mathrm{D}}$, and $\epsilon>0$, they could define the decision functional
\begin{equation}\label{eq:miss}
    \rho(X;\pi,g,\epsilon)=\sup_{\nu\in \mathcal{S}(\pi,g)}\gamma^{\mathrm{KL}}_\epsilon(X|\nu),~~X\in \mathcal{X}_b.
\end{equation}
The parameter $\epsilon$ in \eqref{eq:miss} represents the DM's fear of model misspecification. For example, if it holds that $\epsilon\geq \mathrm{KL}(\mu_0|\mu_{\theta_0})$ and $\gamma$ is Lebesgue continuous, then, by Theorem \ref{prop:consis}, we have
\begin{equation}\label{eq:misscon}\lim_{n\to\infty}\rho(X;\pi_n,g,\epsilon)=\gamma_\epsilon^{\mathrm{KL}}(X|\mu_{\theta_0})\geq \gamma(X|\mu_0),~~X\in \mathcal{X}_b.
\end{equation}
Therefore, the DM does not asymptotically underestimate the true value of their statistic under model misspecification.

    Independently of the asymptotic result, the decision functional in \eqref{eq:miss} is interesting from a conceptual point of view. This follows because the decision functional separates the perception of ambiguity ($\pi$), the attitude toward ambiguity ($g$), and the attitude toward model misspecification ($\epsilon$). The joint incorporation of attitudes toward ambiguity and model misspecification into decision functionals was discussed in \cite{CHMM25}.

    In \cite{SZL23}, the authors also study decision functionals in a Bayesian context that address model misspecification. In that work, given $\pi\in \Delta(\Theta)$ and $\epsilon>0$, the authors define the decision functional
    \begin{equation}\label{eq:SDF}
        \rho(X;\pi,\epsilon)=\int_{\Theta}\mathbb{E}_{\epsilon}^{\mathrm{KL}}(X|\mu_{\theta})\pi(\d\theta),~~X\in \mathcal{X}_b.
    \end{equation}

    One can show that for all $X\in \mathcal{X}_b$, a consistency result analogous to \eqref{eq:misscon} holds for the decision functional in \eqref{eq:SDF}. However, the decision functional \eqref{eq:SDF} has no mechanism to account for attitudes toward ambiguity. Moreover, we have the following proposition. We verify in Appendix \ref{app:6} that, given $X\in \mathcal{X}_b$, $\mathbb{E}_{\epsilon}^{\mathrm{KL}}\circ\iota_X$ is $\Sigma$-measurable.
\begin{proposition}\label{prop:need5}
    If $\tilde\pi,\pi\in \Pi$ satisfy $\tilde\pi\geq_{\mathrm{cx}}\pi$, then
    $$\int_{\Delta}\mathbb{E}_{\epsilon}^{\mathrm{KL}}(X|\nu)\tilde{\pi}(\d\nu)\leq \int_{\Delta}\mathbb{E}_{\epsilon}^{\mathrm{KL}}(X|\nu)\pi(\d\nu),~~X\in \mathcal{X}_b$$
    \end{proposition}
    
    Therefore, the decision functional in \eqref{eq:SDF} reflects an ambiguity-seeking attitude, as discussed in Section \ref{sec:3}. Although the decision functionals in \eqref{eq:miss} and \eqref{eq:SDF} exhibit a similar asymptotic property, they can lead to substantially different decisions in finite samples. 

\section{Conclusion}
Bayesian second-order beliefs can be incorporated into decision functionals without changing the statistic governing pure-risk preferences. The difficulty is that averaging model-specific evaluations need not preserve ambiguity neutrality or strong ambiguity aversion. Ambiguity irons address this difficulty by translating a second-order model and an ambiguity attitude into an ambiguity set, over which the original statistic is evaluated in the worst case. This construction gives Bayesian beliefs and distributionally robust optimization complementary roles in the same decision model.

The distortion ambiguity iron provides a concrete implementation of this approach. It produces convex, compact ambiguity sets and reduces evaluation to the prescribed statistic for acts with an unambiguous distribution. Alternative representations support numerical implementation, and an order-statistic approximation gives pointwise convergence for the underlying Choquet integrals. Portfolio examples illustrate how ambiguity perception and attitude affect evaluations and allocations. When the posterior concentrates on the data-generating model, the stated continuity conditions ensure that ambiguity sets contract and decision functionals recover their pure-risk values.

Posterior concentration does not eliminate misspecification when the model family excludes the data-generating distribution. The proposed KL-based robustification keeps this concern distinct from ambiguity within the model family. Further work may examine alternative ambiguity irons, convergence of optimal decisions, and dynamic decision problems with posterior updating.

\newpage

\appendix

\begin{center}
    \Large Appendices: Omitted proofs and additional results
\end{center}

\section{Additional results needed for the proofs}

Given $n\in \mathbb{N}$, denote by $\mathcal{X}_{n}$ the set of all $X\in \mathcal{X}_b$ such that $n\geq X\geq -n$. Additionally, denote by $\mathcal{C}_n$ the set of all $X\in \mathcal{C}_b$ such that $n\geq X\geq -n$.

\begin{lemma}\label{lemma:closed}
   Given $n\in \N$ and $\mathcal{Y}\subseteq \mathcal{X}_n$, if $\mathcal{Y}$ is closed under pointwise convergence and $\mathcal{C}_n\subseteq \mathcal{Y}$, then $\mathcal{Y}=\mathcal{X}_n$.
\end{lemma}

\begin{proof}
    Define 
    $\mathcal{Y}_0=\{X\in \mathcal{X}:(X\wedge n)\vee (-n)\in \mathcal{Y}\}.$ Since $\mathcal{Y}$ is closed under pointwise convergence, $\mathcal{Y}_0$ is closed under pointwise convergence. Since $\mathcal{C}_n\subseteq \mathcal{Y}$, $\mathcal{C}\subseteq \mathcal{Y}_0$. Therefore, by \citet[Theorem 11.6]{K12A}, $\mathcal{Y}_0=\mathcal{X}$. Therefore, $\mathcal{Y}=\mathcal{X}_n$.
\end{proof}

Denote by $\tau_b$ the coarsest topology on $\Delta$ where the maps $\{\psi_X\}_{X\in \mathcal{X}_b}$ are continuous. 

\begin{proposition}\label{prop:need4}
    Let $\gamma$ be a Lebesgue continuous statistic. Given $X\in\mathcal{C}_b$, $\gamma\circ\iota_X$ is $\tau$-continuous. Given $X\in \mathcal{X}_b$, $\gamma\circ\iota_X$ is $\tau_b$-continuous.
\end{proposition}

\begin{proof}
 Given $a,b\in \mathbb{R}$ satisfying $a\leq b$, denote the set of Borel probability measures on $\mathbb{R}$ with support contained in the interval $[a,b]$ by $\mathcal{D}_{[a,b]}$. Denote by $\tau_{\mathcal{D}_{[a,b]}}$ the subspace topology on $\mathcal{D}_{[a,b]}$ from the topology $\tau_\mathcal{D}$. As $\gamma$ is Lebesgue continuous, $\gamma|_{\mathcal{D}_{[a,b]}}$ is $\tau_{\mathcal{D}_{[a,b]}}$-continuous.

Let $X\in \mathcal{C}_b$. We claim that $\iota_X$ is $\tau/\tau_\mathcal{D}$-continuous. To see this, let $(\mu_k)_{k\in \N}\subseteq \Delta$ be a sequence such that $\mu_k\Rightarrow \mu$ for some $\mu\in \Delta$. Let $f:\mathbb{R}\to\mathbb{R}$ be bounded and continuous. We have
$$\lim_{k\to\infty}\int_{\mathbb{R}}f\d\iota_X(\mu_k)=\lim_{k\to\infty}\mathbb{E}^{\mu_k}[f(X)]=\mathbb{E}^{\mu}[f(X)]=\int_{\mathbb{R}}f\d\iota_X(\mu).$$
Therefore, $\iota_X(\mu_k)\Rightarrow \iota_X(\mu)$, and $\iota_X$ is $\tau/\tau_\mathcal{D}$-continuous. Since $\iota_X(\Delta)\subseteq \mathcal{D}_{[a,b]}$, where $a\leq X\leq b$, we have that $\gamma\circ\iota_X$ is $\tau$-continuous. A similar argument will show that, for all $X\in \mathcal{X}_b$, $\gamma\circ\iota_X$ is $\tau_b$-continuous.
\end{proof}

\begin{proposition}\label{prop:meas2}
    Let $\gamma$ be a Lebesgue continuous statistic and $X\in \mathcal{X}_b$. The mapping $\gamma\circ\iota_X$ is $\Sigma$-measurable.
\end{proposition}

\begin{proof}
    Fix $n\in \N$ and define
$$\mathcal{Y}=\left\{X\in \mathcal{X}_n~|~\gamma\circ\iota_X~\text{is}~\Sigma\text{-measurable}\right\}.$$
By Proposition \ref{prop:need4}, we have that $\mathcal{C}_n\subseteq \mathcal{Y}$.  Let $(X_k)_{k\in \N}\subseteq \mathcal{Y}$ converge pointwise to some $X\in \mathcal{X}_n$. Since $\gamma$ is Lebesgue continuous, $n\geq X_k\geq -n$ for all $k\in \N$, and $(X_k)_{\#}\nu\Rightarrow X_{\#}\nu$ for all $\nu\in \Delta$, we have that $$\lim_{k\to\infty}\gamma\circ\iota_{X_k}(\nu)=\lim_{k\to\infty}\gamma(X_k|\nu)=\gamma(X|\nu)=\gamma\circ\iota_{X}(\nu),~~\nu\in \Delta.$$
Since $\gamma\circ\iota_{X_k}$ is $\Sigma$-measurable for all $k\in \N$, $\gamma\circ\iota_{X}$ is $\Sigma$-measurable. By Lemma \ref{lemma:closed}, we have that $\mathcal{Y}=\mathcal{X}_n$. Since $n\in \N$ was arbitrary, the proof is complete. 
\end{proof}

Proposition \ref{prop:meas2} guarantees that the integrals in the following proposition are well-defined.

\begin{proposition}\label{prop:mac}
    Let $\gamma$ be a Lebesgue continuous statistic and $\varphi:\mathbb{R}\to\mathbb{R}$ be continuous. Assume that for all $\tilde\pi,\pi\in \Pi$ satisfying $\tilde{\pi}\geq_{\mathrm{cx}}\pi$ we have
    $$\int_{\Delta}\varphi(\gamma(X|\nu))\tilde{\pi}(\d\nu)\geq \int_{\Delta}\varphi(\gamma(X|\nu))\pi(\d\nu),~~X\in \mathcal{C}_b.$$
Then, for all $\tilde\pi,\pi\in \Pi$ satisfying $\tilde{\pi}\geq_{\mathrm{cx}}\pi$ we have
$$\int_{\Delta}\varphi(\gamma(X|\nu))\tilde{\pi}(\d\nu)\geq \int_{\Delta}\varphi(\gamma(X|\nu))\pi(\d\nu),~~X\in \mathcal{X}_b.$$
\end{proposition}

\begin{proof}
    Fix $n\in \N$ and define 
    $$\mathcal{Y}=\left\{X\in \mathcal{X}_n~\bigg|~\int_{\Delta}\varphi(\gamma(X|\nu))\tilde{\pi}(\d\nu)\geq \int_{\Delta}\varphi(\gamma(X|\nu))\pi(\d\nu)~\text{for all}~\tilde\pi,\pi\in \Pi~\text{satisfying}~\tilde{\pi}\geq_{\mathrm{cx}}\pi\right\}.$$
    By assumption we have that $\mathcal{C}_n\subseteq \mathcal{Y}$. Let $(X_k)_{k\in \N}\subseteq \mathcal{Y}$ converge pointwise to some $X\in \mathcal{X}_n$. Since $\gamma$ is Lebesgue continuous, $n\geq X_k\geq -n$ for all $k\in \N$, and $(X_k)_{\#}\nu\Rightarrow X_{\#}\nu$ for all $\nu\in \Delta$, we have that 
    $$\lim_{k\to\infty}\gamma(X_k|\nu)=\gamma(X|\nu),~~\nu\in \Delta.$$ Given $\tilde\pi,\pi\in \Pi$ satisfying $\tilde{\pi}\geq_{\mathrm{cx}}\pi$, by the dominated convergence theorem, we have 
    $$\int_{\Delta}\varphi(\gamma(X|\nu))\tilde{\pi}(\d\nu)=\lim_{k\to\infty}\int_{\Delta}\varphi(\gamma(X_k|\nu))\tilde{\pi}(\d\nu)\geq \lim_{k\to\infty}\int_{\Delta}\varphi(\gamma(X_k|\nu))\pi(\d\nu)=\int_{\Delta}\varphi(\gamma(X|\nu))\pi(\d\nu).$$
    Since $\pi,\tilde\pi$ were arbitrary, we have that $X\in \mathcal{Y}$. By Lemma \ref{lemma:closed}, we have that $\mathcal{Y}=\mathcal{X}_n$. Since $n\in \N$ was arbitrary, the proof is complete.
\end{proof}

\begin{corollary}\label{cor:need}
    Given convex $\varphi:\mathbb{R}\to\mathbb{R}$ and $\tilde\pi,\pi\in \Pi$ satisfying $\tilde\pi\geq_{\mathrm{cx}}\pi$, we have
    $$\int_{\Delta}\varphi(\mathbb{E}^{\nu}[X])\tilde\pi(\d\nu)\geq \int_{\Delta}\varphi(\mathbb{E}^{\nu}[X])\pi(\d\nu),~~X\in \mathcal{X}_b.$$
\end{corollary}

\begin{proof}
    Let $X\in \mathcal{C}_b$. Since $\mathbb{E}\circ\iota_X$ is $\tau$-continuous by Proposition \ref{prop:need4}, we have that $\nu\mapsto \varphi(\mathbb{E}^{\nu}[X])$ is convex and $\tau$-continuous. Therefore, by the definition of $\geq_{\mathrm{cx}}$, for all $\tilde\pi,\pi\in \Pi$ satisfying $\tilde\pi\geq_{\mathrm{cx}}\pi$, we have $$\int_{\Delta}\varphi(\mathbb{E}^{\nu}[X])\tilde\pi(\d\nu)\geq \int_{\Delta}\varphi(\mathbb{E}^{\nu}[X])\pi(\d\nu).$$
    Since $X\in \mathcal{C}_b$ was arbitrary, Proposition \ref{prop:mac} completes the proof.
\end{proof}

\section{Proofs accompanying Section \ref{sec:2}}

\label{app:proof}

\begin{verification}
    The mapping $\psi_X$ is $\Sigma$-measurable for all $X\in \mathcal{X}$. 
\end{verification}

\begin{proof}
    If $X\in \mathcal{X}$ has finite range, by \citet[Proposition 7.25]{BS96A}, the mapping $\psi_X$ is $\Sigma$-measurable.

    If $X\in \mathcal{X}$ is non-negative, find an increasing sequence of simple acts $(X_n)_{n\in \N}\subseteq \mathcal{X}$ that converges pointwise to $X$.
    By the monotone convergence theorem, we have
    $$\lim_{n\to\infty}\psi_{X_n}(\mu)=\lim_{n\to\infty}\mathbb{E}^{\mu}[X_n]=\mathbb{E}^{\mu}[X]=\psi_X(\mu),~~\mu\in \Delta.$$
    Since $\psi_{X_n}$ is $\Sigma$-measurable for all $n\in \N$, $\psi_X$ is $\Sigma$-measurable.

    If $X\in \mathcal{X}$ is arbitrary, define 
    $$A_1=\left\{\mu\in \Delta:\mathbb{E}^{\mu}[X_+]=\infty\right\}~~\text{and}~~A_2=\left\{\mu\in \Delta:\mathbb{E}^{\mu}[X_{-}]=\infty\right\}.$$
    Taking $A=A_1\cap A_2$, we have
    $\psi_X(\mu)=\psi_{X_+}(\mu)-\psi_{X_-}(\mu)$ for $\mu\in A^c$, and $\psi_X(\mu)=0$ for $\mu\in A$. This piecewise-defined function is $\Sigma$-measurable.
\end{proof}

\begin{verification}
    For $\pi\in\Pi$ and $X\in L^1(\pi)$, we have
    $\int_{\Delta}\mathbb{E}^{\nu}[X]\pi(\d\nu)=\mathbb{E}^{\mu_{\pi}}[X].$
\end{verification}

\begin{proof}
 If $X=\sum_{k=1}^n a_k\id_{A_k}$, where $\{A_k\}_{k=1}^n\subseteq \mathcal{F}$ partitions $\Omega$ (i.e., $X$ is simple), then we have
    $$\int_{\Delta}\mathbb{E}^{\nu}[X]\pi(\d\nu)=\int_{\Delta}\sum_{k=1}^n a_k\nu(A_k)\pi(\d\nu)=\sum_{k=1}^n a_k\mu_{\pi}(A_k)=\mathbb{E}^{\mu_{\pi}}[X].$$

If $X\in \mathcal{X}$ is non-negative, find an increasing sequence of simple acts $(X_n)_{n\in \N}\subseteq \mathcal{X}$ that converges pointwise to $X$. By three applications of the monotone convergence theorem, we have
    \begin{align*}
        \int_{\Delta}\mathbb{E}^{\nu}[X]\pi(\d\nu)=\int_{\Delta}\lim_{n\to\infty}\mathbb{E}^{\nu}[X_n]\pi(\d\nu)=\lim_{n\to\infty}\int_{\Delta}\mathbb{E}^{\nu}[X_n]\pi(\d\nu)=\lim_{n\to\infty}\mathbb{E}^{\mu_{\pi}}[X_n]=\mathbb{E}^{\mu_{\pi}}[X].
    \end{align*}
    
    If $X\in L^1(\pi)$ is general, first  note that
     $$\max\left\{\int_{\Delta}\mathbb{E}^{\nu}[X_+]\pi(\d\nu),\int_{\Delta}\mathbb{E}^{\nu}[X_-]\pi(\d\nu)\right\}\leq\int_{\Delta}\left(\mathbb{E}^\nu[X_+]+\mathbb{E}^\nu[X_-]\right)\pi(\d\nu)=\int_{\Delta}\mathbb{E}^{\nu}[|X|]\pi(\d\nu)<\infty.$$
     Therefore, $\mathbb{E}^{\nu}[X_+]<\infty$ $\pi$-a.s., $\mathbb{E}^{\nu}[X_-]<\infty$ $\pi$-a.s., and $\mathbb{E}^{\nu}[X]=\mathbb{E}^{\nu}[X_+]-\mathbb{E}^{\nu}[X_-]$ $\pi$-a.s. Thus,
    \begin{align*}
        \int_{\Delta}\mathbb{E}^{\nu}[X]\pi(\d\nu)=\int_{\Delta}\mathbb{E}^{\nu}[X_+]\pi(\d\nu)-\int_{\Delta}\mathbb{E}^{\nu}[X_{-}]\pi(\d\nu)=\mathbb{E}^{\mu_{\pi}}[X_{+}]-\mathbb{E}^{\mu_{\pi}}[X_{-}]=\mathbb{E}^{\mu_{\pi}}[X],
    \end{align*}
    and the proof is complete.
\end{proof}

\section{Proofs accompanying Section \ref{sec:3}}

\label{app:3}

Given a collection $\mathcal{Z}\subseteq \mathcal{F}$, we denote by $\sigma(\mathcal{Z})$ the $\sigma$-algebra generated by $\mathcal{Z}$. Given a collection $\mathcal{Y}\subseteq \mathcal{X}$, we denote by $\sigma(\mathcal{Y})$ the coarsest $\sigma$-algebra making each $X\in \mathcal{Y}$ measurable.

\begin{proof}[Proof of Theorem \ref{theo:big}]
By \citet[Proposition 7.25]{BS96A}, we know that $$\Sigma=\sigma(\{\psi_X\}_{X\in \mathcal{X}_b}).$$

 Let $\pi,\tilde{\pi}\in \Pi$ satisfy $\pi\geq_{\mathrm{cx}}\tilde{\pi}$ and $\tilde{\pi}\geq_{\mathrm{cx}}\pi$. Let $\Gamma\subseteq \Sigma$ denote the $\pi$-system consisting of all finite intersections of sets of the form $\psi_X^{-1}(B)$, where $X\in \mathcal{X}_b$ and $B\in \mathcal{B}(\mathbb{R})$. Clearly, $$\Sigma=\sigma(\{\psi_X\}_{X\in \mathcal{X}_b})=\sigma(\Gamma).$$ If we can show that $\pi$ and $\tilde{\pi}$ agree on $\Gamma$, then, by the $\pi$-$\lambda$ theorem, $\pi=\tilde{\pi}$. To this end, let $\{X_k\}_{k=1}^n\subseteq \mathcal{X}_b$ and $\{B_k\}_{k=1}^n\subseteq \mathcal{B}(\mathbb{R})$. Let $t\in\mathbb{R}^n$ and define $Z(t)=\sum_{k=1}^n t_k\psi_{X_k}$. By Corollary \ref{cor:need}, for all convex $\varphi:\mathbb{R}\to\mathbb{R}$, we have 
\begin{align*}
    \int_{\Delta}\varphi\left(Z(t)\right)\d\pi=\int_{\Delta}\varphi\left( \mathbb{E}^{\nu}\left[\sum_{k=1}^n t_k X_k\right]\right)\pi(\d\nu)=\int_{\Delta}\varphi\left( \mathbb{E}^{\nu}\left[\sum_{k=1}^n t_k X_k\right]\right)\tilde\pi(\d\nu)=\int_{\Delta}\varphi\left(Z(t)\right)\d\tilde\pi.
\end{align*}
Since $\varphi$ was an arbitrary convex function, we have 
$\left(Z(t)\right)_{\#}\pi=\left(Z(t)\right)_{\#}\tilde\pi.$
Since $t\in \mathbb{R}^n$ was arbitrary, 
$(\psi_{X_1},\dots, \psi_{X_n})_{\#}\pi=(\psi_{X_1},\dots, \psi_{X_n})_{\#}\tilde\pi.$
Therefore, 
    $$\pi\left(\bigcap_{k=1}^n\psi_{X_k}^{-1}(B_k)\right)=\tilde{\pi}\left(\bigcap_{k=1}^n\psi_{X_k}^{-1}(B_k)\right).$$
Thus, $\pi$ and $\tilde{\pi}$ agree on $\Gamma$, and the proof is complete. 
\end{proof}

\begin{proof}[Proof of Proposition \ref{prop:check}]
    We will use \citet[Theorem 2]{V17A} to prove the claim. First note that $\Delta$ can be embedded into the vector space of signed measures with finite total variation $\mathcal{M}$. Additionally, we can equip $\mathcal{M}$ with the coarsest topology $\tilde{\tau}$ making the mappings $$\mathcal{M}\to\mathbb{R}:\nu\mapsto \mathbb{E}^{\nu}[X],~~X\in \mathcal{C}_b,$$
    continuous. Clearly, $\tilde\tau|_{\Delta}=\tau$.

    We claim that $\Delta\subseteq\mathcal{M}$ is $\tilde\tau$-closed. To prove the claim, let $(\mu_{\eta})_{\eta\in\mathcal{N}}\subseteq \Delta$ be a net converging to some $\mu\in\mathcal{M}$ in the topology $\tilde\tau$. Clearly, $\mu(\Omega)=1$. Let $d$ be a metric for $\Omega$, $C\in \mathcal{F}$ be closed, and $\epsilon>0$. Define $X^C_\epsilon\in \mathcal{C}_b$ by
    $$X^C_\epsilon(\omega)=\left(1-\frac{d(\omega,C)}{\epsilon}\right)_+,~~\omega\in \Omega,$$
    where $d(\omega,C)=\inf\{d(\omega,\tilde\omega)~|~\tilde\omega\in C\}$. Since $X^C_\epsilon\geq 0$, we have that $\mathbb{E}^\mu[X^C_\epsilon]\geq 0$. Therefore,
    $$\mu(C)=\lim_{\epsilon\to 0^+}\mathbb{E}^{\mu}[X^C_\epsilon]\geq 0.$$
    For the sake of contradiction, assume there exists $A\in \mathcal{F}$ such that $\mu(A)<0$. By inner regularity of $\mu$, there exists a closed $C\in \mathcal{F}$ such that $\mu(C)<0$, which is a contradiction. Therefore, $\mu(A)\geq 0$ for all $A\in \mathcal{F}$. Therefore, $\mu\in \Delta$ and we have proven the earlier claim.
    
   Since $\Delta$ is $\tilde{\tau}$-closed, $\Delta$ is evenly convex, as defined in \cite{V17A}. A consequence of \eqref{eq:pred} is that, for all $\pi\in \Pi$, $\mu_\pi$ is the barycenter of $\pi$; see \cite{V17A} for the definition. Let $\pi\in \Pi$ and $f:\Delta\to\mathbb{R}$ be convex and $\tau$-continuous. Then, by \citet[Theorem 2]{V17A}, we have
    $$\int_{\Delta}f(\nu)\pi(\d\nu)\geq f(\mu_{\pi})=\int_{\Delta}f(\nu)\delta_{\mu_\pi}(\d\nu).$$
    Therefore, since $f$ was arbitrary, $\pi\geq_{\mathrm{cx}}\delta_{\mu_\pi}$.
\end{proof}

\begin{proof}[Proof of Proposition \ref{prop:smooth-properties}]
    Property \ref{property1} is clear. Property \ref{property2} follows from \eqref{eq:pred}. Property \ref{property3} follows by applying Jensen's inequality. Property \ref{property4} follows from Corollary \ref{cor:need}.
\end{proof}

\begin{proof}[Proof of Theorem \ref{theo:mainmot}]
    Define the mapping $\kappa:\Omega\to\Delta$ by $\kappa(\omega)=\delta_{\omega}.$ It is clear that $\kappa$ is continuous and thus measurable. Fix an atomless $\mu\in \Delta$; an atomless first-order model exists because $\Omega$ is uncountable and Polish. Define $\pi_0\in \Pi$ by $\pi_0=\kappa_{\#}\mu$. 
    
    We claim that $\mathfrak{R}(\pi_0)$ is atomless. To see this, let $A\in \mathcal{F}$ satisfy $\mu(A)=0$. Since 
    $$\int_{\Delta}\nu(A)\pi_0(\d\nu)=\int_{\Omega}[\kappa(\omega)](A)\mu(\d\omega)=\int_{\Omega}\id_A(\omega)\mu(\d\omega)=\mu(A)=0,$$
    we have that $\pi_0(\nu\in \Delta:\nu(A)=0)=1$, and $\left[\mathfrak{R}(\pi_0)\right](A)=0$. Therefore, $\mathfrak{R}(\pi_0)\ll \mu$ and, since $\mu$ is atomless, $\mathfrak{R}(\pi_0)$ is atomless.
    
    By the definition of $\kappa$, we have that $$\int_{\Delta}\ell(\gamma(X|\nu))\pi_0(\d\nu)=\int_{\Omega}\ell(\gamma(X|\kappa(\omega)))\mu(\d\omega)=\int_{\Omega}\ell\left(\gamma\left(\delta_{X(\omega)}\right)\right)\mu(\d\omega)=\mathbb{E}^\mu[\ell(X)],~~X\in \mathcal{X}_b.$$ Therefore, by assumption, we have that
    $$\mathbb{E}_\ell(X|\mu)=\ell^{-1}\left(\mathbb{E}^\mu[\ell(X)]\right)=\ell^{-1}\left(\int_{\Delta}\ell(\gamma(X|\nu))\pi_0(\d\nu)\right)=\gamma(X|\mathfrak{R}(\pi_0)),~~X\in \mathcal{X}_b.$$
    Since $\mu$ and $\mathfrak{R}(\pi_0)$ are both atomless, by \citet[Proposition 2 and Theorem 6]{L24A}, $\mathfrak{R}(\pi_0)=\mu$. Therefore, $\mathbb{E}_\ell(X|\mu)=\gamma(X|\mu)$ for all $X\in \mathcal{X}_b$. Since $\mu$ is atomless, we have that $\gamma(Q)=\mathbb{E}_\ell(Q)$ for all $Q\in \mathcal{D}_c$. Thus, $\gamma=\mathbb{E}_\ell$.

    Finally, for all $\pi\in \Pi$, by assumption, we have that
    \begin{align*}
        \mathbb{E}_\ell(X|\mu_\pi)&=\ell^{-1}\left(\int_{\Delta}\ell(\mathbb{E}_\ell(X|\nu))\pi(\d\nu)\right)\\&=\ell^{-1}\left(\int_{\Delta}\ell(\gamma(X|\nu))\pi(\d\nu)\right)=\gamma(X|\mathfrak{R}(\pi))=\mathbb{E}_\ell(X|\mathfrak{R}(\pi)),~~X\in \mathcal{X}_b.
    \end{align*}
    Therefore, for all $A\in \mathcal{F}$, we have that
    $$\ell^{-1}(\mu_\pi(A))=\mathbb{E}_{\ell}(\id_A|\mu_\pi)=\mathbb{E}_\ell(\id_A|\mathfrak{R}(\pi))=\ell^{-1}(\left[\mathfrak{R}(\pi)\right](A)),$$
    and $\mathfrak{R}(\pi)=\mu_\pi$. Since $\pi$ was arbitrary, the proof is complete. 

    Conversely, if $\gamma=\mathbb{E}_{\ell}$ and $\mathfrak{R}(\pi)=\mu_{\pi}$ for every $\pi\in\Pi$, then \eqref{eq:mainmot} follows immediately from \eqref{eq:pred}.
\end{proof}

\begin{verification}
    For all $X\in \mathcal{X}_b$, the mapping $\mathrm{ES}_\alpha\circ\iota_X$ is $\Sigma$-measurable.
\end{verification}

\begin{proof}
     Since $\mathrm{ES}_\alpha$ is Lebesgue continuous, by Proposition \ref{prop:meas2}, $\mathrm{ES}_\alpha\circ\iota_X$ is $\Sigma$-measurable for $X\in \mathcal{X}_b$.
\end{proof}

\begin{proof}[Proof of Proposition \ref{prop:need3}]
    Let $X\in \mathcal{C}_b$. As $\mathrm{ES}_\alpha$ is Lebesgue continuous, by Proposition \ref{prop:need4}, $\mathrm{ES}_\alpha\circ\iota_X$ is $\tau$-continuous. Therefore, $-\mathrm{ES}_\alpha\circ\iota_X$ is convex and $\tau$-continuous. Therefore, by the definition of $\geq_{\mathrm{cx}}$, for all $\tilde\pi,\pi\in \Pi$ satisfying $\tilde\pi\geq_{\mathrm{cx}}\pi$, we have $$\int_{\Delta}-\mathrm{ES}_{\alpha}(X|\nu)\tilde\pi(\d\nu)\geq \int_{\Delta}-\mathrm{ES}_{\alpha}(X|\nu)\pi(\d\nu).$$
    Since $X\in \mathcal{C}_b$ was arbitrary, by Proposition \ref{prop:mac}, we have
    $$\int_{\Delta}-\mathrm{ES}_{\alpha}(X|\nu)\tilde\pi(\d\nu)\geq \int_{\Delta}-\mathrm{ES}_{\alpha}(X|\nu)\pi(\d\nu),~~X\in \mathcal{X}_b,$$
    and the proof is complete.
\end{proof}

\section{Proofs and additional results accompanying Section \ref{sec:4}}

\label{app:distortion}

\begin{proof}[Proof of Proposition \ref{prop:compact}]
    Convexity is clear, as $\mathcal{R}(\pi,g)$ is convex. Let $(\mu_{\eta})_{\eta\in \mathcal{N}}\subseteq \mathcal{S}(\pi,g)$ be a net. We claim that there exists a subnet that converges in $\tau_b$ to some $\mu\in \mathcal{S}(\pi,g)$. Let $(\xi_{\eta})_{\eta\in \mathcal{N}}\subseteq \mathcal{R}(\pi,g)$ be a net such that $\mu_{\xi_\eta}=\mu_{\eta}$ for all $\eta\in \mathcal{N}$. Since $g\circ\pi$ is submodular and continuous, by \cite[Theorem 4.2]{MM04A} and \citet[Theorem 7.14]{Br13A}, there exists a subnet $(\xi_{\varphi(\beta)})_{\beta\in \mathcal{B}}$ and $\xi\in \mathcal{R}(\pi,g)$ such that
    $$\lim_{\beta}\int_{\Delta}f\d\xi_{\varphi(\beta)}=\int_{\Delta}f\d\xi$$
    for all bounded $\Sigma$-measurable $f:\Delta\to\mathbb{R}$. Define $\mu=\mu_\xi\in \mathcal{S}(\pi,g)$. We have that
    $$\lim_{\beta}\mathbb{E}^{\mu_{\varphi(\beta)}}[X]=\lim_{\beta}\int_{\Delta}\mathbb{E}^\nu[X]\xi_{\varphi(\beta)}(\d\nu)=\int_{\Delta}\mathbb{E}^{\nu}[X]\xi(\d\nu)=\mathbb{E}^{\mu}[X],~~X\in \mathcal{X}_b.$$ Therefore, $\mu_{\varphi(\beta)}\to \mu$ in $\tau_b$. Since the original net $(\mu_{\eta})_{\eta\in \mathcal{N}}$ was arbitrary, by \citet[Theorem 7.14]{Br13A}, $\mathcal{S}(\pi,g)$ is $\tau_b$-compact. Since $\tau\subseteq \tau_b$, $\mathcal{S}(\pi,g)$ is $\tau$-compact.
\end{proof}

From the proof of Proposition \ref{prop:compact}, we immediately obtain the following corollary.

\begin{corollary}\label{corr:1}
    Let $\pi\in \Pi$ and $g\in \Lambda_{\mathrm{D}}$. Then $\mathcal{S}(\pi,g)$ is $\tau_b$-compact.
\end{corollary}

\begin{verification}
    The mapping $\nu\mapsto \mathrm{KL}(\mu|\nu)$ is $\Sigma$-measurable for all $\mu\in \Delta$.
\end{verification}

\begin{proof}
    Since $(\Omega,\mathcal{F})$ is a Polish measurable space, by \citet[Equation (2.1)]{DV83A}, we know that for all $\mu,\nu\in \Delta$,
    $$\mathrm{KL}(\mu|\nu)=\sup_{X\in \mathcal{C}_b}\left(\mathbb{E}^{\mu}[X]-\log\left(\mathbb{E}^{\nu}\left[e^X\right]\right)\right).$$
    Given $\mu\in \Delta$ and $X\in \mathcal{C}_b$, since the mapping 
    $$\Delta\to\mathbb{R}:\nu\mapsto \mathbb{E}^{\mu}[X]-\log\left(\mathbb{E}^{\nu}\left[e^X\right]\right)$$
    is $\tau$-continuous, the mapping
    $\nu\mapsto \mathrm{KL}(\mu|\nu)$ is $\tau$-lower semicontinuous. Since $\tau$-lower semicontinuity implies $\Sigma$-measurability, we have shown the original claim.
\end{proof}

\begin{proof}[Proof of Theorem \ref{theo:Main}]
    The fact that $\mathcal{S}$ satisfies the first three conditions of Definition \ref{def:bus-map} is clear. To show the fourth condition, let $\pi,\tilde{\pi}\in \Pi$ with $\tilde{\pi}\geq_{\mathrm{cx}}\pi$. 
    
    Fix $X\in \mathcal{X}_b$. If $\varphi:\mathbb{R}\to\mathbb{R}$ is convex, by Corollary \ref{cor:need}, we have that 
    $$\int_{\mathbb{R}}\varphi(x)(\psi_X)_{\#}\tilde{\pi}(\d x)=\int_{\Delta}\varphi(\psi_X(\nu))\tilde{\pi}(\d\nu)\geq \int_{\Delta}\varphi(\psi_X(\nu))\pi(\d\nu)=\int_{\mathbb{R}}\varphi(x)(\psi_X)_{\#}\pi(\d x).$$
    Therefore, $(\psi_X)_{\#}\tilde{\pi}\geq_{\mathrm{cx}}(\psi_X)_{\#}\pi,$ where $\geq_\mathrm{cx}$ denotes the traditional convex order; see \citet[Section 3.A]{SS07A}. Therefore, we have
    \begin{align*}
        \sup_{\nu\in \mathcal{S}(\tilde{\pi},g)}\mathbb{E}^{\nu}[X]=\sup_{\xi\in \mathcal{R}(\tilde{\pi},g)}\int_{\Delta}\mathbb{E}^{\nu}[X]\xi(\d\nu)&=\int_{\Delta}\mathbb{E}^{\nu}[X]g\circ\tilde\pi(\d\nu)\\&=\int_{\Delta}\psi_X(\nu)g\circ\tilde\pi(\d\nu)\\&=\mathbb{E}^g((\psi_X)_{\#}\tilde\pi)\geq \mathbb{E}^g((\psi_X)_{\#}\pi)= \sup_{\nu\in \mathcal{S}(\pi,g)}\mathbb{E}^{\nu}[X],
    \end{align*}
    where the inequality follows from the fact that $\mathbb{E}^g(Q)\geq \mathbb{E}^g(P)$ whenever $Q\geq_{\mathrm{cx}}P$, since $g$ is concave; see \citet[Theorem 3.A.7]{SS07A}. 
    
    Let $\mu\in \mathcal{S}(\pi,g)$. We have that
\begin{equation}\label{eq:con}
    \mathbb{E}^{\mu}[X]\leq\sup_{\nu\in \mathcal{S}(\pi,g)}\mathbb{E}^{\nu}[X]\leq \sup_{\nu\in \mathcal{S}(\tilde{\pi},g)}\mathbb{E}^{\nu}[X],~~X\in \mathcal{X}_b.
\end{equation}
For the sake of contradiction, assume that $\mu\notin \mathcal{S}(\tilde\pi,g)$. Since $\mathcal{S}(\tilde\pi,g)$ is convex and $\tau$-compact by Proposition \ref{prop:compact}, by the Hahn-Banach separation theorem, there exists $Y\in \mathcal{C}_b$ such that
$$\sup_{\nu\in \mathcal{S}(\tilde\pi,g)}\mathbb{E}^{\nu}[Y]<\mathbb{E}^\mu[Y],$$
which is a contradiction of \eqref{eq:con}. Therefore, $\mu\in \mathcal{S}(\tilde{\pi},g)$, and $\mathcal{S}(\pi,g)\subseteq \mathcal{S}(\tilde\pi,g)$.
\end{proof}

\begin{proof}[Proof of claim in Example \ref{example:beta}]
    Recall that $\Delta$ can be identified with $[0,1]$ and $\Pi$ with distributions supported on $[0,1]$; see Example \ref{ex:ber} for details. We have that
    \begin{align*}
        p_u(a,b,\beta)=\sup\left(\mathcal{S}(\pi_{a,b},g_\beta)\right)&=\sup_{\xi\in \mathcal{R}(\pi_{a,b},g_\beta)}\int_{[0,1]} p\xi(\d p)\\&=\int_{[0,1]} pg_\beta\circ\pi_{a,b}(\d p)=\frac{1}{1-\beta}\int_\beta^1\mathrm{VaR}_\alpha(\pi_{a,b})\d\alpha,
    \end{align*}
    where the last equality follows from the fact that (see, for example, \citet[Theorem 4]{DKLT12A}), given any absolutely continuous distortion function $g$, we have \begin{equation}\label{eq:distortion}
        \mathbb{E}^g(Q)=\int_{[0,1]}\mathrm{VaR}_\alpha(Q)g'(1-\alpha)\d\alpha,~~Q\in\mathcal{D}_c.
    \end{equation} For the lower endpoint, define the distortion function $g^*_\beta$ by $g^*_\beta(x)=1-g_\beta(1-x)$. By \citet[Proposition 5.1]{D94A} we have that $$\int_{[0,1]}-pg_\beta\circ \pi_{a,b}(\d p)=-\int_{[0,1]}pg^*_\beta\circ \pi_{a,b}(\d p).$$
    Therefore,
    \begin{align*}
        p_l(a,b,\beta)=-\sup\left(-\mathcal{S}(\pi_{a,b},g_\beta)\right)&=-\sup_{\xi\in \mathcal{R}(\pi_{a,b},g_\beta)}\int_{[0,1]} -p\xi(\d p)\\&=-\int_{[0,1]} -pg_\beta\circ\pi_{a,b}(\d p)\\&=\int_{[0,1]} pg^*_\beta\circ\pi_{a,b}(\d p)=\frac{1}{1-\beta}\int_0^{1-\beta}\mathrm{VaR}_\alpha(\pi_{a,b})\d\alpha,
    \end{align*}
    where the last equality follows from \eqref{eq:distortion}.
\end{proof}

\begin{proof}[Proof of Theorem \ref{theo:minmax}]

Choose $a\in\mathbb{N}$ such that $g(t)\leq at$ for all
$t\in[0,1]$. Every $\xi\in\mathcal{R}(\pi,g)$ satisfies
$\xi\leq a\pi$. Hence, if
$\nu=\mu_\xi\in\mathcal{S}(\pi,g)$, then
$$
\nu(A)
=
\int_\Delta \eta(A)\,\xi(\d\eta)
\leq
a\int_\Delta \eta(A)\,\pi(\d\eta)
=
a\mu_\pi(A),
~~A\in\mathcal{F}.
$$
Consequently, for every $Z\in L^1(\pi)$,
$
\mathbb{E}^{\nu}[|Z|]
\leq a\mathbb{E}^{\mu_\pi}[|Z|]<\infty
$.

For $m\in\mathbb{R}$, write $Y_m=h(X-m)$ and define
$
R(\nu,m)=m+\mathbb{E}^{\nu}[Y_m]$,
where $\nu\in \mathcal{S}(\pi,g)$ and $m\in \mathbb{R}$.
The integrability assumptions ensure that $R$ is finite.
It is affine in $\nu$ and convex in $m$.

Fix $m\in\mathbb{R}$ and let
$(\mu_n)_{n\in\mathbb{N}}\subseteq\mathcal{S}(\pi,g)$
converge in $\tau$ to $\mu\in\mathcal{S}(\pi,g)$.
By \citet[Lemma 1.33]{K97A}, choose
$(f_k)_{k\in\mathbb{N}}\subseteq\mathcal{C}_b$ such that
$
\lim_{k\to\infty}
\mathbb{E}^{\mu_\pi}[|Y_m-f_k|]=0.
$
We have that
\begin{align*}
\left|
\mathbb{E}^{\mu_n}[Y_m]-\mathbb{E}^{\mu}[Y_m]
\right|
&\leq
\mathbb{E}^{\mu_n}[|Y_m-f_k|]
+
\left|
\mathbb{E}^{\mu_n}[f_k]-\mathbb{E}^{\mu}[f_k]
\right|
+
\mathbb{E}^{\mu}[|f_k-Y_m|]\\
&\leq
2a\mathbb{E}^{\mu_\pi}[|Y_m-f_k|]
+
\left|
\mathbb{E}^{\mu_n}[f_k]-\mathbb{E}^{\mu}[f_k]
\right|.
\end{align*}
 Since $f_k\in \mathcal{C}_b$, $\lim_{n\to\infty}|\mathbb{E}^{\mu_n}[f_k]-\mathbb{E}^{\mu}[f_k]|=0$. Therefore, we have that $$\limsup_{n\to\infty}|\mathbb{E}^{\mu_n}[Y_m]-\mathbb{E}^{\mu}[Y_m]|\leq \lim_{k\to\infty}2a\mathbb{E}^{\mu_\pi}[|Y_m-f_k|]=0.$$
    Thus, $\lim_{n\to\infty}R(\mu_n,m)=R(\mu,m)$. Continuity in $m$ follows from finiteness and convexity.

By Proposition \ref{prop:compact},
$\mathcal{S}(\pi,g)$ is convex and $\tau$-compact.
Sion's minimax theorem
\citep[Theorem I.1.1]{MSZ15A} therefore gives
\begin{align*}
\sup_{\nu\in\mathcal{S}(\pi,g)}
\mathrm{OCE}_h(X|\nu)
=
\sup_{\nu\in\mathcal{S}(\pi,g)}
\inf_{m\in\mathbb{R}}R(\nu,m)
&=
\inf_{m\in\mathbb{R}}
\sup_{\nu\in\mathcal{S}(\pi,g)}R(\nu,m)\\
&=
\inf_{m\in\mathbb{R}}
\left\{
m+
\sup_{\nu\in\mathcal{S}(\pi,g)}
\mathbb{E}^{\nu}[h(X-m)]
\right\}.
\end{align*}
For each $m\in \mathbb{R}$, by
\eqref{eq:Choquet}, we have that
\begin{align*}
\sup_{\nu\in\mathcal{S}(\pi,g)}
\mathbb{E}^{\nu}[h(X-m)]
=
\sup_{\xi\in\mathcal{R}(\pi,g)}
\int_\Delta
\mathbb{E}^{\nu}[h(X-m)]\,\xi(\d\nu)
=
\int_\Delta
\mathbb{E}^{\nu}[h(X-m)]\,g\circ\pi(\d\nu).
\end{align*}
Substituting this identity proves the formula.
\end{proof}

\begin{proof}[Proof of Theorem \ref{prop:var}]
For $x\in\mathbb{R}$, define
$
H(x)=
\sup_{\nu\in\mathcal{S}(\pi,g)}\nu(X>x).
$
Applying 
\eqref{eq:Choquet} gives
\begin{align*}
H(x)
=
\sup_{\xi\in\mathcal{R}(\pi,g)}
\int_\Delta \nu(X>x)\,\xi(\d\nu)&=
\int_\Delta \nu(X>x)\,g\circ\pi(\d\nu)\\
&=
\int_0^1
g\left(
\pi(\nu:\nu(X>x)>t)
\right)\d t.
\end{align*}
Since $\nu(X>x)\downarrow0$ as
$x\to\infty$ for every $\nu\in\Delta$,
for every $t\in(0,1)$,
$$
\lim_{x\to\infty}\pi(\nu:\nu(X>x)>t)=0.
$$
Since $g$ is continuous, by the dominated convergence theorem,
we have that $\lim_{x\to\infty}H(x)=0$.

Choose $b\in\mathbb{R}$ such that $H(b)\leq1-\alpha$.
Then
$$
\nu(X>b)\leq H(b)\leq1-\alpha,~~\nu\in \mathcal{S}(\pi,g).$$
Therefore, for all $\nu\in \mathcal{S}(\pi,g)$, $\mathrm{VaR}_{\alpha}(X|\nu)\leq b$, and
$$
x_0=
\sup_{\nu\in\mathcal{S}(\pi,g)}
\mathrm{VaR}_{\alpha}(X|\nu)
\in\mathbb{R}.
$$
Define
$$
y_0=
\inf\{x\in\mathbb{R}:H(x)\leq1-\alpha\}.
$$
Since $\lim_{x\to\infty}H(x)=0$ and $\mathrm{VaR}_\alpha(X|\mu_{\pi})\leq y_0$, $y_0\in \mathbb{R}$.

Let $\epsilon>0$. Since $H$ is nonincreasing,
the definition of $y_0$ gives
$
H(y_0+\epsilon)\leq1-\alpha.
$
Therefore,
$$
\mathrm{VaR}_{\alpha}(X|\nu)
\leq y_0+\epsilon,~~\nu\in\mathcal{S}(\pi,g),
$$
and $x_0\leq y_0+\epsilon$. Letting $\epsilon\downarrow0$
yields $x_0\leq y_0$.

Conversely, for every $\nu\in\mathcal{S}(\pi,g)$,
$
\mathrm{VaR}_{\alpha}(X|\nu)
\leq x_0<x_0+\epsilon.
$
Therefore,
$$
\nu(X>x_0+\epsilon)\leq1-\alpha,~~\nu\in \mathcal{S}(\pi,g).
$$
Taking the supremum over $\nu$ gives
$H(x_0+\epsilon)\leq1-\alpha$, so
$y_0\leq x_0+\epsilon$. Letting $\epsilon\downarrow0$
yields $y_0\leq x_0$.
\end{proof}

\section{Proofs accompanying Section \ref{sec:7}}

\begin{proof}[Proof of Proposition \ref{prop:meas}]
    Fix $X\in \mathcal{C}$. Let $(\mu_n)_{n\in \N}\subseteq \Delta$ such that $\mu_n\Rightarrow\mu$, and $f:\mathbb{R}\to\mathbb{R}$ be continuous and bounded. We have that $$\lim_{n\to\infty}\int_{\mathbb{R}}fd\iota_X(\mu_n)=\lim_{n\to\infty}\mathbb{E}^{\mu_n}[f(X)]=\mathbb{E}^{\mu}[f(X)]=\int_{\mathbb{R}}f\d\iota_X(\mu).$$
    Therefore, $\iota_X(\mu_n)\Rightarrow\iota_X(\mu)$, and $\iota_X$ is continuous. Define the set
    $$\mathcal{W}=\{X\in \mathcal{X}~|~\iota_X\text{ is }\Sigma/\Sigma_{\mathcal{D}}\text{-measurable}\}.$$
    By the previous observation, $\mathcal{C}\subseteq \mathcal{W}$. It is clear that $\mathcal{W}$ is closed under sequences that converge pointwise. Therefore, by \citet[Theorem 11.6]{K12A}, $\mathcal{X}\subseteq \mathcal{W}$, and $\mathcal{X}= \mathcal{W}$.
\end{proof}

\begin{proof}[Proof of Proposition \ref{prop:lambda}]
    It is clear that $A\in \mathfrak{U}(\pi)$ if and only if there exists $c_A\in [0,1]$ such that $\nu(A)=c_A$ $\pi$-a.s. Therefore, it is straightforward to see that $\mathfrak{U}(\pi)$ is a $\lambda$-system. 
    
    Let $X\in \mathfrak{C}(\pi)$. Then, $(\iota_X)_{\#}\pi=\delta_{Q_{X}^\pi}$ and 
    $$\nu(X^{-1}(B))=X_{\#}\nu(B)=[\iota_X(\nu)](B)=Q_X^\pi(B)~\pi\text{-a.s.,}~~B\in \mathcal{B}(\mathbb{R}).$$
    Therefore, $X^{-1}(B)\in \mathfrak{U}(\pi)$ for all $B\in \mathcal{B}(\mathbb{R}).$

    Conversely, assume that $X\in \mathcal{X}$ satisfies $X^{-1}(B)\in \mathfrak{U}(\pi)$ for all $B\in \mathcal{B}(\mathbb{R})$. Given $q\in \mathbb{Q}$, let $A_q\in \Sigma$ satisfy $\pi(A_q)=1$ and, for all $\nu\in A_q$, $\nu(X\leq q)=c_q$. Let $A=\bigcap_{q\in \mathbb{Q}}A_q$. We have $\pi(A)=1$ and, for all $\nu\in A$ and $q\in \mathbb{Q}$, $\nu(X\leq q)=c_q$. Find the unique $Q\in \mathcal{D}$ such that $Q((-\infty,q])=c_q$ for all $q\in \mathbb{Q}$. Therefore, for all $\nu\in A$, $X_{\#}\nu=Q$. Thus, $(\iota_X)_{\#}\pi=\delta_Q$ and $X\in \mathfrak{C}(\pi)$.
\end{proof}

\begin{proof}[Proof of Proposition \ref{prop:reduc}]
    Let $X\in \mathfrak{C}(\pi)$. We claim that $X_{\#}\mu=Q_X^{\pi}$ for all $\mu\in \mathcal{S}(\pi,g)$. Let $\mu\in \mathcal{S}(\pi,g)$ and $\xi\in \mathcal{R}(\pi,g)$ such that $\mu_{\xi}=\mu$. We know that $ \pi(\iota_X^{-1}(\{Q_X^\pi\}))=1$ and 
    $$\xi\left(\left(\iota_X^{-1}\left(\left\{Q_X^\pi\right\}\right)\right)^c\right)\leq g\left(\pi\left(\left(\iota_X^{-1}\left(\left\{Q_X^\pi\right\}\right)\right)^c\right)\right)=0.$$
    Therefore, $(\iota_X)_{\#}\xi=\delta_{Q_X^{\pi}}$ and
    $$X_{\#}\mu(B)=\int_{\Delta}[\iota_X(\nu)](B)\xi(\d\nu)=\int_{\mathcal{D}}Q(B)\delta_{Q_X^{\pi}}(\d Q)=Q_X^{\pi}(B),~~B\in \mathcal{B}(\mathbb{R}).$$
    Therefore, $X_{\#}\mu=Q_X^{\pi}$ and, since $\mu\in \mathcal{S}(\pi,g)$ was arbitrary, the proof is complete.
\end{proof}

\begin{proof}[Proof of Theorem \ref{prop:LI}]
    Define $\mathcal{A}_X=\{X_{\#}\mu:\mu\in\mathcal{S}(\pi,g)\}$ and $\mathcal{A}_Y=\{Y_{\#}\mu:\mu\in \mathcal{S}(\xi,g)\}.$ Using a similar argument as the proof of Proposition \ref{prop:meas}, we can show that $\iota_Z$ is $\tau_b/\tau_\mathcal{D}$ continuous for all $Z\in \mathcal{X}$. Additionally, $\iota_X$ is affine for all $Z\in \mathcal{X}$. Therefore, by Proposition \ref{prop:compact} and corollary \ref{corr:1}, both $\mathcal{A}_X$ and $\mathcal{A}_Y$ are convex and $\tau_\mathcal{D}$ compact. Let $\eta=(\iota_X)_{\#}\pi=(\iota_Y)_{\#}\xi$. Fix a bounded and continuous $f:\mathbb{R}\to\mathbb{R}$ and define $h_f:\mathcal{D}\to\mathbb{R}$ by $h_f(Q)=\int_{\mathbb{R}}f\d Q$. We have that
    \begin{align*}
        \sup_{Q\in \mathcal{A}_X}\int_{\mathbb{R}}f\d Q=\sup_{\nu\in \mathcal{S}(\pi,g)}\mathbb{E}^\nu[f(X)]&=\int_{\Delta}\mathbb{E}^{\nu}[f(X)]g\circ\pi(\d\nu)\\&=\int_{\Delta}h_f(\iota_X(\nu))g\circ\pi(\d\nu)\\&=\int_{\mathcal{D}}h_f(Q)g\circ\eta(\d Q)\\&=\sup_{Q\in \mathcal{A}_Y}\int_{\mathbb{R}}f\d Q.
    \end{align*}
    Since $(\mathcal{D},\tau_{\mathcal{D}})$ is the restriction of a locally convex topological vector space (as seen by the argument in the proof of Proposition \ref{prop:check}),
    by the Hahn-Banach theorem, we have $\mathcal{A}_Y=\mathcal{A}_X$, which proves the claim.
\end{proof}

\section{Proofs and additional results accompanying Section \ref{sec:5}}

\label{app:4}

\begin{proof}[Proof of Proposition \ref{prop:para}]
    It is clear that
    $\mathfrak{P}_{\#}(\mathcal{R}_0(\pi,g))\subseteq\mathcal{R}(\mathfrak{P}_{\#}\pi,g).$
    Let $\xi\in \mathcal{R}(\mathfrak{P}_{\#}\pi,g)$. It is clear that $\xi\ll \mathfrak{P}_{\#}\pi$. Let $\zeta=\frac{\d\xi}{\d  \mathfrak{P}_{\#}\pi}$. Define 
    $\tilde{\zeta}:\Theta\to [0,\infty)$ by $ \tilde{\zeta}(\theta)=\zeta(\mathfrak{P}(\theta)),$ and $\tilde{\xi}\in \Delta(\Theta)$ by $\tilde{\xi}(A)=\int_{A}\tilde{\zeta}\d\pi$ for all $A\in \mathcal{A}$. 
    Let $A\in \mathcal{A}$. By Condition (3), there exists $B\in \Sigma$ such that $\mathfrak{P}^{-1}(B)=A$. Therefore,
    $$\tilde{\xi}(A)=\int_A\tilde{\zeta}\d\pi=\int_{B}\zeta\d(\mathfrak{P}_{\#}\pi)=\xi(B)\leq g(\mathfrak{P}_{\#}\pi(B))=g(\pi(A)),$$
    and $\tilde\xi\in \mathcal{R}_0(\pi,g)$.
    Since $\mathfrak{P}_{\#}\tilde{\xi}=\xi$, we have that $\xi\in \mathfrak{P}_{\#}(\mathcal{R}_0(\pi,g))$. Therefore, $\mathfrak{P}_{\#}(\mathcal{R}_0(\pi,g))=\mathcal{R}(\mathfrak{P}_{\#}\pi,g).$
    
    Given $\xi\in \mathcal{R}_0(\pi,g)$, since
    $$\int_{\Theta}\mu_\theta(A) \xi(\d\theta)=\int_{\Delta}\nu(A)\mathfrak{P}_{\#}\xi(\d\nu),~~A\in \mathcal{F},$$
    we have that $\mathcal{S}(\mathfrak{P}_{\#}\pi,g)=\mathcal{S}_0(\pi,g)$.
\end{proof}

\begin{proof}[Proof of Theorem \ref{theo:num}]
Define the function $s_n:\mathbb{R}\times \Theta^\N\to [0,1]$ by $s_n(x,\zeta)=\frac{1}{n}\sum_{k=1}^n\id_{(x,\infty)}(Y_k(\zeta)),$ and the function  $s:\mathbb{R}\to[0,1]$ by $s(x)=\pi(\{\theta\in \Theta~|~\mathbb{E}^{\mu_{\theta}}[X]>x\})$. We have that $$\sum_{k=1}^n \left(g\left(\frac{n-k+1}{n}\right)-g\left(\frac{n-k}{n}\right)\right)Y_{k:n}(\zeta)=\int_0^\infty g(s_n(x,\zeta))\d x+\int_{-\infty}^0 (g(s_n(x,\zeta))-1)\d x,~~\zeta\in \Theta^{\N}.$$

For each $q\in \mathbb{Q}$, let $A_q\in\mathcal{A}^{\N}$ be a set with $\kappa(A_q)=1$ and $\lim_{n\to \infty}s_n(q,\zeta)=s(q)$ for all $\zeta\in A_q$. Define $A=\bigcap_{q\in \mathbb{Q}}A_q\in \mathcal{A}^{\N}$. Clearly, $\kappa(A)=1$. We claim that for all $\zeta\in A$, $s_n(\,\cdot\,,\zeta)\to s$ Lebesgue-a.e. To see this, let $x\in \mathbb{R}$ be a continuity point of $s$. Find sequences $(q_m^-)_{m\in \N}\subseteq \mathbb{Q}$ and $(q_m^+)_{m\in \N}\subseteq \mathbb{Q}$ such that $q_m^-\uparrow x$ and $q_m^+\downarrow x$. We have that
$$s(q_m^+)=\lim_{n\to\infty}s_n(q_m^+,\zeta)\leq \liminf_{n\to\infty} s_n(x,\zeta)\leq \limsup_{n\to\infty}s_n(x,\zeta)\leq \lim_{n\to\infty}s_n(q_m^-,\zeta)=s(q_m^-).$$
Therefore, since $x$ is a continuity point of $s$, we have that $$s(x)=\liminf_{n\to\infty} s_n(x,\zeta)= \limsup_{n\to\infty}s_n(x,\zeta).$$ Thus, $\lim_{n\to\infty}s_n(x,\zeta)=s(x)$. Since the set of discontinuity points is countable, the original claim follows. 

For $q\in \mathbb{N}$, let $B_q\in \mathcal{A}^{\mathbb{N}}$ be a set with $\kappa(B_q)=1$ and $$\lim_{n\to \infty}\frac{1}{n}\sum_{k=1}^n(Y_k(\zeta)-q)_+=\int_{\Theta}(\mathbb{E}^{\mu_{\theta}}[X]-q)_+\pi(\d\theta),~~\zeta\in B_q.$$ For $q\in \mathbb{N}$, let $C_q\in \mathcal{A}^{\mathbb{N}}$ be a set with $\kappa(C_q)=1$ and $$\lim_{n\to \infty}\frac{1}{n}\sum_{k=1}^n(-Y_k(\zeta)-q)_+=\int_{\Theta}(-\mathbb{E}^{\mu_{\theta}}[X]-q)_+\pi(\d\theta),~~\zeta\in C_q.$$
Define $D=(\bigcap_{q\in \N} B_q)\cap (\bigcap_{q\in \N} C_q)\cap A\in \mathcal{A}^{\mathbb{N}}$. Clearly, $\kappa(D)=1$.

Fix $\zeta\in D$. We claim that
$$\lim_{n\to\infty}\left(\int_0^\infty g(s_n(x,\zeta))\d x+\int_{-\infty}^0 (g(s_n(x,\zeta))-1)\d x\right)=\int_0^\infty g(s(x))\d x+\int_{-\infty}^0 (g(s(x))-1)\d x.$$ Given $\epsilon>0$, find $M\in \mathbb{N}$ such that 
$$\int_{\Theta}(\mathbb{E}^{\mu_{\theta}}[X]-M)_+\pi(\d\theta)<\epsilon~~\text{and}~~\int_{\Theta}(-\mathbb{E}^{\mu_{\theta}}[X]-M)_+\pi(\d\theta)<\epsilon.$$
Find $N\in \N$ such that for all $n\geq N$, we have 
{\allowdisplaybreaks
\begin{align*}
   &\left|\frac{1}{n}\sum_{k=1}^n(Y_k(\zeta)-M)_+-\int_{\Theta}(\mathbb{E}^{\mu_{\theta}}[X]-M)_+\pi(\d\theta)\right|<\epsilon,\\&\left|\frac{1}{n}\sum_{k=1}^n(-Y_k(\zeta)-M)_+-\int_{\Theta}(-\mathbb{E}^{\mu_{\theta}}[X]-M)_+\pi(\d\theta)\right|<\epsilon,\\&\left|\int_0^M g(s_n(x,\zeta))\d x-\int_0^{M}g(s(x))\d x\right|<\epsilon,\\&\left|\int_{-M}^0 g(s_n(x,\zeta))\d x-\int_{-M}^{0}g(s(x))\d x\right|<\epsilon,
\end{align*}}
where the dominated convergence theorem is invoked to get the last two inequalities. Let $a\in \mathbb{N}$ such that $g(x)\leq ax$ for all $x\in [0,1]$. For all $n\geq N$, we have that

{\allowdisplaybreaks
\begin{align*}
    \Bigg|&\sum_{k=1}^n \left(g\left(\frac{n-k+1}{n}\right)-g\left(\frac{n-k}{n}\right)\right)Y_{k:n}(\zeta)-\int_{\Theta}\mathbb{E}^{\mu_{\theta}}[X]g\circ\pi(\d\theta)\Bigg|\\&= \left|\int_0^{\infty}g(s_n(x,\zeta))\d x+\int_{-\infty}^0(g(s_n(x,\zeta))-1)\d x-\int_0^{\infty}g(s(x))\d x-\int_{-\infty}^0(g(s(x))-1)\d x\right|\\&\leq \left|\int_0^{M}g(s_n(x,\zeta))\d x-\int_0^{M}g(s(x))\d x\right|+\left|\int_{-M}^0(g(s_n(x,\zeta))-1)\d x-\int_{-M}^0(g(s(x))-1)\d x\right|\\&~~~~~+\left|\int_M^{\infty}g(s_n(x,\zeta))\d x-\int_M^{\infty}g(s(x))\d x\right|+\left|\int_{-\infty}^{-M}(g(s_n(x,\zeta))-1)\d x-\int_{-\infty}^{-M}(g(s(x))-1)\d x\right|\\&<\left|\int_{-M}^0g(s_n(x,\zeta))\d x-\int_{-M}^0g(s(x))\d x\right|+\int_M^{\infty}g(s_n(x,\zeta))\d x+\int_M^{\infty}g(s(x))\d x\\&~~~~~+\int_{-\infty}^{-M}(1-g(s_n(x,\zeta)))\d x+\int_{-\infty}^{-M}(1-g(s(x)))\d x+\epsilon\\&\leq a\int_M^{\infty}s_n(x,\zeta)\d x+a\int_M^{\infty}s(x)\d x+\int_{-\infty}^{-M}(1-s_n(x,\zeta))\d x+\int_{-\infty}^{-M}(1-s(x))\d x+2\epsilon\\&=\frac{a}{n}\sum_{k=1}^n(Y_k(\zeta)-M)_++a\int_{\Theta}(\mathbb{E}^{\mu_{\theta}}[X]-M)_+\pi(\d\theta)+\frac{1}{n}\sum_{k=1}^n(-Y_k(\zeta)-M)_+\\&~~~~~+\int_{\Theta}(-\mathbb{E}^{\mu_{\theta}}[X]-M)_+\pi(\d\theta)+2\epsilon\\&\leq \frac{a}{n}\sum_{k=1}^n(Y_k(\zeta)-M)_++\frac{1}{n}\sum_{k=1}^n(-Y_k(\zeta)-M)_++3\epsilon+a\epsilon\\&\leq a\int_{\Theta}(\mathbb{E}^{\mu_{\theta}}[X]-M)_+\pi(\d\theta)+\int_{\Theta}(-\mathbb{E}^{\mu_{\theta}}[X]-M)_+\pi(\d\theta)+4\epsilon+2a\epsilon\leq 5\epsilon + 3a\epsilon=(5+3a)\epsilon.
\end{align*}}
Since $\zeta\in D$ was arbitrary, the proof is complete.
\end{proof}

\section{Proofs accompanying Section \ref{sec:6}}

\label{app:6}

\begin{proposition}\label{prop:max1}
    Let $\mathcal{P}\subseteq \Delta$ be nonempty and $\tau_b$-compact and $\gamma$ be a Lebesgue continuous statistic. Then, for all $X\in \mathcal{X}_b$, there exists $\mu_X\in \mathcal{P}$ such that
    $$\gamma(X|\mu_X)=\sup_{\nu\in \mathcal{P}}\gamma(X|\nu).$$
\end{proposition}

\begin{proof}
    Let $X\in \mathcal{X}_b$. By Proposition \ref{prop:need4}, $\gamma\circ\iota_X$ is $\tau_b$-continuous. Since $\mathcal{P}$ is $\tau_b$-compact, by the extreme-value theorem, there exists $\mu_X\in \mathcal{P}$ such that 
    $$\gamma(X|\mu_X)=\gamma\circ\iota_X(\mu_X)=\sup_{\nu\in \mathcal{P}}\gamma\circ\iota_X(\nu)=\sup_{\nu\in \mathcal{P}}\gamma(X|\nu).$$
    Therefore, the proof is complete.
\end{proof}

\begin{proof}[Proof of Theorem \ref{prop:consis}]
    Fix $X\in \mathcal{X}_b$. By Proposition \ref{prop:max1}, for each $n\in\N$, we can find $\mu_n\in \mathcal{S}(\pi_n,g)$ such that $$\max_{\nu\in \mathcal{S}(\pi_n,g)}\gamma(X|\nu)=\gamma(X|\mu_n).$$ For each $n\in \N$, find $\xi_n\in \mathcal{R}(\pi_n,g)$ satisfying $\mu_{\xi_n}=\mu_n$. We claim that $\xi_n\Rightarrow \delta_{\theta_0}$.

    For $n\in \N$, since $\xi_n\in \mathcal{R}(\pi_n,g)$, we have 
    $$g^*(\pi_n(A))\leq \xi_n(A)\leq g(\pi_n(A)),~~A\in \mathcal{A},$$
    where $g^*(x)=1-g(1-x)$ for all $x\in [0,1]$. Let $h:\Theta\to\mathbb{R}$ be bounded and continuous. It is clear that  $h_{\#}\pi_n\Rightarrow h_{\#}\delta_{\theta_0}$. Thus, for Lebesgue-a.e.\ $x\in \mathbb{R}$, $$\lim_{n\to\infty}\pi_{n}(h>x)=\lim_{n\to\infty}h_{\#}\pi_{n}((x,\infty))=h_{\#}\delta_{\theta_0}((x,\infty))=\id_{(-\infty,h(\theta_0))}(x).$$
    By the dominated convergence theorem, for all continuous distortion functions $\tilde{g}$, we have 
    \begin{align*}
        \lim_{n\to\infty}\int_{\Theta}h\d(\tilde{g}\circ\pi_n)&=\lim_{n\to\infty}\left(\int_0^{\infty} \tilde{g}(\pi_{n}(h>x))\d x+\int_{-\infty}^0 (\tilde{g}(\pi_{n}(h>x))-1)\d x\right)\\&=\int_0^{\infty} \tilde{g}\left(\id_{(-\infty,h(\theta_0))}(x)\right)\d x+\int_{-\infty}^0 \left(\tilde{g}\left(\id_{(-\infty,h(\theta_0))}(x)\right)-1\right)\d x\\&=\int_0^{\infty} \id_{(-\infty,h(\theta_0))}(x)\d x+\int_{-\infty}^0 \left(\id_{(-\infty,h(\theta_0))}(x)-1\right)\d x=h(\theta_0).
    \end{align*}
    Since $$\int_{\Theta}h\d(g^*\circ\pi_n)\leq \int_{\Theta}h\d\xi_n\leq \int_{\Theta}h\d(g\circ\pi_n),$$
    we have that $\lim_{n\to\infty}\int_{\Theta}h\d\xi_n=h(\theta_0)$. Thus, $\xi_n\Rightarrow \delta_{\theta_0}$.

    Since the parametrization is continuous with respect to the total variation metric, for all bounded continuous $f:\mathbb{R}\to\mathbb{R}$, we have that the function $h_f:\Theta\to\mathbb{R}$ defined by
    $$h_f(\theta)=\mathbb{E}^{\mu_\theta}[f(X)]$$
    is continuous and bounded. Therefore, we have that
    \begin{align*}
        \lim_{n\to\infty}\int_{\mathbb{R}}f\d(X_\#\mu_n)&=\lim_{n\to\infty}\mathbb{E}^{\mu_n}[f(X)]\\&=\lim_{n\to\infty}\int_{\Theta}\mathbb{E}^{\mu_\theta}[f(X)]\xi_n(\d\theta)=\lim_{n\to\infty}\int_{\Theta}h_f(\theta)\xi_n(\d\theta)=h_f(\theta_0)=\int_{\mathbb{R}}f\d(X_\#\mu_{\theta_0}),
    \end{align*}
    and $X_{\#}\mu_n\Rightarrow X_{\#}\mu_{\theta_0}$. Since $(X_{\#}\mu_n)_{n\in \N}$ has uniformly bounded support and $\gamma$ is Lebesgue continuous,
    $$\lim_{n\to\infty}\max_{\nu\in \mathcal{S}(\pi_n,g)}\gamma(X|\nu)=\lim_{n\to\infty}\gamma(X|\mu_n)=\gamma(X|\mu_{\theta_0}),$$
    and the proof is complete.
\end{proof}

\begin{proof}[Proof of Proposition \ref{prop:total}]
    Given $\epsilon>0,$ since the parametrization is continuous with respect to the total variation metric, we can find an open neighborhood $U_\epsilon\in \mathcal{A}$ of $\theta_0$ such that 
    $\sup_{\theta\in U_\epsilon}d_{\mathrm{TV}}(\mu_\theta,\mu_{\theta_0})\leq \epsilon.$ Therefore,
    \begin{align*}
        \limsup_{n\to\infty}\sup_{\nu\in \mathcal{S}(\pi_n,g)}d_{\mathrm{TV}}(\nu,\mu_{\theta_0})&=\limsup_{n\to\infty}\sup_{\nu\in \mathcal{S}(\pi_n,g)}\sup_{A\in \mathcal{F}}\left|\nu(A)-\mu_{\theta_0}(A)\right|\\&=\limsup_{n\to\infty}\sup_{\xi\in \mathcal{R}(\pi_n,g)}\sup_{A\in \mathcal{F}}\left|\int_{\Theta}\mu_{\theta}(A)\xi(\d\theta)-\mu_{\theta_0}(A)\right|\\&\leq\limsup_{n\to\infty}\sup_{\xi\in \mathcal{R}(\pi_n,g)}\int_{\Theta}\sup_{A\in \mathcal{F}}\left|\mu_{\theta}(A)-\mu_{\theta_0}(A)\right|\xi(\d\theta)\\&=\limsup_{n\to\infty}\sup_{\xi\in \mathcal{R}(\pi_n,g)}\int_{\Theta}d_{\mathrm{TV}}(\mu_\theta,\mu_{\theta_0})\xi(\d\theta)
        \\&=\limsup_{n\to\infty}\int_{\Theta}d_{\mathrm{TV}}(\mu_\theta,\mu_{\theta_0})g\circ\pi_n(\d\theta)\\&\leq \sup_{\theta\in U_\epsilon}d_{\mathrm{TV}}(\mu_\theta,\mu_{\theta_0})+\limsup_{n\to\infty}\int_{U_\epsilon^c}d_{\mathrm{TV}}(\mu_\theta,\mu_{\theta_0})g\circ\pi_n(\d\theta)\\&\leq \epsilon+2\limsup_{n\to\infty}g(\pi_n(U_\epsilon^c))=\epsilon,
    \end{align*}
where we are using the fact that, by the Portmanteau theorem, $$\limsup_{n\to\infty}\pi_n(U_\epsilon^c)\leq \delta_{\theta_0}(U_\epsilon^c)=0.$$ To guarantee the measurability of $\theta\mapsto d_{\mathrm{TV}}(\mu_\theta,\mu_{\theta_0}),$ note that, for all $A\in \mathcal{F}$, $\theta\mapsto |\mu_\theta(A)-\mu_{\theta_0}(A)|$ is continuous. Therefore, the previously mentioned mapping is lower semicontinuous and it is measurable. As $\epsilon$ was arbitrary, the proof is complete.
\end{proof}

Proposition \ref{prop:KL} is a corollary of Proposition \ref{prop:cor1} and Proposition \ref{prop:cor2}, presented and proved below. Given $\mu\in \Delta$, we define $\Delta(\mu)=\{\nu\in \Delta~|~\nu\ll \mu\}$. Given $\mu\in \Delta$ and $X,Y\in \mathcal{X}$, we say $X\overset{\d}{=}_\mu Y$ if $X_{\#}\mu=Y_{\#}\mu$. Given $\mu\in \Delta$, we say that the decision functional $\rho:\mathcal{X}_b\to\mathbb{R}$ is \emph{$\mu$-law invariant} if $\rho(X)=\rho(Y)$ for all $X,Y\in \mathcal{X}_b$ satisfying $X\overset{\d}{=}_\mu Y$. Clearly, if $\mu\in \Delta$ is atomless, a decision functional $\rho$ is $\mu$-law invariant if and only if there exists a statistic $\gamma$ such that 
$\rho(X)=\gamma(X|\mu)$ for all $X\in \mathcal{X}_b.$
Given $\mu\in \Delta$ and $\nu\in \Delta(\mu)$, define $$\mathcal{T}(\nu|\mu)=\left\{\tilde{\nu}\in \Delta(\mu)~\bigg|~\frac{\d\tilde\nu}{\d\mu}\overset{\d}{=}_{\mu}\frac{\d\nu}{\d\mu}\right\}.$$ Given $\mu\in \Delta$ and  $\mathcal{P}\subseteq \Delta(\mu)$, we say that $\mathcal{P}$ is \emph{$\mu$-law invariant} if for all $\nu\in \mathcal{P}$, $\mathcal{T}(\nu|\mu)\subseteq \mathcal{P}$. Given $\mu\in \Delta$, we say that $X,Y\in \mathcal{X}$ are \emph{$\mu$-comonotonic} if there exists $A\in \mathcal{F}$ with $\mu(A)=1$ such that
    $$(X(\omega)-X(\tilde\omega))(Y(\omega)-Y(\tilde\omega))\geq 0,~~\omega,\tilde{\omega}\in A.$$

\begin{proposition}\label{prop:cor1}
    Let $\mu\in \Delta$ be atomless and $\mathcal{P}\subseteq \Delta(\mu)$ be nonempty and $\mu$-law invariant. Then, for all statistics $\gamma$, the decision functional 
    $$\rho(X)=\sup_{\nu\in \mathcal{P}}\gamma(X|\nu),~~X\in \mathcal{X}_b,$$
    is $\mu$-law invariant.
\end{proposition}

\begin{proof}
    Given $X\in \mathcal{X}_b$, define 
    $$\mathcal{P}(X)=\left\{\nu\in \mathcal{P}~\bigg|~\frac{\d\nu}{\d\mu}\text{ and }X\text{ are $\mu$-comonotonic} \right\}.$$
    Since $\mathcal{P}$ is $\mu$-law invariant, $\mathcal{T}(\nu|\mu)\subseteq \mathcal{P}$ for all $\nu\in \mathcal{P}$. Additionally, since $\mu$ is atomless, given $\nu\in \mathcal{P}$ and $X\in \mathcal{X}_b$, we have that $\mathcal{T}(\nu|\mu)\cap \mathcal{P}(X)\neq \varnothing$; see \citet[Lemma A.25]{FS16A}. Let $\nu\in\mathcal{P}$, $X\in \mathcal{X}_b$, and $\tilde{\nu}\in \mathcal{T}(\nu|\mu)\cap \mathcal{P}(X)$. By the Hardy--Littlewood inequality, we have that
    $$\nu(X>x)=\mathbb{E}^{\mu}\left[\frac{\d\nu}{\d\mu}\id_{(x,\infty)}(X)\right]\leq \mathbb{E}^{\mu}\left[\frac{\d\tilde\nu}{\d\mu}\id_{(x,\infty)}(X)\right]=\tilde{\nu}(X>x),~~x\in \mathbb{R}.$$
    Thus, $X_{\#}\nu\leq_{\mathrm{st}}X_{\#}\tilde\nu$ and $\gamma(X|\nu)\leq \gamma(X|\tilde{\nu})$. Therefore, we have
    \begin{align*}
        \rho(X)=\sup_{\nu\in \mathcal{P}}\gamma(X|\nu)=\sup_{\nu\in \mathcal{P}}\sup_{\tilde{\nu}\in \mathcal{T}(\nu|\mu)}\gamma(X|\tilde\nu)=\sup_{\nu\in \mathcal{P}}\sup_{\tilde{\nu}\in \mathcal{T}(\nu|\mu)\cap\mathcal{P}(X)}\gamma(X|\tilde\nu)=\sup_{\nu\in \mathcal{P}(X)}\gamma(X|\nu),~~X\in \mathcal{X}_b.
    \end{align*}

    Let $X,Y\in \mathcal{X}_b$ satisfy $X\overset{\d}{=}_\mu Y$. Let $\nu\in \mathcal{P}(X)$ and $\tilde{\nu}\in \mathcal{T}(\nu|\mu)\cap\mathcal{P}(Y)$. Since $\left(X,\frac{\d\nu}{\d\mu}\right)\overset{\d}{=}_{\mu}\left(Y,\frac{\d\tilde\nu}{\d\mu}\right)$, we have that
    $$\nu(X>x)=\mathbb{E}^{\mu}\left[\frac{\d\nu}{\d\mu}\id_{(x,\infty)}(X)\right]=\mathbb{E}^{\mu}\left[\frac{\d\tilde\nu}{\d\mu}\id_{(x,\infty)}(Y)\right]=\tilde{\nu}(Y>x),~~x\in \mathbb{R}.$$
    Thus, $X_{\#}\nu=_{\mathrm{st}}Y_{\#}\tilde\nu$ and $\gamma(X|\nu)=\gamma(Y|\tilde{\nu})$. Since $\nu\in \mathcal{P}(X)$ was arbitrary,
    $$\rho(X)=\sup_{\nu\in \mathcal{P}(X)}\gamma(X|\nu)\leq \sup_{\nu\in \mathcal{P}(Y)}\gamma(Y|\nu)=\rho(Y).$$
    A similar argument will show that $\rho(Y)\leq \rho(X)$. Therefore, $\rho(X)=\rho(Y)$ and $\rho$ is $\mu$-law invariant.
\end{proof}

We say that the decision functional $\rho:\mathcal{X}_b\to\mathbb{R}$ is Lebesgue continuous if for bounded sequences $(X_n)_{n\in \N}\subseteq \mathcal{X}_b$ that converge pointwise to some $X\in \mathcal{X}_b$, we have that $\lim_{n\to\infty}\rho(X_n)=\rho(X)$. Clearly, if $\mu\in \Delta$ is atomless, then the decision functional $\rho:\mathcal{X}_b\to\mathbb{R}$ defined by
$\rho(X)=\gamma(X|\mu)$, for some statistic $\gamma$, is Lebesgue continuous if and only if $\gamma$ is Lebesgue continuous.

\begin{proposition}\label{prop:cor2}
    Let $\mathcal{P}\subseteq \Delta$ be nonempty and $\tau_b$-compact and $\gamma$ be a Lebesgue continuous statistic. Then the decision functional
    $$\rho(X)=\sup_{\nu\in \mathcal{P}}\gamma(X|\nu),~~X\in \mathcal{X}_b,$$
    is Lebesgue continuous.
\end{proposition}

\begin{proof}
It is well-known that $\rho$ is Lebesgue continuous if and only if for all uniformly bounded pointwise-increasing or pointwise-decreasing sequences $(X_n)_{n\in \N}\subseteq \mathcal{X}_b$, we have that $\lim_{n\to\infty}\rho(X_n)=\rho(\lim_{n\to\infty}X_n)$. Clearly, since $\gamma$ is Lebesgue continuous, the above holds for all pointwise-increasing sequences. The hard part is showing that it is true for all pointwise-decreasing sequences. Therefore, let $(X_n)_{n\in \N}\subseteq\mathcal{X}_b$ be a uniformly bounded pointwise-decreasing sequence and define $X=\lim_{n\to\infty}X_n$. 

    Let $\Phi_{\mathrm{BL}}$ denote the set of all Lipschitz continuous functions $f:\mathbb{R}\to\mathbb{R}$ that are bounded by, and have Lipschitz constant less than, 1. It is well-known, see \citet[Theorem 11.3.3]{D18A}, that 
    $$d_{\mathrm{BL}}:\mathcal{D}^2\to [0,\infty):(Q,P)\mapsto \sup_{f\in \Phi_{\mathrm{BL}}}\left|\int_{\mathbb{R}}f(x)Q(\d x)-\int_{\mathbb{R}}f(x)P(\d x)\right|,$$
    is a metric and metrizes $\tau_{\mathcal{D}}$. We claim that 
    $$\lim_{n\to \infty}\sup_{\nu\in \mathcal{P}}d_{\mathrm{BL}}((X_n)_{\#}\nu,X_{\#}\nu)=0.$$

    Since $\mathcal{P}$ is $\tau_b$-compact, by \citet[Lemma 4.1]{MM04A} and \citet[Theorem IV.9.2]{DS88A}, there exists $\mu\in \Delta$ such that for all $\epsilon>0$, there exists $\delta>0$ such that
    $$\mu(A)<\delta\implies \sup_{\nu\in \mathcal{P}}\nu(A)\leq \epsilon,~~A\in \mathcal{F}.$$
    Let $\epsilon>0$. By Egorov's theorem, we can find $A\in \mathcal{F}$ and $N\in \N$ such that $\mu(A)>1-\delta$ and for all $n\geq N$, we have 
    $|X_n(\omega)-X(\omega)|\leq \epsilon$ for all $\omega\in A.$
    Therefore, since $\mu(A^c)<\delta$, for all $n\geq N$,
    \begin{align*}
        \sup_{\nu\in \mathcal{P}}d_{\mathrm{BL}}((X_n)_{\#}\nu,X_{\#}\nu)&\leq \sup_{\nu\in \mathcal{P}}\sup_{f\in \Phi_{\mathrm{BL}}}\mathbb{E}^{\nu}[\left|f(X_n)-f(X)\right|]\\&\leq \sup_{\nu\in \mathcal{P}}\sup_{f\in \Phi_{\mathrm{BL}}}\mathbb{E}^{\nu}[\left|f(X_n)-f(X)\right|\id_A]+\sup_{\nu\in \mathcal{P}}\sup_{f\in \Phi_{\mathrm{BL}}}\mathbb{E}^{\nu}[\left|f(X_n)-f(X)\right|\id_{A^c}]\\&\leq \sup_{\nu\in \mathcal{P}}\mathbb{E}^{\nu}[\left|X_n-X\right|\id_A]+2\sup_{\nu\in \mathcal{P}}\nu(A^c)\leq 3\epsilon.
    \end{align*}
    
    Find $a,b\in \mathbb{R}$ such that $a\leq X_n\leq b$ for all $n\in \N$. Let $\mathcal{D}_{[a,b]}$ denote the set of Borel probability measures on $\mathbb{R}$ with support contained in the interval $[a,b]$. Clearly, for all $n\in \N$, $(X_n)_{\#}\mu\in \mathcal{D}_{[a,b]}$ for all $\mu\in \Delta$. Additionally, $X_{\#}\mu\in \mathcal{D}_{[a,b]}$ for all $\mu\in \Delta$. By Prokhorov's theorem, the set $\mathcal{D}_{[a,b]}$ is compact under $\tau_{\mathcal{D}}$. Since $\gamma|_{\mathcal{D}_{[a,b]}}$ is $\tau_{\mathcal{D}}$-continuous (as $\gamma$ is Lebesgue continuous), for all $\epsilon>0$, there exists $\delta>0$ such that 
    \begin{equation}\label{eq:con1}
        d_{\mathrm{BL}}(Q,P)<\delta\implies |\gamma(Q)-\gamma(P)|<\epsilon,~~Q,P\in \mathcal{D}_{[a,b]}.
    \end{equation}
    Given $\epsilon>0$, find $\delta>0$ such that \eqref{eq:con1} holds. Now find $N\in \N$ such that $ \sup_{\nu\in \mathcal{P}}d_{\mathrm{BL}}((X_n)_{\#}\nu,X_{\#}\nu)<\delta$ for all $n\geq N$. For each $n\geq N$, let $\mu_n\in \mathcal{P}$ satisfy 
    $\sup_{\nu\in \mathcal{P}}\gamma(X_n|\nu)=\gamma(X_n|\mu_n).$

    Therefore,
        $$\rho(X_n)-\rho(X)\leq \gamma(X_n|\mu_n)-\gamma(X|\mu_n)<\epsilon.$$
        Thus $\lim_{n\to\infty}\rho(X_n)=\rho(X)$, and $\rho$ is Lebesgue continuous.
\end{proof}

\begin{verification}
    For all $X\in \mathcal{X}_b$, the mapping $\mathbb{E}_{\epsilon}^{\mathrm{KL}}\circ\iota_X$ is $\Sigma$-measurable.
\end{verification}

\begin{proof}
     Since $\mathbb{E}_{\epsilon}^{\mathrm{KL}}$ is Lebesgue continuous, by Proposition \ref{prop:meas2}, $\mathbb{E}_{\epsilon}^{\mathrm{KL}}\circ\iota_X$ is $\Sigma$-measurable for $X\in \mathcal{X}_b$.
\end{proof}

\begin{proof}[Proof of Proposition \ref{prop:need5}]
    Given $\mu\in \Delta$, by \citet[Theorem 3.3]{AJ12A},
    $$\mathbb{E}^{\mathrm{KL}}_\epsilon(X|\mu)=\inf_{z>0}z\left(\log\left(\mathbb{E}^{\mu}\left[e^{X/z}\right]\right)+\epsilon\right),~~X\in\mathcal{X}_b.$$ Therefore, it is clear that, for all $X\in \mathcal{X}_b$, $\mathbb{E}^{\mathrm{KL}}_\epsilon\circ\iota_X$ is concave.

     Let $X\in \mathcal{C}_b$. As $\mathbb{E}^{\mathrm{KL}}_\epsilon$ is Lebesgue continuous, by Proposition \ref{prop:need4}, $\mathbb{E}^{\mathrm{KL}}_\epsilon\circ\iota_X$ is $\tau$-continuous. Therefore, $-\mathbb{E}^{\mathrm{KL}}_\epsilon\circ\iota_X$ is convex and $\tau$-continuous. Therefore, by the definition of $\geq_{\mathrm{cx}}$, for all $\tilde\pi,\pi\in \Pi$ satisfying $\tilde\pi\geq_{\mathrm{cx}}\pi$, we have $$\int_{\Delta}-\mathbb{E}^{\mathrm{KL}}_\epsilon(X|\nu)\tilde\pi(\d\nu)\geq \int_{\Delta}-\mathbb{E}^{\mathrm{KL}}_\epsilon(X|\nu)\pi(\d\nu).$$
    Since $X\in \mathcal{C}_b$ was arbitrary, by Proposition \ref{prop:mac}, we have
    $$\int_{\Delta}-\mathbb{E}^{\mathrm{KL}}_\epsilon(X|\nu)\tilde\pi(\d\nu)\geq \int_{\Delta}-\mathbb{E}^{\mathrm{KL}}_\epsilon(X|\nu)\pi(\d\nu),~~X\in \mathcal{X}_b,$$
    and the proof is complete.
\end{proof}

{
\small
\renewcommand{\refname}{References for the appendices}

}

\newpage

\section{Additional figures}

\label{app:fig}

\begin{figure}[ht]
    \centering
    \includegraphics[width=1\linewidth]{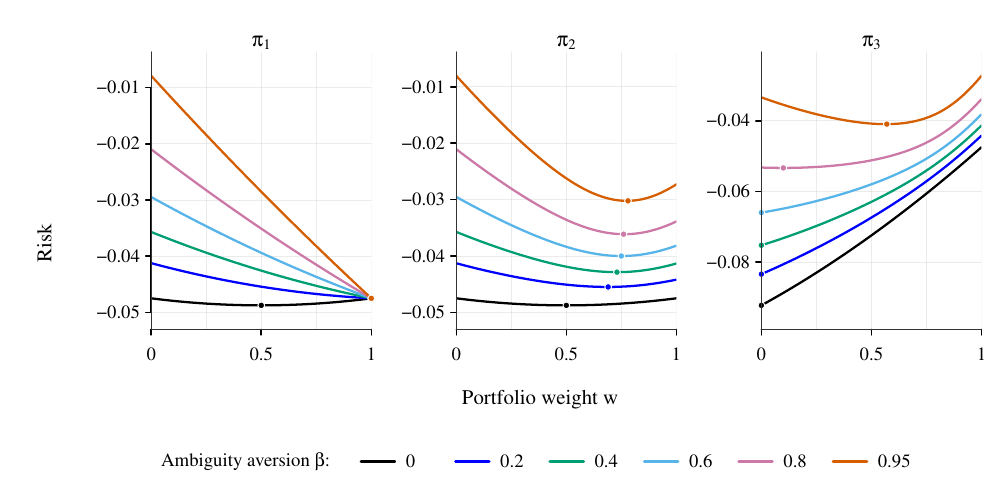}
    \caption{Worst-case entropic risk measure}
    \label{fig:ER}
\end{figure}

\begin{figure}[ht]
    \centering
    \includegraphics[width=1\linewidth]{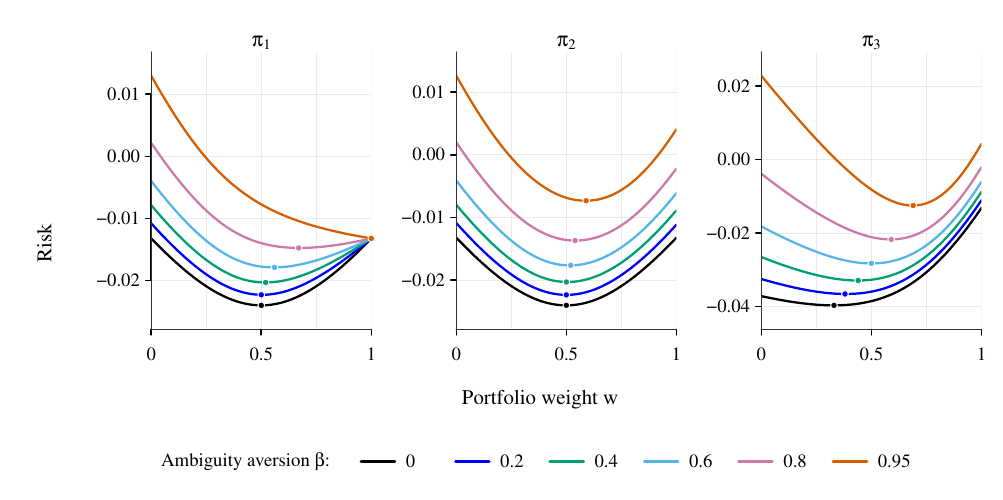}
    \caption{Worst-case value at risk}
    \label{fig:VAR}
\end{figure}
 

{\small
\begin{thebibliography}{10}

\bibitem[\protect\citeauthoryear{Acciaio and Svindland}{2013}]{AS13}
Acciaio, B. and Svindland, G. (2013). Are law-invariant risk functions concave on distributions? \emph{Dependence Modeling}, \textbf{1}, 54--64.

\bibitem[\protect\citeauthoryear{Acerbi and Tasche}{2002}]{AT02}
Acerbi, C. and Tasche, D. (2002). Expected shortfall: A natural coherent alternative to value at risk. \emph{Economic Notes}, \textbf{31}(2), 379--388.

\bibitem[\protect\citeauthoryear{Ahmadi-Javid}{2012}]{AJ12}
Ahmadi-Javid, A. (2012). Entropic value-at-risk: A new coherent risk measure. \emph{Journal of Optimization Theory and Applications}, \textbf{155}(3), 1105--1123.

\bibitem[\protect\citeauthoryear{Ardia}{2008}]{A08}
Ardia, D. (2008). \emph{Financial risk management with Bayesian estimation of GARCH models: Theory and applications.} Berlin, Heidelberg: Springer Berlin Heidelberg.

\bibitem[\protect\citeauthoryear{Ardia et al.}{2017}]{AKT17}
Ardia, D., Kolly, J. and Trottier, D.-A. (2017). The impact of parameter and model uncertainty on market risk predictions from GARCH-type models. \emph{Journal of Forecasting}, \textbf{36}(7), 808--823.


\bibitem[\protect\citeauthoryear{Artzner et al.}{1999}]{ADEH99}
Artzner, P., Delbaen, F., Eber, J.-M. and Heath, D. (1999). Coherent measures of risk. \emph{Mathematical Finance}, \textbf{9}(3), 203--228.

\bibitem[\protect\citeauthoryear{Bartl et al.}{2020}]{BDT20}
Bartl, D., Drapeau, S. and Tangpi, L. (2020). Computational aspects of robust optimized certainty equivalents and option pricing. \emph{Mathematical Finance}, \textbf{30}(1), 287--309.

\bibitem[\protect\citeauthoryear{{Basel Committee on Banking
  Supervision}}{{BCBS}}{2019}]{BASEL19}
{BCBS} (2019).
  {\em  Minimum Capital Requirements for Market Risk.  February 2019.}
 Basel Committee on Banking
  Supervision. Basel: Bank for International Settlements. \url{https://www.bis.org/bcbs/publ/d457.htm} 

\bibitem[\protect\citeauthoryear{Ben-Tal et al.}{2013}]{BTDDMR13}
Ben-Tal, A., den Hertog, D., De Waegenaere, A., Melenberg, B. and Rennen, G. (2013). Robust solutions of optimization problems affected by uncertain probabilities. \emph{Management Science}, \textbf{59}(2), 341--357.

\bibitem[\protect\citeauthoryear{Ben-Tal and Teboulle}{2007}]{BTT07}
Ben-Tal, A. and Teboulle, M. (2007). An old-new concept of convex risk measures: The optimized certainty equivalent. \emph{Mathematical Finance}, \textbf{17}(3), 449--476.

\bibitem[\protect\citeauthoryear{Berk}{1966}]{B66}
Berk, R. H. (1966). Limiting behavior of posterior distributions when the model is incorrect. \emph{The Annals of Mathematical Statistics}, \textbf{37}(1), 51--58.

\bibitem[\protect\citeauthoryear{Bertsekas and Shreve}{1996}]{BS96}
Bertsekas, D. P. and Shreve, S. E. (1996). \emph{Stochastic Optimal Control: The Discrete-Time Case}. Athena Scientific.

\bibitem[\protect\citeauthoryear{Billingsley}{2013}]{B13}
Billingsley, P. (2013). \emph{Convergence of probability measures.} John Wiley \& Sons.

\bibitem[\protect\citeauthoryear{Birghila and Pflug}{2019}]{BP19}
Birghila, C. and Pflug, G. C. (2019). Optimal XL-insurance under Wasserstein-type ambiguity. \emph{Insurance: Mathematics and Economics}, \textbf{88}, 30--43.

\bibitem[\protect\citeauthoryear{Blackwell}{1953}]{B53}
Blackwell, D. (1953). Equivalent comparisons of experiments. \emph{The Annals of Mathematical Statistics}, \textbf{24}(2), 265--272.

\bibitem[\protect\citeauthoryear{Blanchet et al.}{2019}]{BLTY19}
Blanchet, J., Lam, H., Tang, Q. and Yuan, Z. (2019). Robust actuarial risk analysis. \emph{North American Actuarial Journal}, \textbf{23}(1), 33--63.

\bibitem[\protect\citeauthoryear{Cairns}{2000}]{C00}
Cairns, A. J. G. (2000). A discussion of parameter and model uncertainty in insurance. \emph{Insurance: Mathematics and Economics}, \textbf{27}(3), 313--330.

\bibitem[\protect\citeauthoryear{Cerreia-Vioglio et al.}{2026}]{CHMM25}
Cerreia-Vioglio, S., Hansen, L. P., Maccheroni, F. and Marinacci, M. (2026). Making decisions under model misspecification. \emph{The Review of Economic Studies}, \textbf{93}(2), 892--925.

\bibitem[\protect\citeauthoryear{Delage et al.}{2019}]{DKW19}
Delage, E., Kuhn, D. and Wiesemann, W. (2019). ``Dice''-sion--making under uncertainty: When can a random decision reduce risk? \emph{Management Science}, \textbf{65}(7), 3282--3301.

\bibitem[\protect\citeauthoryear{Delage and Ye}{2010}]{DY10}
Delage, E. and Ye, Y. (2010). Distributionally robust optimization under moment uncertainty with application to data-driven problems. \emph{Operations Research}, \textbf{58}(3), 595--612.

\bibitem[\protect\citeauthoryear{Dellaporta et al.}{2025}]{DOD24}
Dellaporta, C., O'Hara, P. and Damoulas, T. (2025). Decision making under the exponential family: Distributionally robust optimisation with Bayesian ambiguity sets. In \emph{Proceedings of the 42nd International Conference on Machine Learning}, \emph{Proceedings of Machine Learning Research}, \textbf{267}, 13011--13043.


\bibitem[\protect\citeauthoryear{Dellaporta et al.}{2026}]{DOD25}
Dellaporta, C., O'Hara, P. and Damoulas, T. (2026). Decision-making under model misspecification: DRO with robust Bayesian ambiguity sets. \emph{Entropy}, \textbf{28}(4), Article 430.

\bibitem[\protect\citeauthoryear{Denneberg}{1994}]{D94}
Denneberg, D. (1994). \emph{Non-Additive Measure and Integral}. Springer Science \& Business Media.

\bibitem[\protect\citeauthoryear{El Ghaoui et al.}{2003}]{GOO03}
El Ghaoui, L., Oks, M. and Oustry, F. (2003). Worst-case value-at-risk and robust portfolio optimization: A conic programming approach. \emph{Operations Research}, \textbf{51}(4), 543--556.

\bibitem[\protect\citeauthoryear{Ellsberg}{1961}]{E61}
Ellsberg, D. (1961). Risk, ambiguity, and the Savage axioms. \emph{The Quarterly Journal of Economics}, \textbf{75}(4), 643--669.

\bibitem[\protect\citeauthoryear{Epstein}{2004}]{E04}
Epstein, L. G. (2004). A definition of uncertainty aversion. In \emph{Uncertainty in Economic Theory}, I. Gilboa, Ed. Routledge, New York, NY, USA. 171--208.

\bibitem[\protect\citeauthoryear{Fadina et al.}{2024}]{FLW24}
Fadina, T., Liu, Y. and Wang, R. (2024). A framework for measures of risk under uncertainty. \emph{Finance and Stochastics}, \textbf{28}(2), 363--390.

\bibitem[\protect\citeauthoryear{Feng and He}{2022}]{FH22}
Feng, G. and He, J. (2022). Factor investing: A Bayesian hierarchical approach. \emph{Journal of Econometrics}, \textbf{230}(1), 183--200.

\bibitem[\protect\citeauthoryear{Föllmer and Knispel}{2011}]{FK11}
Föllmer, H. and Knispel, T. (2011). Entropic risk measures: Coherence vs. convexity, model ambiguity and robust large deviations. \emph{Stochastics and Dynamics}, \textbf{11}(2), 333--351.

\bibitem[\protect\citeauthoryear{Föllmer and Schied}{2002}]{FS02}
Föllmer, H. and Schied, A. (2002). Convex measures of risk and trading constraints. \emph{Finance and Stochastics}, \textbf{6}(4), 429--447.

\bibitem[\protect\citeauthoryear{F\"{o}llmer and Schied}{2016}]{FS16}
F\"ollmer, H. and Schied, A. (2016). \emph{Stochastic Finance: An Introduction in Discrete Time}, 4th ed. Walter de Gruyter, Berlin.

\bibitem[\protect\citeauthoryear{Gajdos et al.}{2008}]{GHTV08}
Gajdos, T., Hayashi, T., Tallon, J.-M. and Vergnaud, J.-C. (2008). Attitude toward imprecise information. \emph{Journal of Economic Theory}, \textbf{140}(1), 27--65.

\bibitem[\protect\citeauthoryear{Ghirardato and Marinacci}{2002}]{GM02}
Ghirardato, P. and Marinacci, M. (2002). Ambiguity made precise: A comparative foundation. \emph{Journal of Economic Theory}, \textbf{102}(2), 251--289.

\bibitem[\protect\citeauthoryear{Ghosal and van der Vaart}{2017}]{GVV17}
Ghosal, S. and van der Vaart, A. W. (2017). \emph{Fundamentals of Nonparametric Bayesian Inference}. Cambridge University Press.


\bibitem[\protect\citeauthoryear{Gilboa and Schmeidler}{1989}]{GS89}
Gilboa, I. and Schmeidler, D. (1989). Maxmin expected utility with a non-unique prior. \emph{Journal of Mathematical Economics}, \textbf{18}(2), 141--153.

\bibitem[\protect\citeauthoryear{Giraud}{2014}]{G14}
Giraud, R. (2014). Second order beliefs models of choice under imprecise risk: Nonadditive second order beliefs versus nonlinear second order utility. \emph{Theoretical Economics}, \textbf{9}(3), 779--816.

\bibitem[\protect\citeauthoryear{Gollier}{2011}]{G11}
Gollier, C. (2011). Portfolio choices and asset prices: The comparative statics of ambiguity aversion. \emph{The Review of Economic Studies}, \textbf{78}(4), 1329--1344.

\bibitem[\protect\citeauthoryear{Ju and Miao}{2012}]{JM12}
Ju, N. and Miao, J. (2012). Ambiguity, learning, and asset returns. \emph{Econometrica}, \textbf{80}(2), 559--591.

\bibitem[\protect\citeauthoryear{Kandel and Stambaugh}{1996}]{KS96}
Kandel, S. and Stambaugh, R. F. (1996). On the predictability of stock returns: An asset-allocation perspective. \emph{The Journal of Finance}, \textbf{51}(2), 385--424.

\bibitem[\protect\citeauthoryear{Kechris}{2012}]{K12}
Kechris, A. S. (2012). \emph{Classical Descriptive Set Theory}. Springer Science \& Business Media.

\bibitem[\protect\citeauthoryear{Klibanoff et al.}{2005}]{KMM05}
	Klibanoff, P., Marinacci, M. and Mukerji, S. (2005).
	A smooth model of decision making under uncertainty. \emph{Econometrica}, \textbf{73}(6), 1849--1892.

\bibitem[\protect\citeauthoryear{Klugman}{1992}]{K92}
Klugman, S. A. (1992). \emph{Bayesian statistics in actuarial science: with emphasis on credibility}.  Kluwer Academic Publishers, Boston.

\bibitem[\protect\citeauthoryear{Knight}{1921}]{K21}
Knight, F. H. (1921). \emph{Risk, Uncertainty and Profit}. Houghton Mifflin.

\bibitem[\protect\citeauthoryear{Liu et al.}{2024}]{LMW24}
Liu, P., Mao, T. and Wang, R. (2024). Quantiles under ambiguity and risk sharing. \emph{arXiv preprint} arXiv:2412.19546.

\bibitem[\protect\citeauthoryear{Maccheroni et al.}{2013}]{MMR13}
Maccheroni, F., Marinacci, M. and Ruffino, D. (2013). Alpha as ambiguity: Robust mean--variance portfolio analysis. \emph{Econometrica}, \textbf{81}(3), 1075--1113.

\bibitem[\protect\citeauthoryear{Maccheroni et al.}{2009}]{MMRT09}
Maccheroni, F., Marinacci, M., Rustichini, A. and Taboga, M. (2009). Portfolio selection with monotone mean--variance preferences. \emph{Mathematical Finance}, \textbf{19}(3), 487--521.


\bibitem[\protect\citeauthoryear{Mao et al.}{Mao et al.}{2026}]{MWW26}
Mao, T., Wang, R. and Wu, Q. (2026). Model aggregation for risk evaluation and robust optimization. \emph{Management Science}, \textbf{72}(7), 6350--6367.



\bibitem[\protect\citeauthoryear{Marinacci}{2002}]{M02}
 Marinacci, M. (2002). Probabilistic sophistication and multiple priors. \emph{Econometrica}, \textbf{70}(2), 755--764.


\bibitem[\protect\citeauthoryear{Marinacci and Montrucchio}{2004}]{MM04}
Marinacci, M. and Montrucchio, L. (2004). Introduction to the mathematics of ambiguity. In \emph{Uncertainty in Economic Theory}, I. Gilboa, Ed. Routledge, New York, NY, USA, 46--107.

\bibitem[\protect\citeauthoryear{Markowitz}{1952}]{M52}
Markowitz, H. (1952). Portfolio selection. \emph{The Journal of Finance}, \textbf{7}(1), 77--91.

\bibitem[\protect\citeauthoryear{McNeil et al.}{McNeil et al.}{2015}]{MFE15}
{McNeil, A. J., Frey, R. and Embrechts, P.} (2015). \emph{Quantitative
Risk Management: Concepts, Techniques and Tools}. Revised Edition.  Princeton, NJ:
Princeton University Press.

\bibitem[\protect\citeauthoryear{Meyn and Tweedie}{2012}]{MT12}
Meyn, S. P., and Tweedie, R. L. (2012). \emph{Markov chains and stochastic stability.} Springer Science \& Business Media. 

\bibitem[\protect\citeauthoryear{Mu et al.}{2024}]{MPST24}
Mu, X., Pomatto, L., Strack, P. and Tamuz, O. (2024). Monotone additive statistics. \emph{Econometrica}, \textbf{92}(4), 995--1031.

\bibitem[\protect\citeauthoryear{Natarajan et al.}{Natarajan et al.}{2008}]{NPS08}
Natarajan, K., Pachamanova, D. and Sim, M. (2008). Incorporating asymmetric distributional information in robust value-at-risk optimization. \emph{Management Science}, \textbf{54}(3), 573--585. 

\bibitem[\protect\citeauthoryear{Phelps}{2001}]{P02}
Phelps, R. R. (2001). \emph{Lectures on Choquet's Theorem}, 2nd ed., Lecture Notes in Mathematics 1757. Springer, Berlin, Heidelberg.

\bibitem[\protect\citeauthoryear{Pratt}{1964}]{P64}
Pratt, J. W. (1964). Risk aversion in the small and in the large. \emph{Econometrica}, \textbf{32}(1/2), 122--136.

\bibitem[\protect\citeauthoryear{Rockafellar and Uryasev}{Rockafellar and Uryasev}{2002}]{RU02}
Rockafellar, R. T. and Uryasev, S. (2002). Conditional value-at-risk for general loss distributions. \emph{Journal of Banking \& Finance}, \textbf{26}(7), 1443--1471.

\bibitem[\protect\citeauthoryear{Rothschild and Stiglitz}{1970}]{RS70}
Rothschild, M. and Stiglitz, J. E. (1970). Increasing risk: I. A definition. \emph{Journal of Economic Theory}, \textbf{2}(3), 225--243.

\bibitem[\protect\citeauthoryear{Savage}{1954}]{S54}
Savage, L. J. (1954). \emph{The Foundations of Statistics}. John Wiley \& Sons.

\bibitem[\protect\citeauthoryear{Shapiro et al.}{2023}]{SZL23}
	Shapiro, A., Zhou, E. and Lin, Y. (2023). Bayesian distributionally robust optimization. \emph{SIAM Journal on Optimization}, \textbf{33}(2), 1279--1304.

\bibitem[\protect\citeauthoryear{Van Oosten and Wang}{2025}]{VW25}
Van Oosten, Z. and Wang, R. (2025). \emph{Aggregate then evaluate.} arXiv preprint arXiv:2512.03396.

\bibitem[\protect\citeauthoryear{von Neumann and Morgenstern}{1947}]{vNM47}
von Neumann, J. and Morgenstern, O. (1947). \emph{Theory of Games and Economic Behavior}. Princeton University Press, Second Edition.

\bibitem[\protect\citeauthoryear{Wang et al.}{Wang et al.}{2020}]{WWW20}
Wang, R., Wei, Y. and Willmot, G. E. (2020). Characterization, robustness and aggregation of signed Choquet integrals. \emph{Mathematics of Operations Research}, \textbf{45}(3), 993--1015.

\bibitem[\protect\citeauthoryear{Wu et al.}{2024}]{WMH24}
Wu, Q., Mao, T. and Hu, T. (2024). Generalized optimized certainty equivalent with applications in the rank-dependent utility model. \emph{SIAM Journal on Financial Mathematics}, \textbf{15}(1), 255--294.

\bibitem[\protect\citeauthoryear{Wu et al.}{2018}]{WZZ18}
Wu, D., Zhu, H. and Zhou, E. (2018). A Bayesian risk approach to data-driven stochastic optimization: Formulations and asymptotics. \emph{SIAM Journal on Optimization}, \textbf{28}(2), 1588--1612.

\bibitem[\protect\citeauthoryear{Yaari}{1987}]{Y87}
Yaari, M. E. (1987). The dual theory of choice under risk. \emph{Econometrica}, \textbf{55}(1), 95--115.

\bibitem[\protect\citeauthoryear{Zhang}{2002}]{Z02}
Zhang, J. (2002). Subjective ambiguity, expected utility and Choquet expected utility. \emph{Economic Theory}, \textbf{20}(1), 159--181.

\bibitem[\protect\citeauthoryear{Zhu and Fukushima}{2009}]{ZF09}
Zhu, S. and Fukushima, M. (2009). Worst-case conditional value-at-risk with application to robust
portfolio management. \emph{Operations Research}, \textbf{57}(5), 1155--1168.


\end{thebibliography}
}


\begin{thebibliography}{99}

\bibitem[\protect\citeauthoryear{Ahmadi-Javid}{2012}]{AJ12A}
Ahmadi-Javid, A. (2012). Entropic value-at-risk: A new coherent risk measure. \emph{Journal of Optimization Theory and Applications}, \textbf{155}(3), 1105--1123.

\bibitem[\protect\citeauthoryear{Bertsekas and Shreve}{1996}]{BS96A}
Bertsekas, D. P. and Shreve, S. E. (1996). \emph{Stochastic Optimal Control: The Discrete-Time Case}. Athena Scientific.

\bibitem[\protect\citeauthoryear{Bredon}{2013}]{Br13A}
Bredon, G. E. (2013). \emph{Topology and Geometry}. Springer Science \& Business Media.

\bibitem[\protect\citeauthoryear{Denneberg}{1994}]{D94A}
Denneberg, D. (1994). \emph{Non-Additive Measure and Integral}. Springer Science \& Business Media.

\bibitem[\protect\citeauthoryear{Dhaene et al.}{2012}]{DKLT12A}
Dhaene, J., Kukush, A., Linders, D. and Tang, Q. (2012). Remarks on quantiles and distortion risk measures. \emph{European Actuarial Journal}, \textbf{2}(2), 319--328.

\bibitem[\protect\citeauthoryear{Donsker and Varadhan}{1983}]{DV83A}
Donsker, M. D. and Varadhan, S. R. S. (1983). Asymptotic evaluation of certain Markov process expectations for large time. IV. \emph{Communications on Pure and Applied Mathematics}, \textbf{36}(2), 183--212.

\bibitem[\protect\citeauthoryear{Dudley}{2018}]{D18A}
Dudley, R. M. (2018). \emph{Real Analysis and Probability}. Chapman and Hall/CRC.

\bibitem[\protect\citeauthoryear{Dunford and Schwartz}{1988}]{DS88A}
Dunford, N. and Schwartz, J. T. (1988). \emph{Linear Operators, Part I: General Theory}. John Wiley \& Sons.

\bibitem[\protect\citeauthoryear{F\"{o}llmer and Schied}{2016}]{FS16A}
F\"ollmer, H. and Schied, A. (2016). \emph{Stochastic Finance: An Introduction in Discrete Time}, 4th ed. Walter de Gruyter, Berlin.

\bibitem[\protect\citeauthoryear{Kallenberg}{1997}]{K97A}
Kallenberg, O. (1997). \emph{Foundations of Modern Probability}. Springer, New York.

\bibitem[\protect\citeauthoryear{Kechris}{2012}]{K12A}
Kechris, A. S. (2012). \emph{Classical Descriptive Set Theory}. Springer Science \& Business Media.

\bibitem[\protect\citeauthoryear{Liebrich}{2024}]{L24A}
Liebrich, F. B. (2024). Are reference measures of law-invariant functionals unique? \emph{Insurance: Mathematics and Economics}, \textbf{118}, 129--141.

\bibitem[\protect\citeauthoryear{Marinacci and Montrucchio}{2004}]{MM04A}
Marinacci, M. and Montrucchio, L. (2004). Introduction to the mathematics of ambiguity. In \emph{Uncertainty in Economic Theory}, I. Gilboa, Ed. Routledge, New York, NY, USA, 46--107.

\bibitem[\protect\citeauthoryear{Mertens et al.}{2015}]{MSZ15A}
Mertens, J.-F., Sorin, S. and Zamir, S. (2015). \emph{Repeated Games}. Cambridge University Press.

\bibitem[\protect\citeauthoryear{Shaked and Shanthikumar}{2007}]{SS07A}
Shaked, M. and Shanthikumar, J. G. (2007). \emph{Stochastic Orders}. Springer, New York.

\bibitem[\protect\citeauthoryear{Vesely}{2017}]{V17A}
Vesely, L. (2017). Jensen's integral inequality in locally convex spaces. \emph{Annals: Series on Science of Mathematics}, \textbf{9}(2), 136--144.

\end{thebibliography}
\end{document}